\documentclass[12pt]{article}
\usepackage[latin1]{inputenc}
\usepackage{blkarray}
\usepackage{lipsum}
\usepackage{amsmath}
\usepackage{physics}
\usepackage{color}
\usepackage{amsfonts}
\usepackage[mathscr]{euscript}
\usepackage{graphicx}
\usepackage{geometry}
\usepackage{amssymb,epsfig}
\usepackage{comment}
\usepackage{tabularx}
\usepackage{bm}
\usepackage{euscript}
\usepackage[dvipsnames]{xcolor}
\usepackage{amsfonts}
\usepackage{exscale}
\usepackage{amsbsy}
\usepackage{textcomp}
\usepackage{hyperref}
\usepackage{slashed}
\usepackage{authblk}
\usepackage{tabularx}
\usepackage{euscript}
\usepackage{graphicx}
\usepackage{color}
\usepackage{exscale}
\usepackage{amsbsy}
\usepackage{textcomp}
\usepackage{hyperref}
\usepackage{doi}
\usepackage{tensor}
\usepackage{wrapfig}
\usepackage{ulem}
\usepackage[font=footnotesize,labelfont=bf]{caption}
\usepackage{ulem} 
\usepackage{makecell}
\usepackage[mathscr]{euscript}

\def\ba#1\ea{\begin{align}#1\end{align}}
\def\bg#1\eg{\begin{gather}#1\end{gather}}
\def\bm#1\em{\begin{multline}#1\end{multline}}
\def\bmd#1\emd{\begin{multlined}#1\end{multlined}}
\newcommand{\ZS}[1]{\textcolor{blue}{\it [ZS: #1]}}

\newcommand{\revision}[1]{\textcolor{red}{[\it Revised: ]}}
\newcommand{\hart}[1]{\textcolor{violet}{\it [HG: #1]}}

\newcommand{\be}{\begin{equation}}
	\newcommand{\ee}{\end{equation}}
\newcommand{\bea}{\begin{eqnarray}}
	\newcommand{\eea}{\end{eqnarray}}

\newcommand{\bs}{\boldsymbol}
\newcommand{\matleft}{\left(\begin{array}}
	\newcommand{\matright}{\end{array}\right)}
\newcommand{\sgn}{\operatorname{sgn}}

\usepackage[numbers,sort&compress]{natbib}
\def\simge{
	\mathrel{\rlap{\raise 0.511ex 
			\hbox{$>$}}{\lower 0.511ex \hbox{$\sim$}}}}

\def\simle{
	\mathrel{\rlap{\raise 0.511ex 
			\hbox{$<$}}{\lower 0.511ex \hbox{$\sim$}}}}

\makeatletter
\renewcommand\section{\@startsection {section}{1}{\z@}%
	{-3.5ex \@plus -1ex \@minus -.2ex}
	{2.3ex \@plus.2ex}%
	{\normalfont\large\bfseries}}
\renewcommand\subsection{\@startsection{subsection}{2}{\z@}%
	{-3.25ex\@plus -1ex \@minus -.2ex}%
	{1.5ex \@plus .2ex}%
	{\normalfont\bfseries}}
\renewcommand\subsubsection{\@startsection{subsubsection}{3}{\z@}%
	{-3.25ex\@plus -1ex \@minus -.2ex}%
	{1.5ex \@plus .2ex}%
	{\normalfont\itshape}}
\makeatother

\def\pplogo{\vbox{\kern-\headheight\kern -29pt
		\halign{##&##\hfil\cr&{\ppnumber}\cr\rule{0pt}{2.5ex}&\ppdate\cr}}}
\makeatletter
\def\ps@firstpage{\ps@empty \def\@oddhead{\hss\pplogo}%
	\let\@evenhead\@oddhead 
}
\hypersetup{
	unicode=false,          
	pdftoolbar=true,        
	pdfmenubar=true,        
	pdffitwindow=false,     
	pdfstartview={FitH},    
	pdftitle={ExcitonicQCP},    
	pdfauthor={},     
	pdfsubject={Subject},   
	pdfcreator={},   
	pdfproducer={}, 
	pdfkeywords={keyword1} {key2} {key3}, 
	pdfnewwindow=true,      
	colorlinks=true,       
	linkcolor=OliveGreen, 
	citecolor=NavyBlue,        
	filecolor=magneta,      
	urlcolor=cyan           
}

\numberwithin{equation}{section}

\newcommand*\samethanks[1][\value{footnote}]{\footnotemark}

\newcommand\beal{\begin{equation}\begin{aligned}}
		\newcommand\eeal{\end{aligned}\end{equation}}

\begin{document}

\normalem

\setcounter{page}0
\def\ppnumber{\vbox{\baselineskip14pt
}}

\def\ppdate{
} 
\date{}

\title{\Large\bf Excitonic quantum criticality: \\ From bilayer graphene to 
narrow Chern bands}
\author{Zhengyan Darius Shi$^1$, Hart Goldman$^{2,3}$, Zhihuan Dong$^1$, and T. Senthil$^1$}
\affil{\it\small $^1$ Department of Physics, Massachusetts Institute of Technology, Cambridge, MA 02139, USA}
\affil{\it\small $^2$ Kadanoff Center for Theoretical Physics, University of Chicago, Chicago, IL 60637, USA}
\affil{\it\small $^3$ School of Physics and Astronomy, University of Minnesota, Minneapolis, MN 55455, USA}
\maketitle\thispagestyle{firstpage}
\begin{abstract}

We study a family of \emph{excitonic quantum phase transitions} describing the evolution of a bilayer metallic state to an inter-layer coherent state where excitons condense. We argue that such transitions can be continuous and exhibit a non-Fermi liquid counterflow response ${\rho_{\mathrm{counterflow}}(\omega)\sim\omega^{2/z}}$ that directly encodes the dynamical critical exponent $z$. Our calculations are performed within a controlled expansion around $z = 2$. This physics is relevant to any system with spin, valley, or layer degrees of freedom.  We consider two contexts for excitonic quantum criticality: (1) a weakly interacting graphene bilayer, and (2) a system of two narrow,  half-filled Chern bands at zero external magnetic field, with total Chern number $C_{\mathrm{tot}}=0$, which may soon be realizable in moir\'{e} 
materials. The latter system hosts a time-reversed pair of composite Fermi liquid states, and the condensation of excitons of the composite fermions leads to an exotic exciton insulator* state with a charge neutral Fermi surface. Our work sheds new light on the physics of inter-layer coherence transitions in 2D materials.


\end{abstract}

\pagebreak
{
\hypersetup{linkcolor=black}
\tableofcontents
}
\pagebreak

\section{Introduction}



In recent years, two-dimensional Van der Waals materials such as graphene and transition metal dichalcogenides (TMD) have emerged as platforms for studying the interplay of interactions and topology in strongly correlated electronic systems. These materials carry with them unprecedented tunability, enabling access to new 
kinds of correlated phases and phase transitions. 
For example, the presence of spin and valley degrees of freedom means that such systems are capable of undergoing Stoner transitions to a range of broken symmetry phases, which have received a great deal of recent attention~\cite{Ma2021_exciton,Gu2021_dipolarexciton,Sun2022_WTe2_exciton,Zeng2023_dipolarexciton_DW,Nguyen2023_dipolarexciton_drag,He2021_IVC_tMBG,Kim2023_TTGIVC,Zhiyu2023_transformerSC,Zhiyu2023_spinvalleyorder}. In particular, the onset of itinerant ferromagnetism has led a rapidly expanding family of moir\'{e} materials, such as twisted bilayer graphene, transition metal dichalcogenide  homobilayers,  and rhombohedral graphene on hexagonal boron nitride (hBN), to    display Chern insulating behavior at integer lattice filling~\cite{Yahui2019_nearlyflat,Yahui2019_TBGhBN,Serlin2020_QAHE,Li2021_QAH_intertwinedMoire,Spanton2018_FCIVdW_finitefield,Sharpe2019_emergentFM,Chen2020_MoireFM,Wilhelm2020_FCI_CDW_TBG,Abouelkomsan2020_FCI,Bultinck2020_TBG_FM,Repellin2020_FCI,Ledwith2020_FCI,Xie2021_FCI_finitefield,foutty2023_bilayerWSe2,Xu2023_FQAH0,cai2023_FQAH1,park2023_FQAH2,lu2023_FQAH_graphene,zeng2023_FQAH_cornell,dong2023theory,zhou2023fractional,dong2023anomalous}. Even without an external magnetic field, these systems have been observed to exhibit incompressible quantum anomalous Hall (FQAH) phases at partial filling~\cite{Xu2023_FQAH0,cai2023_FQAH1,park2023_FQAH2,lu2023_FQAH_graphene,zeng2023_FQAH_cornell}, as well as a compressible state at half filling with a large Hall effect, which was proposed by one of us to be an anomalous composite Fermi liquid (CFL)~\cite{goldman2023_ACFL,dong2023_ACFL}. More recently, 
a transport measurment on tMoTe$_2$ at smaller twist angle and total filling $\nu=3$ has revealed evidence 
for a fractional quantum spin Hall state with zero electric Hall conductivity but fractionally quantized edge conductance~\cite{Kang2024_FQSH}. This experiment then suggests the existence of two spin-degenerate, narrow $C=\pm 1$ bands, which at $\nu=3$ are each half-filled. Understanding the full breadth of exotic phases in these systems is an ongoing effort in both theory and experiment.

While much of the recent progress centers around novel insulating states, metals are by far the more common phase realized at generic lattice filling. At weak interactions, the low energy physics of metals is well-described by Landau Fermi liquid theory. However, the ordinary Fermi liquid phase can be destabilized by strong interactions 
that are accessible in an increasing variety of flat-band materials. This tantalizing possibility motivates us to search for novel non-Fermi liquid states and design measurements that can probe their distinct universal properties.

\begin{figure}[t]
    \centering
    \includegraphics{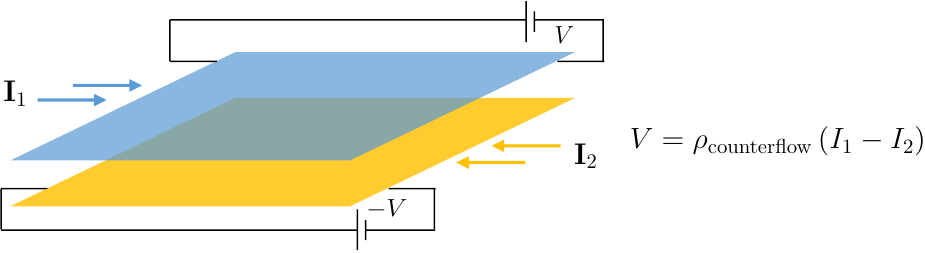}
    \caption{For any electronic system with two charged species (e.g. layer, spin), one can turn on separate voltages $V_1, V_2$ for the two species and measure the induced currents $I_1, I_2$. When $V_1 = -V_2 = V$, it is useful to define a counterflow resistivity $\rho_{\rm counterflow} = V/(I_1-I_2)$, which directly probes inter-species interactions.}
    \label{fig:counterflow}
\end{figure}

In this work, we focus on a class of \textit{itinerant ``excitonic'' quantum phase transitions}, in which particles and holes of different species -- such as spin, valley, or layer -- bind together and condense (the particular interpretation depends on the material system in question). Despite the close resemblance of this transition to XY Stoner ferromagnetism, we argue that within a particular controlled expansion this transition evades the kinds of instabilities that generically cause Hertz-Millis theories of Stoner transitions to become fluctuation-induced first-order transitions~\cite{Belitz2005_RMP,chubukov2004_FM1storder,rech2006_FM1storder,efremov2008_FM1storder,Maslov2009_FM1storder}. Our results therefore indicate that there may be a regime in which excitonic QCPs are continuous. 
Moreover, we demonstrate that these excitonic quantum critical points (QCPs) exhibit a universal, \emph{non-Fermi liquid counterflow},
\begin{align}
\label{eq:NFL_counterflow}
\rho^{xx}_{\mathrm{counterflow}}(\omega) \sim \omega^{2/z}\,,
\end{align}
where $1,2$ denote the species index, $\omega$ denotes frequency, and $z$ is the dynamical critical exponent.  This singular frequency dependence is a direct consequence of critical exciton fluctuations, and sharply contrasts with the Fermi liquid behavior away from the transition. In bilayer systems, the counterflow resistivity can be extracted by passing equal and opposite currents in two independently-contacted layers and measuring the induced voltage in one of the layers (see Fig.~\ref{fig:counterflow}). This technique has played an important role in establishing several key properties of the exciton superfluid in $\nu = 1/2 + 1/2$ quantum Hall bilayers~\cite{Ussishkin1997_CFLbilayer_drag,Lilly1998_Coulombdrag_exp,Kellogg2003_Coulombdrag_exp,Kellogg2004_counterflow_exp,Tutuc2004_counterflow_exp,Eisenstein2004_CFLexciton,Fink2011_counterflow_exp,Liu2017_graphene_drag,Liu2022_exciton_counterflow}. In other setups where the relevant degree of freedom involves spin, the counterflow conductivity corresponds to the current response to circularly polarized light~\cite{Sidler2017, Mak2018, Wang2020}. 

Leveraging the experimental advances in 2D materials, we propose two concrete setups featuring excitonic QCPs with NFL counterflow. In the first, we consider a bilayer of Fermi liquids with ${V(r)\sim1/r}$ Coulomb interactions, such as ordinary graphene doped away from the neutrality point. As the distance between the layers is tuned, the inter-layer Coulomb interaction can lead to condensation of inter-layer excitons. Signatures of this kind of exciton condensation have been observed already in graphene bilayers in the presence of a strong magnetic field, as well as in monolayer TMDs such as WTe$_2$ (which may have condensation of inter-valley excitons)~\cite{Ma2021_exciton,Gu2021_dipolarexciton,Sun2022_WTe2_exciton,Zeng2023_dipolarexciton_DW,Nguyen2023_dipolarexciton_drag,He2021_IVC_tMBG,Kim2023_TTGIVC}. We demonstrate that if the Fermi surfaces of the two layers are identical, then the transition to the exciton condensate will exhibit NFL counterflow as in Eq.~\eqref{eq:NFL_counterflow}. 

\begin{figure}[t]
    \centering
    \includegraphics[width = 0.8\textwidth]{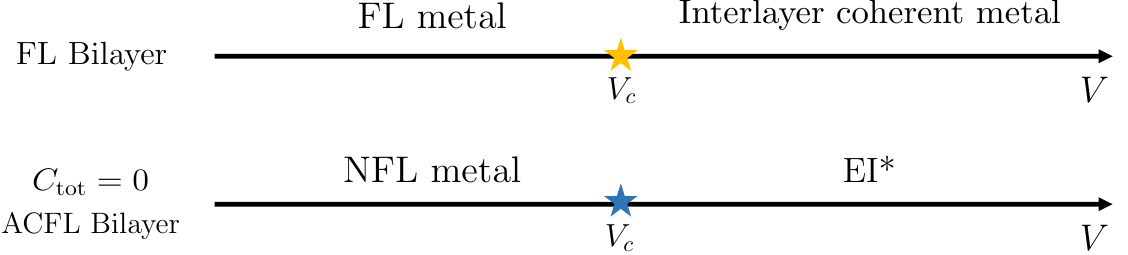}
    \caption{Schematic phase diagrams for two types of ``excitonic" QCPs in bilayer systems. In both cases, the tuning parameter $V$ is the strength of interlayer Coulomb interactions. 
    }
    \label{fig:BilayerACFL_phasediagram}
\end{figure}

Another kind of excitonic QCP can be found in a two-component FQAH system where each layer respectively consists of a half-filled ${C=\pm 1}$ band, such that the total Chern number is ${C_{\mathrm{tot}}=0}$ (see~\cite{zhang2018_oppositefieldbilayer,myersonjain2023conjugate,zhang2024_vortex} for related works on the same setup). This system can be mapped to a bilayer of a CFL with its time-reversed conjugate $\overline{\rm CFL}$. We describe how such a situation could arise in twisted TMD multilayers, although other equally valid setups are likely possible. Unlike the well-studied $\nu=1/2$ bilayers in quantum Hall systems, which are unstable to exciton condensation of the microscopic electrons and holes even for weak Coulomb interactions~\cite{Moon1995_CFLexciton,Bonesteel1996_CFLexciton,Eisenstein2004_CFLexciton,Moller2008_CFLexciton,Milovanovic2009_CFLexciton,Milovanovic2015_CFLexciton,Sodemann2017,Isobe2017}, this $C_{\mathrm{tot}}=0$ system has a stable two-component CFL phase. Therefore, it is meaningful to consider phase transitions to the menagerie of proximate ordered phases. 

One phase of particular interest is an \emph{exciton condensate of composite fermions} (as opposed to electrons). Since the phase resulting from this condensation is electrically insulating, but nevertheless has a neutral Fermi surface~\cite{zhang2018_oppositefieldbilayer,zhang2024_vortex}, we refer to it as an exciton insulator*~(EI*)\footnote{Several other neighboring phases are also discussed in Ref.~\cite{myersonjain2023conjugate}. There, a particle-hole transformation is applied to the composite fermion in one of the two Chern bands. As a result, the exciton condensate in Ref.~\cite{myersonjain2023conjugate} is in fact a superconductor rather than a gapless insulator.}. This state can be realized in a system with spin/valley-contrasting Chern bands. It has been proposed as a candidate ground state for the FQSH phase observed in Ref.~\cite{Kang2024_FQSH}, albeit under the different name of vortex spin liquid~\cite{zhang2024_vortex}. If symmetry ensures that the CFL Fermi surfaces match, then across the bilayer CFL -- EI* transition, the system again exhibits NFL counterflow as in Eq.~\eqref{eq:NFL_counterflow}. However, in contrast to the excitonic transition of ordinary electrons, the optical conductivity will have markedly different behavior due to the large Hall resistivity of each CFL, $\rho^{xy}_{\mathrm{counterflow}}\sim 4(2\pi)$ (in units of $\hbar/e^2$). Consequently, at low frequencies,
\begin{align}\label{eq:intro_mainresult}
\textrm{Fermi liquid bilayer: }\sigma^{xx}_{\mathrm{counterflow}}(\omega)&\sim\frac{1}{\rho^{xx}_{\mathrm{counterflow}}(\omega)}\sim\omega^{-2/z}\,,\\
\rm{CFL}-\overline{\rm CFL} \,\text{bilayer: }\sigma^{xx}_{\mathrm{counterflow}}(\omega)&\sim\rho^{xx}_{\mathrm{counterflow}(\omega)}\sim\omega^{+2/z} \,.
\end{align}
The frequency scaling of the optical conductivity thus gives a striking contrast between the NFL behavior at these two kinds of excitonic transitions. In principle, both transitions can even be accessed in the same system, since displacement field can tune between CFL and electronic Fermi liquid phases. Schematic phase diagrams are depicted in Figure~\ref{fig:BilayerACFL_phasediagram}.

We proceed as follows. In Section~\ref{sec:Bilayer_FL}, we study the interlayer excitonic QCP that arises in a general Fermi liquid bilayer with matching Fermi surfaces (the layer index can also be spin/valley) and argue that the mean-field second order transition survives quantum fluctuations. This discussion is followed by a 
derivation of the NFL counterflow resistivity in Eq.~\eqref{eq:NFL_counterflow}. In Section~\ref{sec:BilayerACFL}, we turn to the Coulomb-coupled $\rm CFL-\overline{\rm CFL}$ bilayer. After 
demonstrating the stability of the Coulomb-coupled $\rm CFL-\overline{\rm CFL}$ bilayer phase via a renormalization group analysis, we show how this system also exhibits NFL counterflow as it evolves to the EI* state when composite fermion excitons condense. 
In Section~\ref{sec:realizations}, we propose several potential realizations of excitonic quantum criticality in 2D Van der Waals materials. We conclude in Section~\ref{sec:discussion}.

\section{Excitonic quantum criticality of electrons}\label{sec:Bilayer_FL}

\subsection{Model and basic signatures of non-Fermi liquid physics}\label{subsec:FLbilayer_setup}

We begin with two identical, spatially parallel Fermi liquid layers in two spatial dimensions, coupled through inter-layer Coulomb interactions. Although we find it useful to regard these layers as physically separated by a distance $d$, the same universal physics can arise in systems where the layer degree of freedom is actually spin or valley, depending on the microscopic system of interest. By varying the spatial distance $d$ between the two layers, the strength of the inter-layer Coulomb interaction can be tuned, and the system can evolve from two decoupled Fermi liquid metals to an inter-layer coherent metal where inter-layer excitons condense. We will argue that such a transition can be continuous, and we will show that this transition can be characterized by NFL counterflow. 

An effective Euclidean action which describes this transition is 
\begin{equation}
    S = S_c + S_{\rm int} + S_{\phi} \,,
\end{equation}
where
\begin{equation}
    \begin{aligned}
        S_c &= \sum_{n=1}^2 \int_{\bs{k}, \omega} c^{\dagger}_n(\bs{k},\omega) \left[i\omega - \epsilon(\bs{k})\right] c_n(\bs{k}, \omega)  \,, \\
        S_{\rm int} &= \int_{\bs{k}, \bs{q}, \omega, \Omega} \frac{1}{2} g(\bs{k})\, \phi(\bs{q}, \Omega)\, c^{\dagger}_2(\bs{k} + \bs{q}/2, \omega + \Omega/2)\, c_1(\bs{k} - \bs{q}/2, \omega - \Omega/2) + \mathrm{h.c.} \,, \\
        S_{\phi} &= \int_{\bs{q}, \Omega} \phi^*(\bs{q}, \Omega) \left[\Omega^2 + |\bs{q}|^2 + M^2\right] \phi(\bs{q},\Omega) + \int_{\bs{x}, \tau} u\, |\phi(\bs{x}, \tau)|^4  \,.
    \end{aligned}
\end{equation}
In the above action, $c_1, c_2$ are spinless fermions in the two layers, and $\phi$ is a complex bosonic order parameter that couples to the exciton operator, $c^{\dagger}_1 c_2$. $\epsilon(\boldsymbol{k})$ denotes the single-particle energy with respect to the Fermi surface, and $g(\boldsymbol{k})$ and $u$ are coupling constants. Throughout this work, we use the notation ${\int_{\bs{x},\tau}=\int d^2\bs{x} d\tau}$ to denote real space integrals and ${\int_{\bs{k},\omega}=\int \frac{d^2\bs{k}d\omega}{(2\pi)^3}}$ to denote momentum space integrals. We also use boldface for spatial vectors. 
The system conserves charge separately on each layer, so the total global symmetry is the product $U(1)\times U(1)$. 

As $M^2$ is tuned across a critical value, $M_c^2$, the system evolves from a pair of weakly interacting Fermi liquids to an inter-layer coherent state where $\phi$ condenses.  In the disordered phase with $M^2>M_c^2$, the quartic self-interaction term in the boson action can be neglected. Integrating out $\phi$ then leads to short-ranged density-density interactions between the two layers that preserve the intra-layer charge conservation, $U(1)\times U(1)$. 

In the inter-layer coherent phase where the excitons condense, $M^2<M_c^2$, and the intra-layer charge conservation symmetry, ${U(1) \times U(1)}$, is spontaneously broken down to a single $U(1)$ symmetry, under which ${c_1\rightarrow e^{i\vartheta}c_1, c_2\rightarrow e^{i\vartheta}c_2}$ simultaneously. 
It is easy to see that the coupling between the Goldstone mode and the Fermi surface vanishes in the low energy limit, and the fermionic quasiparticles survive in the inter-layer coherent phase. 

Approaching the metallic quantum critical point from the disordered phase, the quartic term, $u |\phi|^4$, remains irrelevant, and the gapless fluctuations of $\phi$ lead to non-Fermi liquid self energies for the fermions in both layers. To acquire analytic control over this strongly coupled model, we deform the action by introducing two additional parameters: $\epsilon$, which describes the range of the boson-fermion interactions, and $N$, a new species index,
\begin{align}
    S_c &=\sum_{\alpha = 1}^N \sum_{i=1}^2 \int_{\bs{k}, \omega} c^{\dagger}_{i,\alpha}(\bs{k},\omega) \left[i\omega - \epsilon(\bs{k})\right] c_{i,\alpha}(\bs{k}, \omega)  \,, \nonumber\\
    S_{\rm int} &= \frac{1}{2\sqrt{N}} \sum_{\alpha=1}^N \sum_{i=1}^2 \int_{\bs{k}, \bs{q}, \omega, \Omega} g(\bs{k})\, \phi(\bs{q}, \Omega) \,c^{\dagger}_{2,\alpha}(\bs{k} + \bs{q}/2, \omega + \Omega/2) \,c_{1,\alpha}(\bs{k} - \bs{q}/2, \omega - \Omega/2) + \mathrm{h.c.} \,, \nonumber \\
    \label{eq:UVaction_FLbilayer}
    S_{\phi} &= \int_{\bs{q}, \Omega} \phi^*(\bs{q}, \Omega) \left[\Omega^2 + |\bs{q}|^{1+\epsilon}\right] \phi(\bs{q},\Omega)  \,.
\end{align}
Here $\alpha=1,\dots, N$, with $N=1$ corresponding to the original bilayer system of interest. Following Ref.~\cite{Mross2010}, we work in the double scaling limit where $\epsilon \rightarrow 0$ and $N \rightarrow \infty$ with $\epsilon N \equiv r$ fixed. A controlled expansion of critical singularities can be obtained order by order in $1/N$ (or equivalently $\epsilon$). It was recently shown by one of us that this expansion is also amenable to calculations of quantum critical transport~\cite{Shi2023_mross}. 

To leading order in the $1/N$ expansion, the boson and fermion self-energies satisfy the following self-consistent equations,
\begin{equation}
    \begin{aligned}
    \Pi_{\phi}(\bs{q}, \Omega) &= - \int_{\bs{k}, \omega} \big[g(\bs{k})\big]^2 G^{\alpha\beta}_1(\bs{k} + \bs{q}/2, \omega + \Omega/2) G^{\beta\alpha}_2(\bs{k} - \bs{q}/2, \omega - \Omega/2) \,,\\
    \Sigma_{n}^{\alpha\beta}(\bs{k}, \omega) &= \frac{1}{2N} \int_{\bs{k}, \omega} \big[g(\bs{k} - \bs{q}/2)\big]^2 G_{n}^{\alpha\beta}(\bs{k} - \bs{q}, \omega - \Omega) D_{\phi}(\bs{q}, \Omega) \,,
    \end{aligned}
\end{equation}
where $G^{\alpha\beta}_n, D_{\phi}$, with $n=1,2$ a layer index, are the fermion and boson Green's functions and $\Sigma^{\alpha\beta}_n, \Pi_{\phi}$ are the corresponding self-energies. Repeated indices are taken to be summed over. Due to the layer exchange symmetry, $\Sigma_1 = \Sigma_2$ and $G_1 = G_2$. Therefore, the above equations reduce to the Eliashberg equations for a single layer. The Eliashberg solutions are well-known,
\begin{align}
    \Sigma^{\alpha\beta}_n(\bs{k}, i\omega) &= \Sigma(i\omega)\,\delta^{\alpha\beta} = i C(g, r) \sgn(\omega) |\omega|^{\frac{2}{2+\epsilon}}\,\delta^{\alpha\beta} \,, \\ \Pi_{\phi}(\bs{q}, i \Omega) &= - \gamma(g) \frac{|\Omega|}{\sqrt{q^2 + c^2 |\Sigma(i\Omega)|^2}} \,.
\end{align}
Note that we have set the coupling, $g(\bs{k})\equiv g$, to be a uniform constant, since the IR singularities are not sensitive to its precise momentum dependence. 
We also introduce the smooth, real functions, $C(g,r)$ and $\gamma(g)$, the precise forms of which are unimportant.

The self-energies quoted above directly determine the critical singularities at the excitonic phase transition. When extrapolated to the physical value $\epsilon = 1$ and $N = 1$, we recover the standard non-Fermi liquid features of Hertz-Millis models: (1) A sharp Lorentzian momentum distribution curve (MDC) and a $\omega^{2/3}$ broadened energy distribution curve (EDC) in angle-resolved photoemission spectroscopy (ARPES); (2) A fermionic contribution to the specific heat which scales as $C_{\rm fermion} \sim T^{2/3}$; (3) Scale-invariant layer pseudospin-pseudospin correlation function with dynamical critical exponent $z = 3$.


\subsection{On the continuity of the excitonic phase transition}\label{subsec:FL_bilayer_continuity} 


Up to this point, we have implicitly assumed that the excitonic quantum phase transition is continuous. Within the Hertz-Millis framework, this assumption is correct at the mean-field level, but it can in principle be violated by quantum fluctuations. Even in the disordered phase where the order parameter field is gapped, the gapless fluctuations on the Fermi surface can generate singularities in the bosonic sector through the Yukawa interaction. We now examine these effects for a large class of Hertz-Millis models \emph{within the double expansion introduced in Section~\ref{subsec:FLbilayer_setup}}, and we demonstrate that within this framework the mean-field continuous transitions survive the leading nontrivial fluctuation corrections. 

The excitonic phase transition considered in this work can be embedded in a broader class of itinerant ferromagnetic transitions (i.e. Stoner transitions). To make the generalization, let us interpret the layer index as a pseudospin and rewrite the complex boson as $\phi = \phi_x + i \phi_y$. In the spin language, the interaction term can be recast as
\begin{equation}
    S_{\rm int} = g \int_{\bs{x},\tau} \sum_{i=x,y} \phi_i S^i \,, \qquad S^i = \frac{1}{2} c^{\dagger}_n \tau^i_{nm} c_m \,,
\end{equation}
where $\tau^i$, $i=x,y$ are Pauli matrices. Therefore, the excitonic transition is in the same universality class as an easy-plane $U(1)$ Stoner transition. The more general Stoner transitions correspond to replacing $U(1)$ with other subgroups of the full spin $SU(2)$ symmetry. 

What is the fate of the itinerant Stoner transition for an arbitrary subgroup $G \subset SU(2)$? Around two decades ago, this question was addressed in a series of pioneering works using an augmented version of the Eliashberg approximation~\cite{Belitz2005_RMP,chubukov2004_FM1storder,rech2006_FM1storder,efremov2008_FM1storder,Maslov2009_FM1storder}. These works reached the surprising conclusion that mean-field-continuous itinerant ferromagnetic transitions in the symmetry class $G = SU(2)$ or $G = U(1)$ are necessarily destabilized by strong quantum fluctuations. The source of the instability is a negative quantum correction to the static boson susceptibility $\Pi_{\phi}(\bs{q}, \Omega = 0) \sim - |\bs{q}|^{3/2}$ which either induces a direct first order transition or a continuous transition to a distinct, non-uniform ordered phase where $\phi(\bs{Q} \neq 0)$ condenses. After this work, Ref.~\cite{Lee2009} identified that in the deep IR limit where $\bs{q}, \Omega$ become much smaller than all energy and momentum scales in the model, higher loop diagrams can proliferate and potentially compete with the leading corrections to $\Pi_{\phi}(\bs{q}, \Omega)$ extracted from the augmented Eliashberg approximation. Even though such corrections to the Eliashberg approximation are numerically small, this motivates us to revisit the boson susceptibility calculation with an alternative controlled expansion where a small parameter is declared \emph{a priori}.



We choose to adopt the double expansion introduced in Section~\ref{subsec:FLbilayer_setup}, which involves $N$ fermion species interacting with a critical boson with bare kinetic term deformed from $|\bs{q}|^2$ to $|\bs{q}|^{1+\epsilon}$. The expansion is performed by taking $N$ to be large and $\epsilon$ small, holding their product, $N\epsilon$, fixed. 
As reviewed in Section~\ref{subsec:FLbilayer_setup}, the leading, $\mathcal{O}(N^0)$ boson self energy is given by the Eliashberg approximation and contains no singular correction in the static limit. At $\mathcal{O}(N^{-1})$, the boson self energy receives contributions from several Feynman diagrams that can be explicitly evaluated using the techniques developed in Ref.~\cite{rech2006_FM1storder} (see Appendix~\ref{app:1storder} for details). After accounting for the spin structure factors in each symmetry class, we arrive at the general result,
\begin{align}
    D^{-1}(\bs{q}, \Omega = 0) &= |\bs{q}|^{1+\epsilon} + \frac{1}{N}\, \delta \Pi(\bs{q}) + \mathcal{O}(N^{-2}) \,, 
    \end{align}
    where
    \begin{align}
    \delta\Pi(\bs{q}) &=\begin{cases}
        - A(\epsilon)\, |\bs{q}|^{1+\epsilon/2} & G = SU(2) \\ - B(\epsilon)\, |\bs{q}|^{1+\epsilon/2} & G = U(1) \\ 0 & G = Z_2/I 
    \end{cases}\,.
\end{align}
Here $A(\epsilon), B(\epsilon)$ are positive functions in the range $0 \leq \epsilon \leq 1$. 
For the $Z_2/I$ case, the corrections to the boson susceptibility vanish at this order, in agreement with the results of Ref.~\cite{rech2006_FM1storder}. 

For the isotropic ferromagnet with $G = SU(2)$ and the easy-plane ferromagnet with $G = U(1)$, we recover the results in Ref.~\cite{rech2006_FM1storder} if the parameters $\epsilon, N$ are extrapolated to 1. However, within the scope of the controlled expansion, our calculations are only accurate to $\mathcal{O}(\epsilon)$. Expanding all expressions to this order and keeping $\epsilon N\sim\mathcal{O}(1)$ fixed, we find 
\begin{align}\label{eq:Pi_final_doubleexp}
    D^{-1}(\bs{q}, \Omega = 0) &= |\bs{q}| +  \epsilon\,\left[ |\bs{q}| \log |\bs{q}| - \frac{C(0)}{N\epsilon}\, |\bs{q}|\right] + \mathcal{O}(\epsilon^2) \,,
\end{align}
where
\begin{align}
C(0)&=\begin{cases}
        A(0) & G = SU(2) \\ B(0) & G = U(1) \\ 0 & G = Z_2/I
    \end{cases}
\end{align}
is a constant determined by symmetry factors. Crucially, the $G$-dependent $\mathcal{O}(\epsilon)$ correction to the boson susceptibility is subleading relative to the term proportional to $|\bs{q}| \log |\bs{q}|$, which just comes from expanding the factor of $|\bs{q}|^{1+\epsilon}$ in the tree level  boson kinetic term. Therefore, at leading nontrivial order in the double expansion, the mean-field continuous transition survives quantum fluctuations for every choice of $G$. 

Determining the relevance of these stability results for the physical Stoner transitions requires a detailed understanding of the extrapolation from $\epsilon = 0$ to $\epsilon = 1$ and $N$ to some value depending on $G$. Although such understanding remains beyond reach, there are two conceptually distinct scenarios. The first possibility is that for a given $N$ the $\epsilon \ll 1$ limit is smoothly connected to the $\epsilon = 1$ limit, and corrections to $\delta \Pi(\bs{q}, \Omega = 0)$ that are higher order in $\epsilon$ simply renormalize the critical exponents of the transition. In this scenario, the transition is continuous even at $\epsilon=1$. 
In the second scenario, there is some critical $\epsilon_c < 1$ such that fluctuations destabilize the fixed point for $\epsilon > \epsilon_c$, and the conclusions of the earlier Eliashberg analysis would hold. It would be interesting to use a combination of higher order diagrammatic calculations and large scale numerical simulations to distinguish between these two scenarios and obtain a reliable answer at $\epsilon = 1$, but this is beyond the scope of the present work. 

\subsection{NFL counterflow: General motivation}\label{subsec:FLbilayer_nonpert}


In Sections~\ref{subsec:FLbilayer_setup} and \ref{subsec:FL_bilayer_continuity}, we established the continuity of itinerant excitonic transitions 
to leading nontrivial order in a particular controlled expansion 
and described some of its basic critical properties. Now, let us turn to transport, which is often the simplest experimental probe in 2D Van der Waals materials. Our goal is to find universal transport coefficients that (1) are intrinsic to the infrared (IR) fixed point of the theory, and (2) are amenable to controlled calculations. 
In other words, we are after transport quantities which exhibit universal behavior in the scaling limit, where operators that are irrelevant in the renormalization group (RG) sense vanish. Using general arguments, we will see in this section that one quantity naturally capable of satisfying both requirements is the counterflow conductivity. We will follow this up with a direct calculation deriving NFL counterflow in Section~\ref{subsec:FLbilayer_pert}.

In Refs.~\cite{shi2022_gifts,shi2023_loop}, it was shown that the universal transport at metallic quantum critical points can be tightly constrained by emergent symmetries and anomalies. 
At low energies, compressible metals should generally display an infinite number of conserved densities, $n_\theta$, associated with each angle $\theta$ on the Fermi surface\footnote{Although Refs.~\cite{shi2022_gifts,shi2023_loop} work with a ``mid-IR'' effective theory where the Fermi surface is decomposed into discrete patches such that charge conservation in each patch is explicit, the symmetry and anomaly constraints discussed in this section should emerge in the scaling limit even without adopting that starting point~\cite{Else2020}. We further emphasize that we \textit{do not make use of patch theories} for the diagrammatic calculations in this work.},
\begin{align}
\partial_tn_\theta+\bs{\nabla}\cdot\bs{j}_\theta&=0\,,
\end{align}
where $\bs{j}_\theta$ is the current at angle $\theta$ and we have Wick rotated back to real time, $t$. Note that $n_\theta$ may carry species indices corresponding e.g. to layer,  spin, valley, etc., but we suppress these indices for now. On coupling the Fermi surface to a set of external fields, $\{\mathcal{A}_\mu\}$, or fluctuating bosonic fields, $\{\varphi\}$, the charges at each angle $\theta$ feel an electric field, $\mathcal{E}_\theta[\{\mathcal{A}_\mu\},\{\varphi\}]$ which breaks this conservation law anomalously, in analogy with the chiral anomaly in 1d. In general,
\begin{align}
D_tn_\theta+\bs{D}\cdot\bs{j}_\theta&=\mathcal{E}_\theta[\{\mathcal{A}_\mu\},\{\varphi\}]\,.
\end{align}
Here $D_\mu$, $\mu=t,x,y$, is taken to be a covariant derivative, which is necessary if $\mathcal{A}$ and/or $\varphi$ couple to charge densities of non-Abelian symmetries, as will be the case for the excitonic quantum critical point of interest. This anomaly equation provides a non-perturbative constraint in the low energy limit relating density, current, and order parameter fluctuations that can in turn be leveraged to constrain transport.

We now consider the symmetry and anomaly constraints on transport at the excitonic QCP in Eq.~\eqref{eq:UVaction_FLbilayer}, assuming the two Fermi surfaces on each layer are identical. We will sketch the arguments here, leaving a more detailed discussion to Appendix~\ref{app:gifts}. We introduce conserved densities associated with each layer, ${n_\theta^{m}=c_m^\dagger c_m(\theta)}$, $m = 1,2$ (no summation is assumed). Expanding the dispersion near the Fermi surface, $\epsilon_\theta(\bs{k})= v_F(\theta)k_\perp+\kappa(\theta)k_\parallel^2+\dots$, where $v_F$ is the Fermi velocity and $\kappa$ is the Fermi surface curvature,  one sees immediately that the current on each layer can be decomposed into components perpendicular to and along the Fermi surface, 
\begin{align}
\bs{j}^m_\theta=(j_{\perp,\theta}^m,j_{\parallel,\theta}^m)\,,\qquad j_{\perp,\theta}^m=v_F(\theta)\,n^m_\theta\,,\qquad j_{\parallel,\theta}^m=\kappa(\theta)\, c^\dagger_{m}(\theta)(-i\partial_\parallel)c_m(\theta)\,.
\end{align}
Here $\partial_\parallel$ denotes a spatial derivative projected along the Fermi surface. We further denote the fermion currents integrated over the Fermi surface by capital letters, i.e. $\bs{J}^m=\int_\theta\bs{j}^m_\theta$, $J^m_\perp=\int_\theta j^m_{\perp,\theta}$, and $J^m_\parallel=\int_\theta j^m_{\parallel,\theta}$.


We start with the total electromagnetic (EM) response. Because the order parameter is charge neutral, the total EM density and current at angle $\theta$ are $n^{\mathrm{EM}}_\theta=n^1_\theta+n^2_\theta$ and $\bs{j}^{\mathrm{EM}}_\theta=\bs{j}_\theta^1+\bs{j}_\theta^2$. To study the EM response, we turn on a vector potential, $\bs{A}$, that minimally couples to the fermions on both layers, 
\begin{align}
\bs{A}\cdot\bs{J}^\mathrm{EM} = A_\perp(J_\perp^1+J_\perp^2)+A_{\parallel}(J_\parallel^1+J_\parallel^2)\,.
\end{align}
Because the excitonic order parameter, $\phi$, does not couple to the total charge density, $n^{\mathrm{tot}}_\theta$, it does not source an electric field felt by the EM charges at angle $\theta$. Hence, only the electric field generated by the external vector potential, $\bs{E}=-\partial_t\bs{A}$, will enter the anomaly equation. Following the arguments in Ref.~\cite{shi2022_gifts}, the anomaly equation satisfied by the EM densities and currents is
\begin{align}\label{eq:EM_anomaly}
\partial_t n^{\mathrm{EM}}_\theta+\bs{\nabla}\cdot \bs{j}^{\mathrm{EM}}_\theta = -2\,\frac{\Lambda(\theta)}{(2\pi)^2}\,\partial_tA_\perp\,,
\end{align}
where $\Lambda(\theta)= |\partial_\theta \bs{k}_F|$. For a spatially uniform vector potential, the gradient term vanishes, and this equation can be recast in frequency space as an exact expression for the optical conductivity,
\begin{align}
J^{\mathrm{EM}}_\perp(\omega)=2\Pi_0\,\frac{i}{\omega}\,E_\perp(\omega)\,,
\end{align}
where $\Pi_0=\int_\theta \Lambda(\theta)/(2\pi)^2$. There is no conductivity associated with $J_\parallel$ in the IR limit, as it can be argued to satisfy an emergent conservation law~\cite{shi2022_gifts}. Thus, the EM conductivity can be exactly fixed in the scaling limit to a Drude form,
\begin{equation}\label{eq:charge_conductivity_FLbilayer}
    \operatorname{Re}\sigma_{\rm EM}(\bs{q}= 0, \omega) = 2\pi\,\Pi_0\,\delta(\omega) \,,
\end{equation}
with no universal quantum critical corrections. Ref.~\cite{shi2022_gifts} demonstrated that this result is generic for Hertz-Millis theories, with the particular value of the Drude weight depending on symmetry. 
It can be understood intuitively as follows: The process that potentially gives rise to an incoherent conductivity is the scattering of fermions from one layer to the other. When both layers are subject to the same electric field, the incoherent scattering from layer 1 to layer 2 is balanced by the backscattering from layer 2 to layer 1. Therefore, the measured total charge conductivity is insensitive to the interlayer scattering rate. 

Although quantum critical singularities do not enter the EM conductivity, they can in fact enter \emph{counterflow}, or the response to fields of opposite sign on each layer,
\begin{align}
2(\bs{A}_z\cdot \bs{J}^{z})=\bs{A}_z\cdot(\bs{J}^1-\bs{J}^2)\,.
\end{align}
The factor of $2$ is a normalization choice that will be useful below. The first distinguishing property of counterflow is that the order parameter, $\phi$, must carry charge under $\bs{A}_z$. Therefore, the total counterflow conductivity must include a bosonic contribution
\begin{equation}
    \sigma_{\rm counterflow} = \tilde{\sigma}_{c} + \sigma_{\phi} \,,
\end{equation}
where 
\begin{equation}
    \tilde{\sigma}_{c}(\bs{q}=0, \Omega) = \frac{i}{\Omega} G_{J^1 - J^2, J^1 - J^2}(\bs{q}=0,\Omega) \,, \quad \sigma_{\phi} = \frac{i}{\Omega} G_{J_{\phi} J_{\phi}}(\bs{q}=0,\Omega) \,.
\end{equation}
Here $\bs{J}_{\phi} = \phi^* (-i\nabla)^{\epsilon} \phi$ is the bosonic current operator. 
Since the correlation function, $G_{J_{\phi} J_{\phi}}$, of the bosonic current is not constrained by any anomaly, critical singularities may enter through it. 

\begin{figure}[t]
    \centering
    \includegraphics[width = 0.8\textwidth]{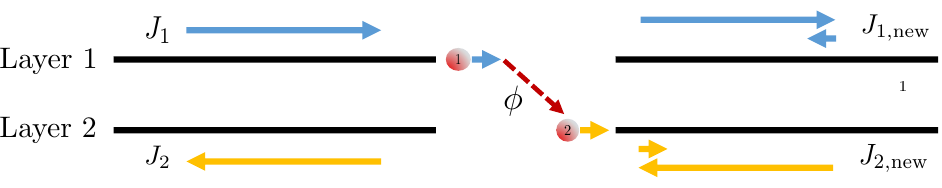}
    \caption{At the excitonic phase transition, a small-momentum boson $\phi$ annihilates a right-moving layer-1 electron and creates a right-moving layer-2 electron with nearly identical momentum, thereby reducing the counterflow current. On the right side, $\bs{J}_{1, \rm new}$ and $\bs{J}_{2, \rm new}$ are the vector sums of the two blue/yellow arrows respectively. }
    \label{fig:intuition_counterflow}
\end{figure}

Furthermore, the anomaly equations cannot fix the fermionic part of the counterflow conductivity, $\Re \tilde{\sigma}_c(\Omega)$, to be a Drude peak $\delta(\Omega)$ with no dissipation (a formal treatment is provided in Appendix~\ref{app:gifts}). Intuitively, this is because the counterflow current $\bs{J}_1 - \bs{J}_2$ does not overlap with the total conserved momentum $\bs{P}_1 + \bs{P}_2$ and can relax without any momentum bottleneck. To estimate the relaxation rate, let us turn on equal and opposite spatially uniform electric fields in the two layers $\bs{E}_1 = - \bs{E}_2$ and study the dynamics of the induced currents $\bs{J}_i$. As shown in Fig.~\ref{fig:intuition_counterflow}, the most efficient way to degrade the counterflow current $\bs{J}_1 - \bs{J}_2$ is through small-angle scattering events mediated by the critical boson $\phi$. If we neglect more complicated virtual processes, then this picture suggests that the current decay rate is directly proportional to the single-fermion decay rate. In other words, we expect the counterflow resistivity to scale with the fermion self-energy derived in Section~\ref{subsec:FLbilayer_setup}
\begin{equation}
    \rho_{\rm counterflow}(\omega) \sim \Sigma(\omega) \sim \omega^{2/z} \,.
\end{equation}
As we will see in Section~\ref{subsec:FLbilayer_pert}, this simple scaling estimate agrees with a more complete calculation that accounts for higher-order vertex corrections. 

\subsection{Calculation of NFL counterflow}\label{subsec:FLbilayer_pert}

We now present the results of a controlled calculation of the counterflow conductivity at the excitonic quantum critical point in Eq.~\eqref{eq:UVaction_FLbilayer}, leaving some technical details to Appendix~\ref{app:transport}. 
In particular, we compute the fermion conductivity matrix in layer space,
\begin{align}
    \sigma_c^{nm} = \frac{1}{i\Omega}\, G_{J^n_\perp J_\perp^m}(\Omega,\bs{q}=0)\,&,\qquad
    J_\perp^n(\Omega, \bs{q}=0)=\int_{\nu,\bs{p}} v_F(\bs{p}) \,c^\dagger_n(\Omega+\nu,\bs{p})c_n(\nu,\bs{p})\,. 
\end{align}
The matrix $\sigma_c^{nm}$ gives the the fermion current response in layer $m$ due to an electric field in layer $n$. We neglect the contribution of the current component parallel to the Fermi surfaces, $J^n_\parallel$, as it does not include any singular contributions to transport. 

Because we take the two layers to be identical, the layer-exchange symmetry swapping $c_1$ with $c_2$ ensures that $\sigma_{12} = \sigma_{21}$ and $\sigma_{11} = \sigma_{22}$. 
Combining this observation with the exact (in the scaling limit) result for the EM conductivity, Eq.~\eqref{eq:charge_conductivity_FLbilayer},  leads to a constraint relating the diagonal and off-diagonal components of $\sigma_c^{nm}$,
\begin{equation}\label{eq:total_conductivity_FLbilayer}
    \sigma^{11}_c + \sigma^{12}_c
    = \frac{i \Pi_0}{\Omega} \,.
\end{equation}
Therefore, to determine the full matrix $\sigma_c^{nm}(\bs{q}=0,\Omega)$, it suffices to compute the conductivity for layer 1 alone, $\sigma_{11}(\bs{q}=0,\Omega)$. 

Before sketching the calculation, we first state the final scaling result, 
\begin{equation}\label{eq:conductivity_FLbilayeranswer}
     \sigma_c^{11}(\bs{q}=0, \Omega) = i \frac{\Pi_0}{2\Omega} + \frac{\beta \Pi_0}{4}  (-i\Omega)^{-\frac{2}{2+\epsilon}} \,, \quad \sigma_c^{12}(\bs{q}=0, \Omega) = i \frac{\Pi_0}{2\Omega} - \frac{\beta \Pi_0}{4}  (-i\Omega)^{-\frac{2}{2+\epsilon}} \,,
\end{equation}
where $\beta$ is some real constant. 
To obtain the full conductivity matrix, this result must be combined with the order parameter conductivity matrix, $\sigma_\phi$. However, the anomaly equations fix $\phi_m$ and $n_{i,\theta}$ to have the same scaling dimension (see Appendix~\ref{app:gifts}). Hence, the scaling dimension of $J_{\phi} \propto \phi^* \nabla \phi$ is much larger than the scaling dimension of $J_\perp$, making the bosonic contribution, $G_{J_{\phi} J_{\phi}}(\bs{q} = 0, \Omega) \ll G_{J_\perp J_\perp}(\bs{q} = 0, \Omega)$ in the IR limit\footnote{This heuristic argument can be checked by a diagrammatic calculation of $G_{J_{\phi} J_{\phi}}(\bs{q} = 0, \Omega)$ in Appendix~\ref{app:bosonic_conductivity}.}, $\Omega \rightarrow 0$. 

Recalling from the previous subsection that the fermion contribution to the counterflow conductivity, $\sigma_{\mathrm{counterflow}}$, is $\tilde{\sigma}_{c} = 2 (\sigma_c^{11} - \sigma_c^{12})$, we then arrive at the main result of this section,
\begin{equation}\label{eq:main_result_FLbilayer_pert}
    \begin{aligned}
    \sigma_{\rm counterflow}(\bs{q}=0,\Omega) &= \tilde{\sigma}_{c} + \sigma_{\phi} \\
    &= \beta \Pi_0 (-i\Omega)^{-\frac{2}{2+\epsilon}} + \text{subleading} \,.
    \end{aligned}
\end{equation}
As promised, this final form of the counterflow conductivity at the excitonic QCP 
satisfies quantum critical scaling and is sensitive to the dynamical critical exponent, $z=2+\epsilon$. 

\begin{figure}[t]
    \centering
    \includegraphics[width = 0.4\textwidth]{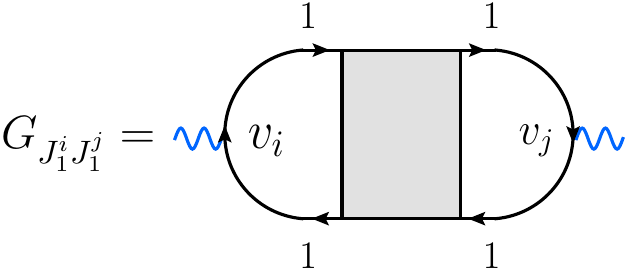}
    \includegraphics[width = 0.8\textwidth]{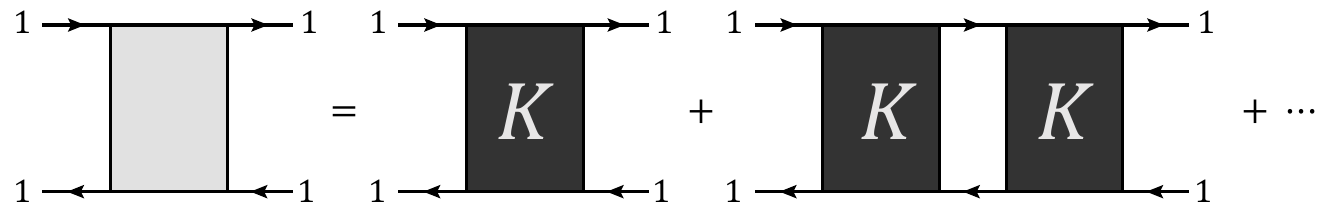}
    \caption{The structure of Feynman diagrams for the conductivity associated with fermion in layer 1 (more generally a particular species) in the excitonic QCP. The solid lines are fully dressed fermion propagators and the grey box is the fully dressed four-point vertex. The grey box can be written as a geometric sum of particle-hole irreducible diagrams with four external layer-1 fermion legs. }
    \label{fig:ph_reducible_FLbilayer}
\end{figure}
We now sketch the calculation leading to this result. 
The defining feature of Hertz-Millis models is that quantum critical interactions are mediated by Yukawa-type couplings between fermion bilinears and gapless bosons. 
In the excitonic QCP, the fermionic current-current correlation function can always be decomposed 
as the geometric sum of a particle-hole irreducible kernel $K$ as shown in Fig.~\ref{fig:ph_reducible_FLbilayer}. $K$ contains all diagrams with four external fermion legs that cannot be factorized by cutting a pair of parallel particle and hole lines in layer 1 (see Section 3 of Ref.~\cite{Shi2023_mross} for a precise definition). Each perturbative scheme gives a recipe for calculating the kernel $K$ in powers of some small parameter $1/N$:
\begin{equation}
    K = \sum_{n=0}^{\infty} N^{-n} K^{(n)} \,.
\end{equation}
In order for a perturbative expansion to be tractable, each $K^{(n)}$ should only contain a finite number of diagrams. 

\begin{figure}[t]
    \centering
    \includegraphics[width = \textwidth]{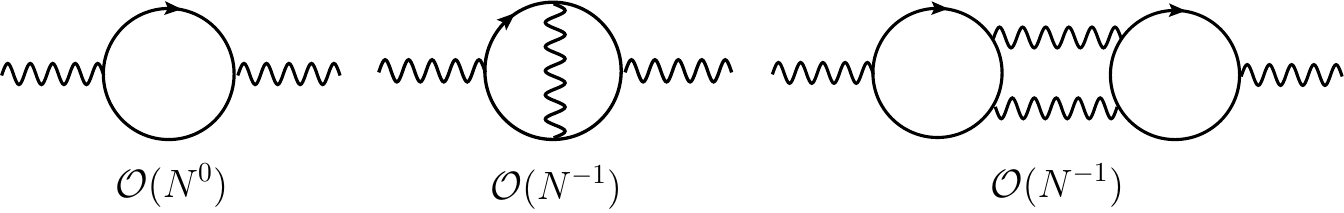}
    \caption{Examples of diagrams for $G_{J_1J_1}$: bubble diagram (left), Maki-Thompson diagram (middle), Aslamazov-Larkin diagram (right). Naive combinatorial arguments predict that a diagram with $n_L$ fermion loops and $n_V$ interaction vertices scales as $N^{k}$ with $k = n_L - n_V/2$. However, this scaling is not correct due to additional powers of $N$ that can appear in loop integrals.}
    \label{fig:combinatorial_scaling}
\end{figure}

With this general discussion in mind, we specialize to the excitonic QCP, Eq.~\eqref{eq:UVaction_FLbilayer} and work within the double expansion scheme laid out in Section~\ref{subsec:FLbilayer_setup}, wherein the critical boson $\phi$ has a deformed kinetic term $q^{1+\epsilon}$ and the fermions have $2N$ flavors with $\epsilon N = r$ fixed. Our goal is to determine the set of diagrams that can contribute to $K^{(0)}$ and then perform the geometric sum to extract $G_{J^1J^1}$ at leading order in $1/N$. Na\"{i}vely, this task appears simple because (up to differences in form factors) the grey box in Fig.~\ref{fig:ph_reducible_FLbilayer} should contain the same set of diagrams as the boson self energy in Section~\ref{subsec:FLbilayer_setup}. Within the double expansion, a Feynman diagram with $n_L$ fermion loops and $n_V$ interaction vertices comes with a combinatorial prefactor $N^k$ where $k = n_L - n_V/2$. Based on this scaling, one would conclude that the only $\mathcal{O}(N^0)$ diagram for $G_{J^1J^1}$ is a one-loop bubble diagram in Fig.~\ref{fig:combinatorial_scaling} with $n_L = 1, n_V = 2$. All other irreducible diagrams are suppressed by at least a factor of $1/N$ and can be neglected at leading order (see Fig.~\ref{fig:combinatorial_scaling} for two such examples).

The na\"{i}ve argument above is actually incorrect. The key reason is that, since the internal boson propagators in each diagram depend explicitly on $\epsilon \sim 1/N$, the internal loop integrals can carry additional factors of $N$. As shown in Ref.~\cite{Mross2010,shi2022_gifts}, for the correlation function $G_{J^1J^1}(\bs{q}, \Omega)$, these additional factors do not appear in the quantum critical regime $\Omega \ll q$, but do appear in the transport regime $q \ll \Omega$. For example, consider the Maki-Thompson diagram in Fig.~\ref{fig:combinatorial_scaling}. In the quantum critical regime, it is indeed $\mathcal{O}(N^{-1})$ suppressed. But in the transport regime, the integral over the internal boson momentum gives a factor of $1/\epsilon \sim N$, thereby enhancing the diagram to $\mathcal{O}(N^0)$. Many higher loop integrals come with higher powers of $1/\epsilon$, generating an infinite number of diagrams for the conductivity.

The surprising result from Ref.~\cite{Shi2023_mross} is that the diagrams which violate combinatorial scaling can be \textit{systematically classified}. In Appendix~\ref{app:diagram_FLbilayer}, we use this classification to write down an explicit formula for $K^{(0)}$ which only involves a small number of diagrams. Using techniques developed in Refs.~\cite{shi2023_loop,guo2022_largeN,guo2023migdal,guo2023fluctuation}, we then compute the geometric sum and arrive at a formula for $G_{J_1J_1}$ valid in the small $\Omega$ limit
\begin{equation}\label{eq:maintext_FLbilayer_JJ}
    G_{J^1 J^1}(\bs{q}=0, i \Omega) = - \frac{\Pi_0}{2} \left[1 + \frac{\beta}{2} |\Omega|^{\frac{\epsilon}{2+\epsilon}} \right] \,.
\end{equation}
Upon an analytic continuation $i\Omega \rightarrow \Omega + i 0^+$, we find the layer-1 conductivity to be
\begin{equation}\label{eq:layer1_conductivity_FLbilayer}
    \sigma^{11}_c(\bs{q}=0,\Omega) = \frac{G_{J^1J^1}(\bs{q}=0,\Omega)}{i\Omega} = \frac{i\Pi_0}{2\Omega} + \frac{\beta \Pi_0}{4} (-i\Omega)^{-\frac{2}{2+\epsilon}} \,.
\end{equation}
Combining the layer-1 conductivity in Eq.~\eqref{eq:layer1_conductivity_FLbilayer} with the total charge conductivity in Eq.~\eqref{eq:total_conductivity_FLbilayer}, we immediately infer the interlayer conductivity 
\begin{equation}
    \sigma^{12}_c(\bs{q}=0,\Omega) = \frac{i\Pi_0}{2\Omega} - \frac{\beta\Pi_0}{4} (-i\Omega)^{-\frac{2}{2+\epsilon}}\,.
\end{equation}
Therefore, the conductivity associated with the counterflow current $J_1 - J_2$ is free of Drude weight and sees only the incoherent scattering processes mediated by the critical boson,
\begin{equation}
    \sigma_{\rm counterflow}(\bs{q}=0,\Omega) = 2 \left[\sigma_{11}(\bs{q}=0,\Omega)- \sigma_{12}(\bs{q}=0,\Omega)\right] = \beta \Pi_0 (-i \Omega)^{-\frac{2}{2+\epsilon}}\,.
\end{equation}
When extrapolated to the physical parameter $\epsilon = 1$, this incoherent counterflow conductivity has the familiar $\omega^{-2/3}$ scaling. Physically, the absence of a Drude weight originates from the absence of momentum drag: a counterflow current carries equal but opposite momentum in the two layers. Hence, the relaxation of this current is not constrained by momentum conservation and is directly driven by the critical scattering between two layers mediated by the gapless boson $\phi$. 


\section{Excitonic quantum criticality of composite fermions}\label{sec:BilayerACFL}

We now proceed to study a related excitonic quantum critical point involving \emph{composite fermions}, which can arise in a system of two half-filled Chern bands of Chern numbers $C=\pm 1$, such that the total Chern number is $C_{\mathrm{tot}}=0$. Although such a system is not possible in traditional quantum Hall multilayers, we expect it to be realistic in flat Chern band systems at zero magnetic field, especially as the landscape of materials exhibiting fractional quantum anomalous Hall (FQAH) phases continues to expand.

We start with a microscopic model of half-filled, narrow Chern bands with Chern number $C = \pm 1$, where we assume the absence of any external magnetic field. If these bands sufficiently resemble Landau levels, each will support composite Fermi liquid phases~\cite{goldman2023_ACFL,dong2023_ACFL} related to one another by time-reversal conjugation, which we denote by $\text{CFL}$ (for the $C=+1$ band) and $\overline{\text{CFL}}$ (for the $C=-1$ band). 

Due to the absence of an external magnetic field, traditional flux attachment is not the appropriate framework for describing these states. Instead, the long wavelength physics of these states is best captured by a parton construction~\cite{Barkeshli2012}, which can naturally incorporate the presence of Chern bands (see the Supplemental Material of Ref.~\cite{goldman2023_ACFL} for a discussion of this approach in the context of the anomalous CFL in tMoTe$_2$). In this approach, the electron operator on each layer is written as a product, $e_n=c_n\,b_n$, $n=1,2$, where $c_n$ is a neutral composite fermion on layer $n$ and $b_n$ is a bosonic holon carrying the electric charge. Because this decomposition is redundant, the composite fermions and holons each couple with opposite charge to an emergent U$(1)$ gauge field, $a_\mu$. Because the physical electrons on each layer half-fill a narrow Chern band, the holons do as well, and they form a $\nu=\pm1/2$ bosonic fractional Chern insulator state. Integrating out the holons leads to an effective theory at long wavelengths which closely resembles the Halperin-Lee-Read (HLR) theories\footnote{Because the physically relevant Chern band systems strongly break particle-hole symmetry, there is no reason at this point to expect the emergence of Dirac composite fermions~\cite{Son2015}, despite the importance of such a description for Landau level quantum Hall systems.} of Fermi surfaces coupled to Chern-Simons gauge fields~\cite{Halperin1992}. We remark that although the parton procedure yields the correct effective theory, a microscopic construction of composite fermions projected into a Chern band is an essential open problem in the field.

With weak Coulomb interactions, we show in Section~\ref{subsec:ACFLbilayer_RGstability} that the ground state of this system features an extended NFL phase with singular single-particle Green's functions but Fermi-liquid like transport properties. This is in sharp contrast to ordinary quantum Hall bilayers of half-filled Landau levels with total filling $\nu_{\mathrm{tot}}=1$, which in the presence of Coulomb interactions is unstable to condensation of inter-layer excitons made from the microscopic electrons and holes, leading to a gapped ground state~\cite{Moon1995_CFLexciton,Bonesteel1996_CFLexciton,Eisenstein2004_CFLexciton,Moller2008_CFLexciton,Milovanovic2009_CFLexciton,Milovanovic2015_CFLexciton,Sodemann2017,Isobe2017}. As the strength of the Coulomb interactions is increased in the $C_{\mathrm{tot}}=0$ system, formation of \emph{excitons of composite fermions} becomes energetically favored. 
Upon coupling to external gauge fields in the two layers, the exciton condensate of composite fermions shows several unique experimental signatures. As originally shown in Ref.~\cite{zhang2018_oppositefieldbilayer}, the exciton condensate features a residual neutral Fermi surface, implying a vanishing linear response to the total electric field $E_+ = E_1 + E_2$ and normal quantum oscillations under the total magnetic field $B_+ = B_1 + B_2$. For these reasons, we dub this phase an exciton insulator* (EI*), although we note that earlier references have referred to the same phase as a ``vortex spin liquid''~\cite{zhang2024_vortex} or a ``composite fermion insulator''~\cite{zhang2018_oppositefieldbilayer}. Despite its metallic transport properties under the relative electric field $E_- = E_1 - E_2$, the EI* state does not show quantum oscillations under the relative magnetic field $B_- = B_1 - B_2$. The origin of these unusual properties will be reviewed in Section~\ref{subsec:ACFLbilayer_proximate}.

The key proposal of this Section is that the evolution between the bilayer NFL and EI* phases can be described by a direct, continuous excitonic quantum phase transition characterized by NFL counterflow. Unlike the electronic QCP discussed in Section~\ref{sec:Bilayer_FL}, here gapless fluctuations of the exciton order parameter $\phi$ coexist with gauge field and composite fermion fluctuations intrinsic to the CFL-$\overline{\rm CFL}$ bilayer. 
It is the interplay between these gapless fluctuations that gives rise to NFL counterflow. Note that to obtain NFL counterflow we will again assume that the two composite Fermi surfaces are identical -- this can be ensured by a judicious choice of microscopic symmetries that we will discuss in Section~\ref{sec:realizations}. 

We summarize our NFL counterflow calculation here. Due to 
the flux attachment, it is helpful to derive a generalized Ioffe-Larkin rule that relates the physical counterflow resistivity, $\rho_{\mathrm{counterflow}}$, to the 
counterflow conductivity of the composite fermions, again denoted $\tilde{\sigma}_c$, and of the excitonic order parameter, $\sigma_\phi$ (precise definitions will be given in Section~\ref{subsec:ACFLbilayer_setup}),
\begin{equation}\label{eq:counterflow_Ioffe_Larkin}
    (\rho_{\mathrm{counterflow}})_{ij} = 4(2\pi)\,\varepsilon_{ij} + (\tilde{\sigma}_c + \sigma_{\phi})_{ij}^{-1} \,,
\end{equation}
where $i,j$ are spatial indices. Remarkably, we find that despite the presence of gauge fluctuations the composite fermion conductivity, $\tilde{\sigma}_c$, is identical to 
that of the fermion contribution at the electronic exciton QCP, Eq.~\eqref{eq:conductivity_FLbilayeranswer}, up to constant numerical prefactors,
\begin{align}
\tilde{\sigma}_c(\bs{q}=0,\Omega)&\sim(-i\Omega)^{-\frac{2}{2+\epsilon}}\,.
\end{align}
Because $\sigma_{\phi}$ is subleading relative to $\sigma_c$ in the IR limit, these results together imply that in the counterflow (CF) channel, the QCP features a quantum critical incoherent resistivity
\begin{equation}\label{eq:main_result_sec3}
    (\rho_{\rm counterflow})_{xx}(\bs{q}=0, \Omega) \approx 
    \mathcal{C}(-i \Omega)^{+\frac{2}{2+\epsilon}} + \text{subleading} \,,
\end{equation}
where $\mathcal{C}$ is a constant factor. The presence of a large Hall component to the counterflow resistivity further implies that the counterflow conductivity also scales as
\begin{align}
(\sigma_{\rm counterflow})_{xx}(\bs{q}=0, \Omega) \sim (-i \Omega)^{+\frac{2}{2+\epsilon}}\,. 
\end{align}
The exponent in this expression has opposite sign from the analogous expression, Eq.~\eqref{eq:main_result_FLbilayer_pert}, for the excitonic QCP of electrons discussed in Section~\ref{subsec:FLbilayer_pert}. This implies that the NFL counterflow conductivity falls to zero as $\Omega\rightarrow 0$ for the bilayer CFL to EI* transition, while at the Fermi liquid to inter-layer coherent metal transition the NFL counterflow conductivity diverges as $\Omega\rightarrow 0$. In the rest of this Section, we will derive these results in detail.



\subsection{Bilayer composite Fermi liquid at \texorpdfstring{$C_{\mathrm{tot}}=0$}{}}\label{subsec:ACFLbilayer_setup}


At long wavelengths, we describe the single layer CFL state in a narrow $C=1$ band using a generalized Halperin-Lee-Read (HLR) action~\cite{Halperin1992,goldman2023_ACFL}, 
\begin{align}
    S_{\rm CFL} &= \int_{\bs{x},t} c^{\dagger} \left[i\partial_t + a_t + A_t - \epsilon(\bs{k} + \bs{a} + \bs{A})\right] c + \int_{\bs{x},\bs{x}',t} [\rho(\bs{x},t)-\overline\rho]\, V(\bs{x}-\bs{x}')\, [\rho(\bs{x}',t)-\overline\rho]\nonumber\\ &\quad +\int_{\bs{x},t}\left(-\frac{1}{2\pi} a db - \frac{2}{4\pi} b db-a_t\overline\rho\right) \,.
\end{align}
Here $a$ is the emergent fluctuating gauge field, and $V(r) = \frac{\alpha}{r}$ is a repulsive, long-ranged Coulomb potential. 
The auxiliary hydrodynamic gauge field $b_{\mu}$ is introduced so that all Chern-Simons terms are properly quantized. 
Throughout this Section we will use the notation, $AdB=\varepsilon_{\mu\nu\lambda}A_\mu\partial_\nu B_\lambda$. The equation of motion for $a_t$ locks the density to the emergent gauge flux as ${c^\dagger c=\overline\rho-\frac{\bs{\nabla} \times \bs{a}}{4\pi}}$, such that when the composite fermions feel no magnetic field their density is $\overline\rho$. Furthermore, $\overline\rho$ is chosen such that the composite fermions are at half filling of the Chern band, i.e. ${\overline\rho\times (\textrm{unit cell area})=1/2}$.

Time-reversal symmetry ($\mathcal{T}$) flips the sign of the Chern number to $C=-1$. Under $\mathcal{T}$, the spatial components $a_i, b_i, A_i$ reverse sign while the temporal components $a_t, b_t, A_t$ remain invariant. As a result, the action transforms to
\begin{align}
    S_{\rm \overline{CFL}} &= \int_{\bs{x},t} c^{\dagger} \left[i\partial_t + a_t + A_t - \epsilon(-\bs{k} - \bs{a} - \bs{A})\right] c + \int_{\bs{x},\bs{x}',t} [\rho(\bs{x},t)-\overline\rho]\, V(\bs{x}-\bs{x}')\, [\rho(\bs{x}',t)-\overline\rho] \nonumber\\ &\quad +\int_{\bs{x},t}\left(\frac{1}{2\pi} a db + \frac{2}{4\pi} b db-a_t\overline\rho\right) \,.
\end{align}
When the dispersion relation $\epsilon(\bs{k})$ is inversion-symmetric, the actions for CFL and $\overline{\rm CFL}$ differ only in the sign of the Chern-Simons term. 
Indeed, in 
the $\overline{\rm CFL}$ state, ${\rho = c^{\dagger} c = \overline\rho+\frac{\bs{\nabla} \times \bs{a}}{4\pi}}$. 
The major contrasting feature of this state is that positive statistical gauge flux reduces the charge density in the CFL theory, while it increases the charge density in the $\overline{\mathrm{CFL}}$ theory. 

We now combine the CFL and $\overline{\mathrm{CFL}}$ to form a $C_{\mathrm{tot}}=0$ bilayer state. As noted above, the flux attachment constraint locks density to flux, meaning that the density-density interactions can be replaced with flux-flux interactions. Consequently, the intra-layer Coulomb interactions become repulsive flux-flux interactions. On the other hand, more importantly, the inter-layer repulsive Coulomb interaction becomes an \emph{attractive} flux-flux interaction. In real and momentum space, these interactions take the form
\begin{equation}
    V_{\rm intra}(\bs{r}) = \frac{4\pi}{g^2 |\bs{r}|} \,, \quad V_{\rm inter}(\bs{r}) = - 2\, \frac{4\pi}{g^2\sqrt{r^2 + d^2}} \,,
\end{equation}
\begin{equation}
    V_{\rm intra}(\bs{q}) = \frac{8 \pi^2}{g^2 |\bs{q}|} \,, \quad V_{\rm inter}(\bs{q}) = - 2\, \frac{8 \pi^2}{g^2|\bs{q}|} e^{-|\bs{q}| d} \,. 
\end{equation}
Here we have assumed that the CFL and $\overline{\mathrm{CFL}}$ are physically separated by a distance $d$, although this is not strictly necessary since they could equally well correspond to e.g. different valleys in a material. Using these momentum space interactions, we can simplify the effective action for the coupled bilayer to
\begin{equation}
    S = S_{\rm CFL}[c_1, a_1, b_1, A_1] + S_{\rm \overline{CFL}}[c_2, a_2, b_2, A_2] + S_{\rm int}[a_1, a_2]  \,,
\end{equation}
where
\begin{align}
    S_{\rm CFL}[c_1, a_1, b_1, A_1] &= \int_{\bs{x},t} c^{\dagger}_1 \left[i \partial_t + a_{1,t} + A_{1,t} - \epsilon(\bs{k} + \bs{a}_1 + \bs{A}_1)\right] c_1 \nonumber\\ &\quad+ \int_{\bs{q},\Omega}\frac{|\bs{q}|}{2g^2} \left|a_1^T(\bs{q}, \Omega)\right|^2 + \int_{\bs{x},t}\left(-\frac{1}{2\pi} a_1 db_1 - \frac{2}{4\pi} b_1 db_1\right)   \,, \\
    S_{\rm \overline{CFL}}[c_2, a_2, b_2, A_2] &= \int_{\bs{x},t} c^{\dagger}_2 \left[i \partial_t + a_{2,t} + A_{2,t} - \epsilon(\bs{k} + \bs{a}_2 + \bs{A}_2)\right] c_2 \nonumber\\&\quad+ \int_{\bs{q},\Omega}\frac{|\bs{q}|}{2g^2}\left|a_2^T(\bs{q}, \Omega)\right|^2 + \int_{\bs{x},t}\left(\frac{1}{2\pi} a_2 db_2 + \frac{2}{4\pi} b_2 db_2\right)  \,,\\
    S_{\rm int}[a_1,a_2] &= - \frac{1}{2g^2} \int_{\bs{q},\Omega} \left[a_1^T(\bs{q}, \Omega) a_2^T(-\bs{q}, - \Omega) + a_2^T(\bs{q}, \Omega) a_1^T(-\bs{q}, - \Omega)\right] |\bs{q}| e^{-|\bs{q}| d} \,.
\end{align}
Note that we have dropped the terms fixing the density to $\overline\rho$, equivalently opting to fix the scalar potentials, $A_{1,t}$ and $A_{2,t}$, such that ${\langle c_1^\dagger c_1\rangle = \langle c_2^\dagger c_2\rangle = \overline\rho}$. 

Now the density-density interactions have become quadratic kinetic terms for the transverse fluctuations, $a^T_i(\bs{q}) = \varepsilon^{ij} \hat q_i a_j$, of the emergent statistical gauge fields. 
Via a change of basis, $a_{\pm} = a_1 \pm a_2$, we can then diagonalize $S_{\rm int}$ as 
\begin{equation}
    S_{\rm int}[a_+, a_-] = \frac{1}{2g^2} \int_{\bs{q},\Omega} \Big[\left(- |\bs{q}| e^{-|\bs{q}| d}\right) |a^T_+(\bs{q}, \Omega)|^2 + \frac{1}{2g^2} \left(|\bs{q}| e^{-|\bs{q}| d} \right) |a^T_-(\bs{q}, \Omega)|^2\Big] \,. 
\end{equation}
If we perform a similar transformation on the hydrodynamic gauge fields $b_{\pm} = b_1 \pm b_2$, then the Chern-Simons terms become purely off-diagonal in $\pm$ space. The total quadratic action for the gauge fields then takes the form
\begin{align}
    S_{\rm gauge}
    &= \frac{1}{2g^2} \int_{\bs{q},\Omega} \Big[ \left(|\bs{q}|- |\bs{q}| e^{-|\bs{q}| d}\right) |a^T_+(\bs{q}, \Omega)|^2 + \frac{1}{2g^2} \left(|\bs{q}| + |\bs{q}| e^{-|\bs{q}| d} \right) |a^T_-(\bs{q}, \Omega)|^2\Big] \nonumber\\ &\quad+ \int_{\bs{x},t} \left(- \frac{2}{2\pi} a_+ d b_- + \frac{2}{2\pi} a_- d b_+ - \frac{4}{2\pi} b_+ d b_- \right)  \,. 
\end{align}
In the long wavelength limit, the exponential factors can be expanded order by order in $|\bs{q}|$. Keeping only the leading term (which controls the IR singularities) and absorbing some 
constant factors into the definitions of $g_{\pm}$, the quadratic action further simplifies to
\begin{equation}\label{eq:Sgauge}
    \begin{aligned}
    S_{\rm gauge} &= \frac{1}{2g_+^2} \int |\bs{q}|^2 \cdot |a_+^T(\bs{q}, \Omega)|^2 + \frac{1}{2g_-^2} \int |\bs{q}| \cdot |a_-^T(\bs{q}, \Omega)|^2 \\
    &\quad+ \int_{\bs{x},t} \left(- \frac{2}{2\pi} a_+ d b_- + \frac{2}{2\pi} a_- d b_+ - \frac{4}{2\pi} b_+ d b_- \right)\,.
    \end{aligned}
\end{equation}

The key feature of this action is the asymmetry between $a_+$ and $a_-$. The emergent short-ranged power-law interactions mediated by $a_+$ generate a NFL self energy for the fermions, 
\begin{align}
\Sigma_{n}(\omega) \sim (i\omega)^{2/3}\,,\qquad n=1,2\,,
\end{align}
which is much more singular than the marginal Fermi liquid self energy $\Sigma(\omega) = \omega \log \frac{E_F}{\omega} + i \omega$ of a single CFL layer. Hence, the $C_{\mathrm{tot}}=0$ bilayer with Coulomb interactions is in a distinct NFL phase from two separate, decoupled CFL layers. Given this basic description of the phase, we can begin to study its stability, proximate phases and transport properties.

\subsection{Stability of the \texorpdfstring{$\rm CFL-\overline{CFL}$}{} bilayer phase}\label{subsec:ACFLbilayer_RGstability}

To demonstrate the stability of the $C_{\mathrm{tot}}=0$ bilayer phase, we generalize the $\epsilon$-expansion developed by Nayak and Wilczek to allow for two species of emergent gauge fields~\cite{Nayak1994}. 
Since only the $a_+$ gauge field mediates short-range interactions, we will replace its kinetic term $|\bs{q}|^2$ with $|\bs{q}|^{1+\epsilon_+}$ while keeping the kinetic term of $a_-$ as $|\bs{q}|$. We perform a perturbative RG calculation keeping $0 < \epsilon_+ \ll 1$ throughout, only extrapolating to the physical scenario $\epsilon_+ = 1$ in the end. 

In analogy with Ref.~\cite{Nayak1994}, we derive one-loop RG equations for the dimensionless running fine structure constants, $\alpha_{\pm} = \frac{g_{\pm}^2 v_F}{4\pi^2}$, 
\begin{align}
    \frac{d\alpha_+}{dl} &= \alpha_+ \left[\epsilon_+ - 2 (\alpha_+ + \alpha_-) \right] \,,\\
    \frac{d\alpha_-}{dl} &= - 2 \alpha_- (\alpha_+ + \alpha_-) \,,
\end{align}
where $l$ is a running length scale. The detailed derivation of these equations is presented in Appendix~\ref{app:RG}. These coupled RG equations have two solutions,
\begin{enumerate}
    \item $\alpha_+ = \alpha_- = 0$ : A Gaussian fixed point with two unstable directions. 
    \item  $\alpha_+ = \frac{\epsilon_+}{2}, \alpha_- = 0$ : A fully IR-stable fixed point, corresponding to a NFL phase. The theory will naturally flow to this solution at small $\alpha_+, \alpha_-$. 
\end{enumerate}
This simple RG analysis shows that in the absence of additional interactions, the NFL phase that we have identified is stable. 
Moreover, near the fixed point, the tree-level marginal coupling $\alpha_-$ picks up an anomalous dimension $-\epsilon$ and becomes irrelevant. This means that $a_-$ decouples at low energy, a simplification that will be used in the transport calculation in Section~\ref{subsec:ACFLbilayer_pert}.

In order to truly demonstrate the stability of this bilayer NFL fixed point, we must consider weak BCS pairing of composite fermions, both within each layer and between them, 
\begin{equation}
    H_{\rm intra, BCS} = \sum_i \int_{\bs{k}, \bs{k}'} V_{\rm intra}(\bs{k} - \bs{k}') \,c_{i}^{\dagger}(\bs{k})\,c_{i}^{\dagger}(-\bs{k})\, c_{i}(\bs{k}') \,c_{i}(-\bs{k}') \,,
\end{equation}
\begin{equation}
    H_{\rm inter, BCS} = \sum_i \int_{\bs{k}, \bs{k}'} V_{\rm inter}(\bs{k} - \bs{k}') \,c_{1}^{\dagger}(\bs{k})\,c_{2}^{\dagger}(-\bs{k})\, c_{1}(\bs{k}') \,c_{2}(-\bs{k}') \,.
\end{equation}
In quantum Hall bilayers with $\nu_{\mathrm{tot}}=1$, the classic inter-layer exciton condensation of electrons and holes can be explained in terms of inter-layer pairing of composite fermions, which can be demonstrated in that case to be relevant in the RG sense~\cite{Moon1995_CFLexciton,Bonesteel1996_CFLexciton,Eisenstein2004_CFLexciton,Moller2008_CFLexciton,Milovanovic2009_CFLexciton,Milovanovic2015_CFLexciton,Sodemann2017,Isobe2017}.  
However, this is not necessarily the case for the $C_{\mathrm{tot}}=0$ CFL bilayer, as repulsive interactions with gapless order parameters and gauge fields 
can have strong effects and lead to metallic fixed points with finite BCS interactions~\cite{Metlitski2014,Raghu2015}. 

Generalizing the RG analysis of Ref.~\cite{Sodemann2017}, we find that the RG equations for pairing in the 
$C_{\mathrm{tot}}=0$ bilayer gives the one-loop RG flow,
\begin{equation}
    \begin{aligned}
    \frac{d V_{\rm intra}}{dl} &= - V_{\rm intra}^2 + \alpha_+ + \alpha_- \,, \\
    \frac{d V_{\rm inter}}{dl} &= - V_{\rm inter}^2 + \alpha_+ - \alpha_-  \,.
    \end{aligned}
\end{equation}
These equations can be understood intuitively. Indeed, $a_+$ couples with the same sign to the currents in the two layers, while $a_-$ couples with opposite signs. As a result, for antipodal fermions on the same layer, both $a_+,a_-$ mediate a repulsive interaction. For antipodal fermions on opposite layers, $a_+$ continues to mediate a repulsive interaction, but $a_-$ mediates an attractive interaction. This explains why a finite $\alpha_-$ enhances the interlayer BCS instability. 
However, the stable $C_{\mathrm{tot}}=0$ CFL bilayer fixed point has $\alpha_-=0$. Hence, the RG equations for $V_{\rm intra}$ and $V_{\rm inter}$ about this NFL fixed point become identical, and both BCS couplings flow to a nontrivial fixed-point value, 
\begin{equation}
    V_{\rm intra,*} = V_{\rm inter,*} = \sqrt{\alpha_{+,*}} = \sqrt{\frac{\epsilon_+}{2}} \,.
\end{equation}
Therefore, we reach the surprising conclusion that the coupled $\rm CFL-\overline{CFL}$ bilayer system realizes a stable NFL phase. The presence of finite BCS couplings in the IR fixed point does not leave a direct imprint on 
the leading singularities in transport, and will be neglected for the rest of the paper. 

\subsection{Evolution from a \texorpdfstring{$\rm CFL-\overline{\rm CFL}$}{} bilayer to an exciton insulator*}\label{subsec:ACFLbilayer_proximate}

Having established that the $C_{\mathrm{tot}}=0$ CFL bilayer supports a stable NFL phase for weak Coulomb interactions, we can now consider the possibility of continuous quantum phase transitions out of this state. One natural possibility is that, as Coulomb interactions are turned up, inter-layer excitons of composite fermions, described by the operator $c^\dagger_1 c_2$, can form and condense. Such condensation is the fractionalized analogue of the excitonic transition studied in Section~\ref{sec:Bilayer_FL}, and the resulting phase has a radically different character from its electronic counterpart: it is an insulating phase with a gapless, electrically neutral Fermi surface, or exciton insulator* (EI*).

Since $c_1$ and $c_2$ are respectively charged under $a_1+A_1$ and $a_2 + A_2$, 
the inter-layer exciton, $c_1^\dagger c_2$, is charged under $(a_1+A_1)-(a_2+A_2)$. The composite fermion exciton condensate Higgses this combination of gauge fields, pinning $a_1-a_2=A_2-A_1$ at low energies but leaving the composite fermions gapless. After integrating out $b_{\pm}$, the effective action involving $a_+$ and $A_\pm$ takes the form 
\begin{equation}
    \begin{aligned}
    S &= 
    \sum_{n=1,2}\int_{\bs{x},t} c_n^\dagger\left[i\partial_t+ a_{+,t} + A_{+,t}-\epsilon\left(\bs{k}+ \bs{a}_+ +\bs{A}_+\right)\right]c_n\\
    &\quad+ \frac{1}{2g_+^2} \int_{\bs{q},\Omega} |\bs{q}|^2 \cdot |a^T_+(\bs{q},\Omega)|^2 + \frac{1}{2g_-^2} \int_{\bs{q},\Omega} |\bs{q}| \cdot |A^T_-(\bs{q},\Omega)|^2 - \int_{\bs{x},t} \frac{1}{2\pi} A_- d a_+ \,.
    \end{aligned}
\end{equation}
The action above implies several exotic properties of the EI* state in response to the external gauge fields $A_{\pm}$. Microscopically, $A_+$ couples to the EM current, while $A_-$ couples to the counterflow current. If $A_- = 0$, the dependence of the action on $A_+$ can be eliminated up to a gradient term by shifting the fluctuating gauge field $a_+ \rightarrow a_+ - A_+$. This means that as $\bs{q}\rightarrow0$ the DC conductivity matrix will vanish, meaning that this state is an EM insulator. 
On the other hand, when $A_+ = 0$ but $A_-\neq 0$, the spatial component $A_-^i$ 
remains and couples to the emergent gauge flux, which can conduct. Therefore, this phase is an insulating state with an exotic neutral Fermi surface 
exhibiting metallic counterflow! However, despite its metallic response under $\bs{E}_- = - \partial_t \bs{A}_-$, the EI* phase does not exhibit quantum oscillations under a relative external magnetic field $B_- = \nabla \times \bs{A}_-$ since $B_-$ only shifts the density of the composite fermions.
We note that Ref.~\cite{zhang2018_oppositefieldbilayer,yahui2020_bibilayer} considered the same state, albeit under the name of ``composite fermion insulator.'' 

The quantum phase transition between the $\rm CFL-\overline{\rm CFL}$ bilayer and the EI* state is an excitonic quantum critical point capable of exhibiting NFL counterflow. The effective theory describing 
this phase transition is identical to the bilayer exciton QCP, except for the addition of two emergent gauge fields with Chern-Simons terms. The full action therefore takes the form
\begin{align}
    S &= \sum_{n=1,2}\int_{\bs{x},t} c_n^\dagger\Big[i\partial_t+a_{n,t}+A_{n,t}-\epsilon\left(\bs{k}+\bs{a}_n+\bs{A}_n\right)\Big]c_n \nonumber\\
    &\quad+ \int_{\bs{q},\Omega} \frac{|\bs{q}|^{2}}{2 g_+^2}\, |a_+(\bs{q})|^2 + \int_{\bs{q},\Omega}\frac{|\bs{q}|}{2 g_-^2}\, |a_-(\bs{q})|^2 - \int_{\bs{x},t} \left( \frac{2}{2\pi} a_+ d b_- + \frac{2}{2\pi} a_- d b_+ + \frac{4}{2\pi} b_+ d b_- \right) \nonumber \\
    \label{eq:fullaction_QCP}
    &\quad+ \int_{\bs{x},t} [\phi\, c_1^\dagger c_2+\mathrm{h.c.}] + \int \frac{1}{2g_{\phi}^2}\, \phi^*(\bs{r}, t) \left(-i \nabla + \bs{a}_-\right)^{2} \phi(\bs{r}, t)  \,,
\end{align}
where we remind the reader that $a_{\pm}= a_1 \pm a_2$ and $b_{\pm} = b_1 \pm b_2$. Since $a_+, a_-$ and $\phi$ all mediate power-law interactions, the fermion self-energy now includes singular corrections from a total of three gapless bosonic modes! Nevertheless, a slight generalization of the analysis in Appendix~\ref{app:1storder} shows that a continuous transition survives in the presence of these additional gapless fluctuations. 

To analyze the gapless fluctuations, we follow the double expansion scheme used in Section~\ref{sec:Bilayer_FL}, setting $\epsilon_- = 0$ with $\epsilon_+ = \epsilon_{\phi} \equiv \epsilon$. The physical limit is approached as $\epsilon \rightarrow 1$. To leading order in the double expansion, the fermion and boson self energies are determined by the self-consistent Eliashberg equations,
\begin{equation}
    \begin{aligned}
    \Sigma_{n}(\bs{k}, i\omega) &= \frac{g_+^2}{2N} \int_{\bs{q}, \Omega} D_{a_+}(\bs{q}, i\Omega) \left[G_1(\bs{k}-\bs{q}, i\omega-i\Omega) + G_2(\bs{k}-\bs{q}, i\omega-i\Omega)\right] \\
    &\hspace{0.5cm} + \frac{g_-^2}{2N} \int_{\bs{q}, \Omega} D_{a_-}(\bs{q}, i\Omega) \left[G_1(\bs{k}-\bs{q}, i\omega-i\Omega) + G_2(\bs{k}-\bs{q}, i\omega-i\Omega)\right] \\
    &\hspace{0.5cm} + \frac{g_{\phi}^2}{2N} \int_{\bs{q}, \Omega} \sum_{m=1,2} s_{n,m}\, D_{\phi}(\bs{q}, i\Omega) G_{m}(\bs{k}-\bs{q}, i\omega-i\Omega)\,, \\
    \Pi_{\phi}(\bs{q}, i\Omega) &= - g_{\phi}^2 \int_{\bs{k}, \omega} G_1(\bs{k}+\bs{q}/2, i \omega + i \Omega/2) G_2(\bs{k}-\bs{q}/2, i \omega - i \Omega/2) \,,\\
    \Pi_{a_{\pm}}(\bs{q}, i\Omega) &= - \frac{g_{\pm}^2}{2} \sum_{i=1}^2 \int_{\bs{k}, \omega} G_i(\bs{k}+\bs{q}/2, i \omega + i \Omega/2) G_i(\bs{k}-\bs{q}/2, i \omega - i \Omega/2) \,.
    \end{aligned}
\end{equation}
Here we have defined $s_{n,m}=0$ for $n=m$ and $s_{n,m}=1$ for $n\neq m$. Due to the layer-exchange symmetry, we identify $\Sigma_1 = \Sigma_2 \equiv \Sigma$ and $G_1 = G_2 \equiv G$. Therefore, the above equations simplify to
\begin{equation}
    \begin{aligned}
        \Sigma(\bs{k}, i\omega) &= \frac{g_+^2}{N} \int_{\bs{q}, \Omega} D_{a_+}(\bs{q}, i\Omega) G(\bs{k}-\bs{q}, i\omega-i\Omega) + \frac{g_-^2}{N} \int_{\bs{q}, \Omega} D_{a_-}(\bs{q}, i\Omega) G(\bs{k}-\bs{q}, i\omega-i\Omega) \\
        &\hspace{0.5cm}+ \frac{g_{\phi}^2}{2N} \int_{\bs{q}, \Omega} D_{\phi}(\bs{q}, i\Omega) G(\bs{k}-\bs{q}, i\omega-i\Omega) \,,\\
        \Pi_{\phi}(\bs{q}, i\Omega) &= - g_{\phi}^2 \int_{\bs{k}, \omega} G(\bs{k}+\bs{q}/2, i \omega + i \Omega/2) G(\bs{k}-\bs{q}/2, i \omega - i \Omega/2) \,, \\
        \Pi_{a_{\pm}}(\bs{q}, i\Omega) &= - g_{\pm}^2 \int_{\bs{k}, \omega} G(\bs{k}+\bs{q}/2, i \omega + i \Omega/2) G(\bs{k}-\bs{q}/2, i \omega - i \Omega/2) \,.
    \end{aligned}
\end{equation}
The solution of these equations has a simple structure. The fermion self-energy is a sum of NFL singularities induced by $a_+, \phi$ fluctuations and marginal Fermi liquid (MFL) singularities induced by $a_-$ 
\begin{equation}
    \Sigma(\bs{k}, i\omega) = i \left[C_{a_+} + C_{\phi}\right] \sgn(\omega) |\omega|^{\frac{2}{2+\epsilon}} + i C_{a_-} |\omega| \log \frac{E_F}{|\omega|} \,,
\end{equation}
where $C_{a_+}, C_{a_-}, C_{\phi}$ are real constants that depend on $g_+, g_-, g_{\phi}$. The boson and gauge field self energies follow the same Landau damping form with different numerical prefactors
\begin{equation}
    \Pi_{a_+, a_-, \phi}(\bs{q}, i\Omega) = -\gamma_{a_+, a_-, \phi} \frac{|\Omega|}{\sqrt{q^2 + c^2 |\Sigma(i\Omega)|^2}} \,,
\end{equation}
where $c$ is a universal constant and $\gamma_{a_+}, \gamma_{a_-}, \gamma_{\phi}$ are real positive constants depending on $g_+, g_-, g_{\phi}$. The mixture of NFL and MFL scalings in the fermion spectral function induces a fermionic contribution to the specific heat,
\begin{equation}
    C_{V, \,\rm fermion}(T) \sim c_0 \,T^{2/(2+\epsilon)} + c_1\, T \log (E_F/T) \,.
\end{equation}

\subsection{NFL counterflow of composite fermions}\label{subsec:ACFLbilayer_pert}


We are now in a position to study charge and counterflow transport at the bilayer CFL -- EI* critical point. The structure of the transport analysis will parallel the analysis of the Fermi liquid bilayer in Section~\ref{subsec:FLbilayer_nonpert} and Section~\ref{subsec:FLbilayer_pert}, with additional subtleties arising from the gauge fluctuations. Despite these subtleties, we will arrive at identical scaling results:
\begin{equation}
    \sigma_{\rm charge}(\bs{q}=0, \Omega) = \frac{2i \Pi_0}{\Omega} \,, \quad \rho_{\rm counterflow}(\bs{q}=0, \Omega) \sim (-i \Omega)^{\frac{2}{2+\epsilon}} \,. 
\end{equation}

Let us define the matrices, $\sigma_c, \sigma_\phi$ to be the conductivities of the composite fermions and the excitonic order parameter field in both layer and real space. These $4\times4$ matrices in layer and real space are respectively the linear response of $c_n$ to $a_n+A_n$ and the linear response of $\phi$ to $a_- + A_-$.  
In Appendix~\ref{app:ioffe_larkin}, we derive an Ioffe-Larkin composition rule relating these responses to the physical resistivity, $\rho=\sigma^{-1}$, or the measured response of the bilayer system to separate potentials, $A_1$ and $A_2$, on each layer,
\begin{equation}
    \rho = \rho_{\mathrm{CS}} + (\sigma_c + \sigma_{\phi})^{-1} \,,
\end{equation}
We emphasize that this relation is exact, simply being a consequence of the flux attachment constraint. Indeed, the first term is a consequence of flux attachment, endowing each layer with a $\pm2\frac{h}{e^2}$ Hall resistivity,
\begin{align}
\rho_{\mathrm{CS}}&=\matleft{cc} 2\,(2\pi)\varepsilon_{ij} & 0 \\ 0 & -2\,(2\pi) \varepsilon_{ij}\matright\,,
\end{align}
where $i,j=x,y$ are spatial indices.

We now proceed to compute $\sigma_c$ and $\sigma_\phi$. 
Starting with $\sigma_c$, it follows from anomaly arguments discussed in Appendix~\ref{app:gifts} that the full matrix $\sigma_c$ can be fixed by a perturbative evaluation of the layer-1 conductivity alone. Following Section~\ref{subsec:FLbilayer_pert}, we again work in the double expansion and deform the kinetic terms of both $\phi$ and $a_+$ from $q^2$ to $q^{1+\epsilon}$. Moreover, since $a_-$ only leads to less singular corrections ($\alpha_-$ is driven irrelevant near the NFL fixed point), we will drop $a_-$ altogether and relabel $a_+$ as $a$ to simplify the perturbative calculation. In this new notation, the layer-1 conductivity takes the form
\begin{equation}\label{eq:gauge_invariant_current_correlator}
    \sigma_{11}(\bs{q}=0, \Omega) = \frac{1}{i\Omega} G_{J^1_{c,\,\rm gauge}, \, J^1_{c,\,\rm gauge}}(\bs{q} = 0, \Omega) \,,
\end{equation}
where the gauge-invariant fermion current $J^1_c$ can be decomposed as
\begin{equation}
    \bs{J}^1_{c,\,\rm gauge} = \bs{J}^1 + \bs{a} \rho_1 \,, \quad \bs{J}^1(\bs{q}) = \int_{\bs{k}} v_F(\bs{k}) \, c^{\dagger}_1(\bs{k} + \bs{q}/2) c_1(\bs{k}-\bs{q}/2) \,. 
\end{equation}
We first evaluate the correlation functions of $\bs{J}^1$. As explained in Section~\ref{subsec:FLbilayer_pert}, $G_{J^1J^1}$ can always be expressed as the geometric series of a particle-hole irreducible kernel $K$. However, in the current setting, the inclusion of the emergent gauge field $a$ complicates the structure of $K$. A naive expectation is that the kernel is simply a sum of $K_a$ and $K_{\phi}$ where $K_a, K_{\phi}$ correspond to the leading order particle-hole irreducible kernels for a single spinless Fermi surface coupled to a $U(1)$ gauge field $a$/critical boson $\phi$. However, this expectation is incorrect. 
In Appendix~\ref{app:diagram_ACFLbilayer}, we identify the correct kernel $K$ to leading order in the $1/N$ expansion and then perform the geometric sum. The final result takes the form
\begin{equation}\label{eq:maintext_ACFLbilayer_JJ}
    G_{J^1J^1}(\bs{q}=0, i\Omega) = - \frac{\Pi_0}{2} \left[1 + \frac{\beta'}{2} |\Omega|^{\frac{\epsilon}{2+\epsilon}}\right] \,, 
\end{equation}
where $\beta'$ is a real constant. 

Next we turn to the $\bs{a}$-dependent term $G_{a \rho_1, a \rho_1}$ in Eq.~\eqref{eq:gauge_invariant_current_correlator}. Using the equations of motion for $a_0$, the fermion density $\rho_1(\bs{q})$ can be traded for the gauge flux $q a^T(\bs{q})$. Therefore, the lowest loop diagram contains a single gauge field loop with two $\bs{a}$ propagators scaling as
\begin{equation}
    \begin{aligned}
    G_{a \rho_1, \, a \rho_1}(\bs{q}=0, \Omega) &= \int_{\bs{q}', \Omega'} |\bs{q}'|^2 \, D_a(\bs{q}', i \Omega' - i \Omega/2) \, D_a(\bs{q}', i\Omega' + i \Omega/2)  \\
    &\sim \Omega^{\frac{2}{2+\epsilon} + 1} \cdot \Omega^{\frac{2}{2+\epsilon}} \cdot \Omega^{- \frac{2+2\epsilon}{2+\epsilon}} \sim \Omega^{\frac{4-\epsilon}{2+\epsilon}} \,. 
    \end{aligned}
\end{equation}
Extrapolating to $\epsilon = 1$ gives $G_{a \rho_1, \, a \rho_1}(\bs{q}=0, \Omega) \sim \Omega$, which is subleading relative to Eq.~\eqref{eq:maintext_ACFLbilayer_JJ}. 

Putting together the calculations above, and applying the analytic continuation $i\Omega \rightarrow \Omega$, we can extract the layer-resolved conductivity in the fermionic sector
\begin{equation}
    \sigma_{c,11}(\bs{q}=0,\Omega) = \frac{i\Pi_0}{2\Omega} + \frac{\beta' \Pi_0}{4} (-i\Omega)^{-\frac{2}{2+\epsilon}} \,,
\end{equation}
\begin{equation}
    \sigma_{c,12}(\bs{q}=0,\Omega) = \frac{i\Pi_0}{2\Omega} - \frac{\beta' \Pi_0}{4} (-i\Omega)^{-\frac{2}{2+\epsilon}} \,,
\end{equation}
\begin{equation}\label{eq:fermion_counterflow_cflbilayer_finalresult}
    \sigma_{c, \rm counterflow}(\bs{q}=0,\Omega) = 2 \left[\sigma_{11}(\bs{q}=0,\Omega)- \sigma_{12}(\bs{q}=0,\Omega)\right] = \beta' \Pi_0 (-i\Omega)^{-\frac{2}{2+\epsilon}} \,.
\end{equation}
In summary, the inclusion of gauge fluctuations modifies the numerical coefficient in front of the incoherent conductivity in the fermionic sector, but does not change its frequency scaling. In hindsight, this should not be surprising, since anomaly arguments imply that gauge fields alone cannot generate any critical incoherent conductivity. 

Finally, we analyze the contributions to the counterflow conductivity from the boson sector $\sigma_{\phi}$. In Appendix~\ref{app:bosonic_conductivity}, we identify the gauge invariant boson current operator $J^{\rm gauge}_{\phi}$ and show that its two-point correlator scales as
\begin{equation}
    G_{J^{\rm gauge}_{\phi}, J^{\rm gauge}_{\phi}}(\bs{q}=0, \Omega) \sim \Omega \,.
\end{equation}
This scaling result directly implies that
\begin{equation}\label{eq:boson_counterflow_cflbilayer_finalresult}
    \sigma_{\phi, \rm counterflow}(\bs{q}=0,\Omega) = \frac{1}{i\Omega} G_{J^{\rm gauge}_{\phi}, J^{\rm gauge}_{\phi}}(\bs{q}=0, \Omega) \sim \mathcal{O}(1) \ll \sigma_{c, \rm counterflow}(\bs{q} = 0, \Omega) \,. 
\end{equation}
Combining Eq.~\eqref{eq:boson_counterflow_cflbilayer_finalresult} with Eq.~\eqref{eq:fermion_counterflow_cflbilayer_finalresult} and setting $\epsilon = z - 2$, we recover the total counterflow resistivity in the longitudinal channel as promised in Eq.~\eqref{eq:main_result_sec3}
\begin{equation}
    \sigma^{-1}_{\rm counterflow}(\bs{q}=0,\Omega) \approx \rho_{\rm CS} + (\beta' \Pi_0)^{-1} (-i\Omega)^{2/z} \,. 
\end{equation}

\section{Realizations in 2D materials}\label{sec:realizations}

In this section, we propose realizations of the two different types of excitonic quantum phase transitions discussed in Section~\ref{sec:Bilayer_FL} and Section~\ref{sec:BilayerACFL}. Our focus is on situations in which microscopic symmetries ensure the matching of the Fermi surfaces associated with both layers or valleys, such that nonvanishing NFL counterflow is realized. 


\subsubsection*{Excitonic QCP of electrons: Graphene bilayers without moir\'{e}}

We start with the excitonic QCPs for electrons described in Section~\ref{sec:Bilayer_FL}. We recall that the basic physical setup for this type of transition is a parallel bilayer of identical, clean Fermi liquids coupled by interlayer Coulomb interactions. The strength of the Coulomb interactions can be tuned by varying the distance, $d$, between the two layers. At large $d$, the two Fermi liquid layers are essentially decoupled. As $d$ decreases past some critical threshold $d_c$, excitons formed by electrons and holes in opposite layers condense, giving rise to an inter-layer coherent metallic state. 

The most promising realization of this transition is perhaps in the graphene-hBN-graphene heterostructure, where inter-layer excitons have already been observed through Coulomb drag measurements at large magnetic fields~\cite{Gorbachev2012_graphene_drag,Liu2017_graphene_drag}. Due to the presence of an insulating hBN layer, direct tunneling between the two graphene layers is strongly suppressed, leaving Coulomb interactions as the dominant interlayer coupling. Moreover, advances in fabrication techniques have decreased the thickness of the hBN layer to the nanometer scale, thereby enhancing Coulomb interactions and enabling the transition into an exciton-condensed phase. 

If the graphene layers are identical and the system has a layer-exchange symmetry, their Fermi surfaces will match, and NFL counterflow will be possible as the system approaches the transition to inter-layer coherence. The counterflow resistivity can be extracted by simply measuring the resistivity on each layer individually and taking the difference.

Another system pertinent to this excitonic QCP is the rhombohedral graphene. In this setup, an out-of-plane electric field opens up a band gap near charge neutrality and creates a relatively flat band bottom for the conduction electrons. Instead of having Fermi surfaces in two physically separated layers, here we consider the interplay of Fermi surfaces in two valleys.
Note that due to trigonal warping, the Fermi surfaces in two valleys do not match in general. But at low density and large displacement field, the warping effect diminishes and the Fermi surface becomes approximately rotationally symmetric. In recent experiments~\cite{arp2023intervalley}, the inter-valley exciton condensation phase, which is often dubbed the inter-valley coherent phase (IVC), has been observed at low electron density in both bilayer and trilayer ABC graphene, giving us hope to search for the exciton QCP.

\subsubsection*{Excitonic QCP of composite fermions: moir\'{e} graphene and TMD multilayers}

With the advent of fractional quantum anomalous Hall (FQAH)  systems in moir\'{e} materials, including transport measurements strongly suggestive of anomalous CFL phases~\cite{Xu2023_FQAH0,park2023_FQAH2,lu2023_FQAH_graphene}, 
preparing systems of two degenerate, half-filled $C=\pm 1$ bands will likely become possible. In such systems, the bilayer CFL-EI* quantum critical point described in Section~\ref{sec:BilayerACFL} should be realizable. 
To obtain nonvanishing NFL counterflow at the transition, the key ingredients are opposite Chern bands with perfectly nested Fermi surfaces.

A simple realization is the twisted transition metal dichalcogenide bilayer (t-TMD). As demonstrated by recent studies \cite{cai2023_FQAH1, Xu2023_FQAH0, zeng2023_FQAH_cornell, Reddy2023fractional, Dong2023CFL, wang2023fractional} in twisted MoTe$_2$ bilayer, at the single-particle level, the conduction electrons in this class of systems see a pair of valley-contrasting Chern bands 
that are related by time-reversal. Here, the spin is locked to the valley by a strong Ising spin-orbit coupling.
We consider total filling $\nu_T=1$ electron per moir\'e unit cell. In the non-interacting limit, the two opposite Chern bands are each half-filled, in which case we get two Fermi liquids. In the opposite limit where interaction completely dominates over the dispersion, the system becomes valley-polarized. It is conceivable that between the two extreme limits, there may be an intermediate regime, where the interaction is just enough to favor a composite Fermi liquid over the Fermi liquid in each valley, and at the same time not too strong, so that the dispersion, which penalizes the valley polarization, is still respectable.
Note that $C_2$ symmetry is absent in TMD systems, and the time reversal operation that maps $\bs{k}$ to $-\bs{k}$ does not enforce perfect nesting of composite Fermi surfaces in the two valleys. However, at low doping (corresponding to a small twist angle), the trigonal warping effect of the TMD continuum band is extremely weak. As a result, there is an approximate $C_6$ symmetry within each valley, which is evident from both the dispersion and the quantum geometry~\cite{Reddy2023fractional, dong2023_ACFL, wang2023fractional}. For this reason, deviations from the bilayer CFL-EI* quantum critical point will only be observable at very low temperatures.

As the landscape of FQAH materials continues to grow, we anticipate that this illustrative example may give way to even simpler realizations in the future.

\section{Discussion}\label{sec:discussion}

In this paper, we explored a family of metallic quantum critical points associated with exciton condensation in Fermi liquid bilayers as well as composite Fermi liquid bilayers with total Chern number $C_{\rm tot} = 0$ (in the latter case, the excitons are made of composite fermions and composite holes). We showed via a controlled calculation that, in contrast to the conventional lore, these excitonic transitions can be continuous even in the presence of strong quantum fluctuations, 
and lead to striking non-Fermi liquid physics. 
Moreover, we proposed a simple experimental probe, the counterflow resistivity, which exhibit non-Fermi liquid scaling $\rho_{\rm counterflow}(\omega) \sim \omega^{2/z}$ where $z$ is the dynamical critical exponent. Physically, the counterflow conductivity measures the linear response of the counterflow current $\bs{J}_1 - \bs{J}_2$ to equal and opposite electric fields in the two layers $\bs{E}_1 = - \bs{E}_2 = \bs{E}$. Since the total electric field vanishes, the counterflow current carries no momentum and its relaxation is not constrained by momentum conservation. This simple observation explains why critical interlayer scattering mediated by the fluctuating exciton order parameter can lead to rapid decay of the counterflow current, thereby generating a non-Fermi liquid resistivity.

Although previous works based on partially controlled approximations argued that a metallic excitonic transition is necessarily first order, we reached the opposite conclusion using a controlled perturbative expansion in which an $\mathcal{O}(1)$ parameter $\epsilon$ in the physical model is taken to be small. 
Of course, we do not have analytic control over the theory   as $\epsilon$ is extrapolated to 1 but nevertheless we hope that our analysis will motivate an experimental study of these excitonic transitions in two dimensional Van der Waals heterostructures (some concrete realizations were proposed in Section~\ref{sec:realizations}). If the transition is indeed continuous, then a  measurement of the counterflow resistivity as a function of temperature or frequency determines the dynamical critical exponent $z$, which can be compared with the theoretical prediction $z = 3$. Though the theory of metallic quantum criticality of Hertz-Millis type has existed for decades, we are not aware of any previous measurement that was able to extract this exponent in two dimensional systems. In principle, in a layered quasi-two dimensional material tuned to such a quantum critical point, $z$ can be extracted from the heat capacity but there do not seem to be any clear-cut examples of such criticality. Therefore, counterflow resistivity on these excitonic transitions may provide the most reliable estimate of $z$ in this class of Hertz-Millis QCPs. 

Looking beyond excitonic transitions, counterflow transport in multi-layer/multi-species fermionic systems can be a useful probe for more exotic quantum critical points. The key reason is that the counterflow current carries zero momentum and can relax rapidly as a result of inter-species critical scattering. Even for metallic quantum critical points beyond the Hertz-Millis paradigm, inter-species scattering should be a generic feature, and critical exponents associated with these scattering processes lead to sharp features in the scaling of counterflow resistivity. Thinking along these lines, it may also be fruitful to search for analogous quantities in thermoelectric transport that can potentially capture a different set of critical exponents. It is our hope that the high tunability of two dimensional Van der Waals materials will facilitate the discovery of such novel non-Fermi liquids that have so far eluded our imagination.



\section*{Acknowledgements}

We thank Jason Alicea, Luca Delacr\'{e}taz, Zhiyu Dong, Dominic Else, Haoyu Guo, Bertrand Halperin, Adarsh Patri, \'{E}tienne Lantagne-Hurtubise, Dam Thanh Son, Alex Thomson, Ashvin Vishwanath, and Ya-Hui Zhang for related discussions. We are especially grateful to Andrey Chubukov for pointing out a misunderstanding of the existing literature on continuous ferromagnetic transitions in a previous version of this paper. ZDS and TS are supported by the Department of Energy under grant DE-SC0008739. TS was  also supported partially
through a Simons Investigator Award from the Simons Foundation, and by the
Simons Collaboration on Ultra-Quantum Matter, which is a grant from the Simons Foundation (651446, TS). HG is supported by the Kadanoff Fellowship from the University of Chicago. 

\appendix

\section{On the stability of quantum Stoner transitions}\label{app:1storder} 

In this Appendix, we elaborate upon the discussion in Section~\ref{subsec:FL_bilayer_continuity} and analyze the role of quantum fluctuations near itinerant ferromagnetic transitions in two spatial dimensions. In particular, we consider 
order parameters in 
four different symmetry classes -- $I, Z_2, U(1),$ and $SU(2)$ -- and diagnose the instability by calculating the static order parameter susceptibility $\Pi_{\phi}(\bs{q}, \Omega = 0)$ \emph{at criticality}. Note that here we use $I$ to denote the trivial subgroup of $SU(2)$, meaning that it includes $SU(2)$-singlet order parameters that have non-trivial momentum-dependent form factors. In contrast, $Z_2$ is meant to denote the case of an Ising order parameter with uniform form factors.

Within the controlled large-$N$, small $\epsilon=z-2$ expansion introduced in Section~\ref{subsec:FLbilayer_setup}, we find that, for all four symmetry classes, the mean-field continuous transition survives fluctuations to leading nontrivial order. We discuss these conclusions in light of previous works based on augmented Eliashberg approximations (e.g. Refs.~\cite{chubukov2004_FM1storder,rech2006_FM1storder,Maslov2009_FM1storder}), which find  instabilities for the $U(1)$ and $SU(2)$ cases at $\epsilon = 1$ ($z=3$). We explain the origin of this difference in detail below and discuss possible scenarios for reconciliation. 



\subsection{Instability within the Eliashberg approximation}\label{app:1storder_I}

We begin by reviewing approaches to Hertz-Millis models based on the Migdal-Eliashberg approximation, which was initially motivated 
in the context of the electron-phonon problem. We start by considering a general Hertz-Millis model with Yukawa coupling, $g$, and bare Euclidean boson/fermion propagators respectively of the form,
\begin{equation}
    D_0(\bs{q}, i \Omega) = \frac{1}{|\bs{q}|^{z-1} + \lambda\, \Omega^2} \,, \quad G_0(\bs{k}, i\omega) = \frac{1}{i \omega - \epsilon(\bs{k})} \,.
\end{equation}
One may then posit the following set of self-consistent Schwinger-Dyson equations for the boson/fermion self energies,
\begin{equation}
    \begin{aligned}
    \Pi(\bs{q}, i\Omega) &= - g^2 \int_{\bs{k}, \omega} G\left(\bs{k} + \frac{q}{2}, i \omega + i \frac{\Omega}{2}\right) \, G\left(\bs{k} - \frac{q}{2}, i \omega - i \frac{\Omega}{2}\right) \,, \\ \Sigma(\bs{k}, i\omega) &= g^2 \int_{\bs{q}, \Omega} G(\bs{k} - \bs{q}, i\omega - i\Omega) D_{\phi}(\bs{q}, i \Omega) \,,
    \end{aligned}
\end{equation}
where the renormalized propagators $G, D_{\phi}$ are related to the bare propagators $G_0, D_0$ by
\begin{equation}
    G^{-1} = G_0^{-1} - \Sigma \,, \quad D_{\phi}^{-1} = D_0^{-1} - \Pi \,. 
\end{equation}
The standard approach to solving these self-consistent equations is via iteration. Concretely, one starts with an ansatz $\Pi_0, \Sigma_0$ for $\Pi, \Sigma$ and plugs it into the right hand side of the self-consistent equations to generate new approximations, $\Pi_1, \Sigma_1$, which become the input in the next around of iteration. The process is repeated until convergence.

In the low energy limit, the conventional assumption is that $\Sigma(\bs{k}, i\omega)$ depends strongly on $\omega$ but weakly on $\bs{k}$. This assumption motivates the following initial ansatz,
\begin{equation}\label{eq:appA_naiveEliashberg}
    \Pi_0(\bs{q}, i\Omega) = \gamma \frac{|\Omega|}{\sqrt{q^2 + c^2 |\Sigma_0(i \Omega)|^2}} \,, \quad \Sigma_0(\bs{k}, i\omega) = i C(g) \sgn(\omega) |\omega|^{2/z} \,.
\end{equation}
With an appropriate choice of constants, $C(g)$ and $c$, this ansatz \emph{almost} reproduces itself under one round of iteration:
\begin{equation}
\label{eq: momentum-dependent self energy}
    \Pi_1(\bs{q}, i\Omega) \approx \Pi_0(\bs{q}, i\Omega) \,, \quad \Sigma_1(\bs{k}, i\omega) \approx \Sigma_0(i\omega) + C_1(g)\, \epsilon(\bs{k})\, \Sigma_0(i\omega) \log (\frac{M}{|\omega|})+\dots \,,
\end{equation}
Here the new contribution is a \emph{sub-leading, momentum-dependent correction} to the fermion self energy. Since 
the second term is proportional to the fermion dispersion, $\epsilon(\boldsymbol{k})$, the difference $\Sigma_1-\Sigma_0$ becomes vanishingly small near the Fermi surface. 
Therefore, Eq.~\eqref{eq:appA_naiveEliashberg} is typically treated as an IR-exact solution to the self-consistent equations.

This conclusion, however, elides a singular consequence. Indeed, an important observation in Ref.~\cite{rech2006_FM1storder} is that after a second round of iteration, the momentum-dependent part of the fermion self-energy, $\Sigma_1(\bs{k}, i\omega)$, generates a singular, \emph{negative} correction to the boson self-energy:
\begin{equation}
\label{eq: Pi2}
    \Pi_2(\bs{q}, i\Omega) \approx \Pi_1(\bs{q}, i\Omega) - A_G |\bs{q}|^{\frac{z}{2}} \,, \quad \Sigma_2(\bs{k}, i\omega) \approx \Sigma_1(\bs{k},i\omega) \,, 
\end{equation}
where $A_G$ is a positive constant that depends on the subgroup $G \subset SU(2)$. Since the singular negative term, $|\bs{q}|^{z/2}$, dominates over the tree-level boson kinetic term, $|\bs{q}|^{z-1}$, for all $2 < z$, the conventional solution, Eq.~\eqref{eq:appA_naiveEliashberg}, to the self-consistent equations appears unstable. Restoring stability thus requires inclusion of quantum corrections beyond traditional Eliashberg theory, such as Aslamasov-Larkin diagrams.

We identify two possible strategies for including the necessary corrections systematically. The first strategy, adopted in Ref.~\cite{rech2006_FM1storder}, is to fix $z=3$ and augment Eliashberg theory by introducing a large number $N$ of identical fermion species such that the $N \rightarrow \infty$ result for the boson susceptibility already involves additional higher loop diagrams. 
The second strategy, which is the perspective taken in the main text of this paper, is to use an alternative large-$N$ expansion in which the ${N \rightarrow \infty}$ limit corresponds to Eq.~\eqref{eq:appA_naiveEliashberg} 
with $z = 2$. In the following two sections, we describe these approaches in detail and contrast their merits and drawbacks. 

Before explaining both of these strategies in more detail, we pause to make two comments. First, the basic physics identified here is not unique to the quantum critical point: The general form of Eq.~\eqref{eq: momentum-dependent self energy} is also valid for Fermi liquids, and the presence of momentum-dependent corrections to the self-energy is actually crucial for obtaining the correct momentum-dependence of the density susceptibility of a Fermi liquid, $\chi(\bs{q})-\chi(\bs{q}=0)\sim|\bs{q}|$~\cite{chubukov2004_singularFL}. 

Our second remark is that the instability of the traditional Eliashberg saddle point presents a challenge to any proposed 
large-$N$ expansion in which this saddle point (with $z > 2$) is exact in the $N \rightarrow \infty$ limit. A prominent example is the Yukawa-SYK model studied in Refs.~\cite{Esterlis2021,Patel2022,guo2022_largeN}. It is not clear to us how this issue can be cured naturally in such approaches without introducing additional (possibly uncontrolled) ingredients.

\subsection{Augmented Eliashberg approximation}

Now we review how augmenting the traditional Migdal-Eliashberg procedure can cancel off the instability identified above, following the calculations of Ref.~\cite{rech2006_FM1storder}. 
To motivate the augmented Eliashberg approximation, consider a deformed Hertz-Millis action with $N$ fermion species, 
\begin{equation}
    S = S_c + S_{\rm int} + S_{\phi} \,,
\end{equation}
where
\begin{equation}
    \begin{aligned}
        S_c &= \sum_{n=1}^N \sum_{\alpha} \int_{\bs{k}, \omega} c^{\dagger}_{n,\alpha}(\bs{k},\omega) \left[i\omega - \epsilon(\bs{k})\right] c_{n,\alpha}(\bs{k}, \omega)  \,, \\
        S_{\rm int} &= \sum_{n = 1}^N \sum_{\alpha, \beta} \sum_i \int_{\bs{k}, \bs{q}, \omega, \Omega} \frac{g}{2\sqrt{N}} \, \phi_i(\bs{q}, \Omega)\, c^{\dagger}_{\alpha}(\bs{k} + \bs{q}/2, \omega + \Omega/2)\,\Gamma^i_{\alpha\beta}\, c_{\beta}(\bs{k} - \bs{q}/2, \omega - \Omega/2) \,, \\
        S_{\phi} &= \sum_i \int_{\bs{q}, \Omega}\Bigg( \phi_i(\bs{q}, \Omega) \left[\Omega^2 + |\bs{q}|^2 + M^2\right] \phi_i(\bs{q},\Omega) + V[\phi]\Bigg) \,.
    \end{aligned}
\end{equation}
Here $n = 1, \ldots, N$ is a $U(N)$ fermion flavor index, $\alpha,\beta$ are fermion spin indices, and $i$ is a boson species index, with $\phi_i$ taken to be in a vector representation of a subgroup $G\subset SU(2)$. $g>0$ is taken to be a (possibly momentum-dependent) Yukawa coupling, and $\Gamma^i_{\alpha\beta}$ is a set of spin structure factors. 
The order parameter is always taken to be a fermion flavor singlet. 

In the large-$N$ limit, quantum corrections 
can be organized in powers of $1/N$. 
At $\mathcal{O}(N^0)$, the boson self-energy appears to be given solely by a one-loop bubble diagram of bare fermion propagators, and the fermion self-energy should vanish. The na\"{i}ve $N \rightarrow \infty$ solution takes the form,
\begin{equation}
    D_0(\bs{q}, i\Omega) = \frac{1}{|\bs{q}|^2 + \gamma \frac{|\Omega|}{\sqrt{q^2+\Omega^2}}} \,, \quad G_0(\bs{k}, i\omega) = \frac{1}{i\omega - \epsilon(\bs{k})} \,.
\end{equation}

\begin{figure}
    \centering
    \includegraphics[width = 0.8\textwidth]{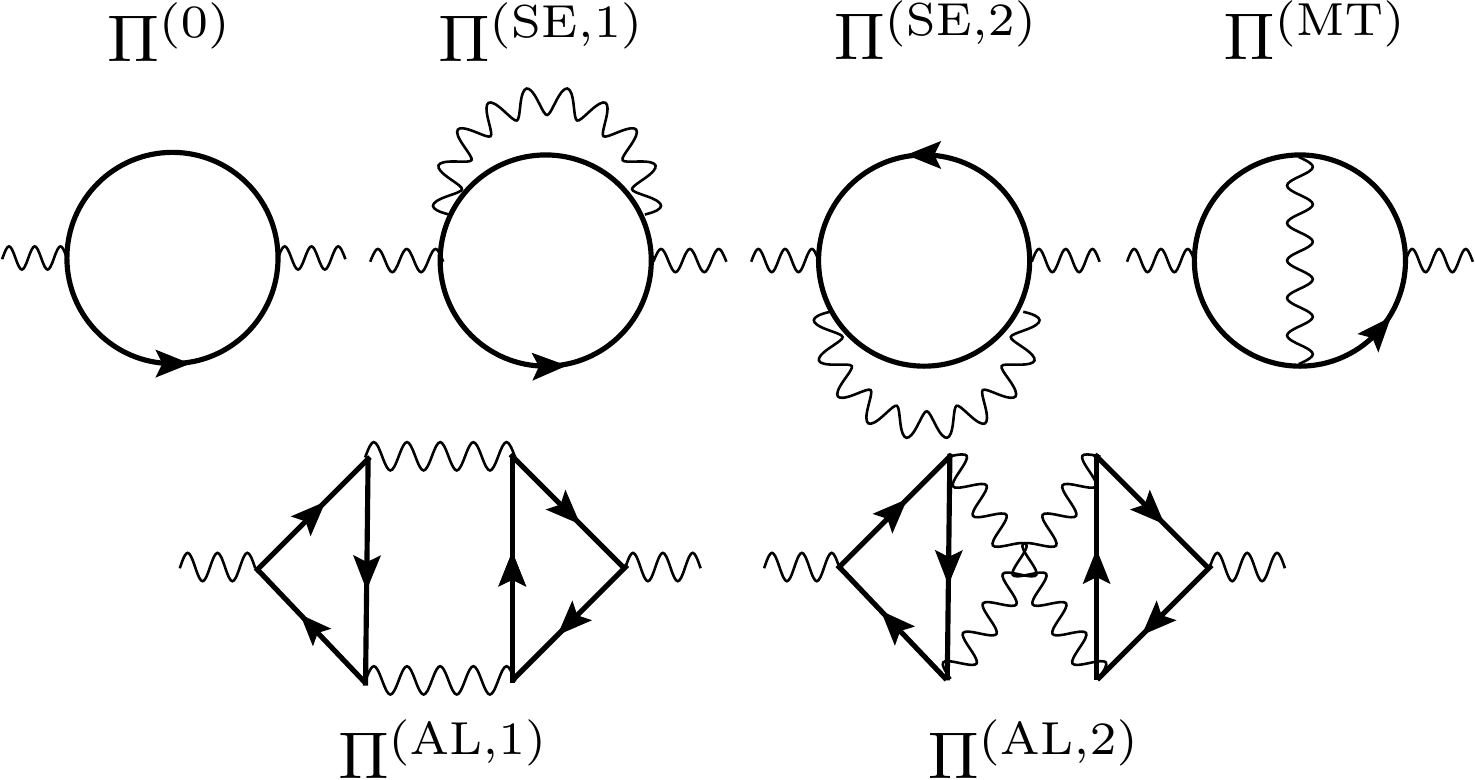}
    \caption{Diagrams that contribute to the boson self-energy $\Pi(\bs{q}, i\Omega)$ at $\mathcal{O}(N^{-1})$ according to combinatorial scaling.}
    \label{fig:RPA_diagrams}
\end{figure}

Now consider corrections to the boson and fermion propagators at $\mathcal{O}(N^{-1})$. For the fermion self-energy, the only relevant diagram is the minimal one-loop diagram with internal propagators $G_0, D_0$, leading to the familiar result,
\begin{equation}
    \Sigma(i\omega) \sim \frac{1}{N}\, \sgn(\omega) |\omega|^{2/3} \,. 
\end{equation}
For the boson-self energy, 
there are five diagrams contributing at $\mathcal{O}(1/N)$, as shown in Fig.~\ref{fig:RPA_diagrams}. 
However, with bare fermion propagators, these diagrams contain IR divergences. It is therefore necessary to resum $\Sigma(i\omega)$ into the fermion propagator before evaluating these diagrams:
\begin{equation}
    G_0(\bs{k}, i\omega) \rightarrow G(\bs{k}, i\omega) = \frac{1}{i\omega + \frac{i C(g)}{N} \sgn(\omega) |\omega|^{2/3} - \epsilon(\bs{k})} \,.
\end{equation}
As was understood in detail in Ref.~\cite{Lee2009}, although the inclusion of $\Sigma(i\omega)$ regulates the IR divergences, this comes at a cost of rendering the $1/N$ expansion uncontrolled. Indeed, in the regime $\Sigma(i\omega) \gg \omega$, fermion propagators become enhanced by a factor of $N$.  
A basic consequence of this enhancement is that the five formerly $\mathcal{O}(1/N)$ corrections to the boson propagator in Fig.~\ref{fig:RPA_diagrams} are promoted to $\mathcal{O}(N^0)$ and compete with the one-loop bubble diagram. The inclusion of this larger set of diagrams in the $N = \infty$ solution defines what we term the \emph{augmented Eliashberg approximation}.

Working within this augmented Eliashberg approximation, Ref.~\cite{rech2006_FM1storder} then proceeded to calculate all the diagrams in Fig.~\ref{fig:RPA_diagrams}. For a given subgroup $G \subset SU(2)$, each of these diagrams (labeled by an index $i$) is the product of a spin structure factor $\Gamma_{G,i}$ and a multi-loop integral $I_{i}(\bs{q}, i\Omega = 0)$ which is independent of the group $G$. The total contribution to the boson susceptibility is therefore
\begin{equation}
    \delta \Pi_G(\bs{q}, i\Omega = 0) = \sum_{i \,= \, \rm SE, MT, AL1, AL2} \Gamma_{G,i} \cdot I_{i}(\bs{q},i\Omega = 0) \,.
\end{equation}
The spin structure factors are traces of Pauli matrices. For the groups $G = I, Z_2, U(1), SU(2)$, they can be evaluated to yield
\begin{equation}
    \Gamma_{SU(2), \rm MT} = -2 \quad \Gamma_{SU(2), \rm SE} = 6 \quad \Gamma_{SU(2), \rm AL1} = 8 \quad \Gamma_{SU(2), \rm AL2} = -8  \,,
\end{equation}
\begin{equation}
    \Gamma_{U(1), \rm MT} = 0 \quad \Gamma_{U(1), \rm SE} = 4 \quad \Gamma_{U(1), \rm AL1} = 0 \quad \Gamma_{U(1), \rm AL2} = 0 \,,
\end{equation}
\begin{equation}
    \Gamma_{Z_2, \rm MT} = 2 \quad \Gamma_{Z_2, \rm SE} = 2 \quad \Gamma_{Z_2, \rm AL1} = 0 \quad \Gamma_{Z_2, \rm AL2} = 0  \,,
\end{equation}
\begin{equation}
    \Gamma_{I, \rm MT} = 2 \quad \Gamma_{I, \rm SE} = 2 \quad \Gamma_{I, \rm AL1} = 4 \quad \Gamma_{I, \rm AL2} = 4 \,. 
\end{equation}
As for the integrals, the leading singular contributions at small $\bs{q}$ were explicitly calculated in Ref.~\cite{rech2006_FM1storder} and found to be
\begin{equation}
    \begin{aligned}
    I_{\rm SE}(\bs{q},i\Omega=0) &= - I_{\rm MT}(\bs{q}, i\Omega = 0) = -\alpha |\bs{q}|^{3/2} \,, \\
    I_{\rm AL1}(\bs{q},i\Omega=0) &= - I_{\rm AL2}(\bs{q}, i\Omega =0) = - \beta |\bs{q}|^{3/2}  \,,
    \end{aligned}
\end{equation}
where $\alpha, \beta$ are positive constants independent of $N$. These terms can in principle cause an instability and produce a fluctuation-induced first-order transition.

Plugging these integrals into the boson susceptibility, one finds
\begin{equation}\label{eq:Rech_integrals}
    \delta \Pi_G(\bs{q}, i\Omega = 0) = \begin{cases}
        - 8 (\alpha + 2 \beta) |\bs{q}|^{3/2} & G = SU(2) \\ -4 \alpha |\bs{q}|^{3/2} & G = U(1) \\ 0 & G = Z_2/I 
    \end{cases} \,.
\end{equation}
As anticipated from the preceding discussion of the augmented Eliashberg approach from the point of view of the large-$N$ expansion, these singular corrections do not carry any factor of $N$ and compete at the same order with the result in Eq.~\eqref{eq: Pi2}. Since a negative boson susceptibility at the critical point signals an instability, these results led the authors of Ref.~\cite{rech2006_FM1storder} to conclude that the $SU(2)$ and $U(1)$ phase transitions are first order, while the $Z_2/I$ phase transitions remain continuous. 

Although we motivated the particular choice of diagrams in Fig.~\ref{fig:RPA_diagrams} using a large-$N$ expansion, the enhancement of the resummed fermion propagator by a factor of $N$ noted above actually implies that 
the diagrams in Fig.~\ref{fig:RPA_diagrams} are not complete~\cite{Lee2009}. 
Even at $\mathcal{O}(N^0)$, one needs to in principle account for an infinite set of diagrams
, which is a formidable task. One optimistic conjecture is that the higher loop diagrams have a momentum scaling that is less singular than $|\bs{q}|^{3/2}$, so that their corrections can be neglected in the IR limit. Additional analytic insights are needed to settle this conjecture. 


\subsection{Large-$N$, small $z-2$ double expansion}

We now describe the second strategy outlined in Appendix~\ref{app:1storder_I}: 
the double expansion introduced in Ref.~\cite{Mross2010}, which carries an advantage of being manifestly under perturbative control, despite introducing non-locality into the order parameter action. 
In this framework, one only needs to systematically account for a finite set of diagrams at each order in $\epsilon\sim 1/N$. In what follows, we will perform calculations to $\mathcal{O}(\epsilon)$ and derive the results \eqref{eq:Pi_final_doubleexp} stated in the main text. 

The double expansion is defined in Section~\ref{subsec:FLbilayer_setup}. At $\mathcal{O}(\epsilon^0)$, the boson and fermion propagators indeed satisfy the self-consistent equations, 
but with dynamical critical exponent $z = 2 + \epsilon \rightarrow 2$. As a result, the issues identified in Appendix~\ref{app:1storder_I} for $z > 2$ do not apply, and the fermion and boson propagators in the $N \rightarrow \infty$ limit take the standard form,
\begin{equation}
    G^{-1}(\bs{k}, i\omega) = i\, \mathcal{C}_1(g, r)\, \omega - \epsilon(\bs{k}) \,, \quad D_{\phi}^{-1}(\bs{q}, i\Omega) = |\bs{q}| + \gamma \frac{|\Omega|}{\sqrt{q^2 + \Omega^2}} \,,
\end{equation}
where $\mathcal{C}_1$ is a constant, and we remind the reader that $r=N\epsilon$. Our goal is to compute the static order parameter susceptibility at $\mathcal{O}(\epsilon)$. 

As could be anticipated based on the calculations in Appendix~\ref{app:1storder_I}, even the one-loop boson and fermion self energies possess new contributions at $\mathcal{O}(\epsilon)$. Indeed, the same kind of momentum-dependent corrections occur in the fermion self-energy, leading to a negative correction to the boson self-energy that $\sim-|\bs{q}|$,
\begin{equation}
    \begin{aligned}
    G^{-1}(\bs{k}, i\omega) &= i\, \mathcal{C}_1(g,r)\, \omega \left(1 - \frac{\epsilon}{2} \log |\omega|\right) - \epsilon(\bs{k}) \left[1 + \mathcal{C}_2(g,r)\, \omega \left(1 - \frac{\epsilon}{2} \log |\omega|\right) \right]  \,, \\
    D_{\phi}^{-1}(\bs{q}, i\Omega) &= |\bs{q}| + \Big( |\bs{q}| \log |\bs{q}| -  \alpha(\epsilon)\,\mathcal{A}(G)\, |\bs{q}|\Big)\,,
    \end{aligned}
\end{equation}
where
\begin{align}
\mathcal{A}(G)&=\begin{cases}
        6 & G = SU(2) \\ 4 & G = U(1) \\ 2 & G = Z_2/I 
    \end{cases} \,,
\end{align}
and $\mathcal{C}_2$ is another constant. Importantly, $\alpha(\epsilon)$ is a positive constant for $\epsilon \in [0,1]$ that reduces to $\alpha$ (the constant defined in Eq.~\eqref{eq:Rech_integrals}) when $\epsilon = 1$. Note that the boson self-energy above is calculated using dressed fermion propagators, as mandated by the double expansion. 

The agreement with the traditional Migdal-Eliashberg approximation in Appendix~\ref{app:1storder_I} at $\epsilon=1$ is not a coincidence. 
Recall that the integral defining $\alpha(\epsilon)$ corresponds to the diagrams $\Pi^{(\rm SE, 1)}$ and $\Pi^{(\rm SE,2)}$ in Fig.~\ref{fig:RPA_diagrams}. In evaluating these diagrams, it is helpful to decompose the fermion self-energy as $\Sigma(\bs{k}, i\omega) = \Sigma_0(i\omega) + \delta \Sigma(\bs{k}, i\omega)$. The dressed fermion legs in these diagrams can be written as 
\begin{equation}
    G(\bs{k}, i \omega) \Sigma(\bs{k}, i\omega) G(\bs{k}, i\omega) = \left[\frac{1}{-\epsilon(\bs{k}) - \Sigma_0(i\omega) - \delta \Sigma(\bs{k}, i\omega)}\right]^2 \left[\Sigma_0(i\omega) + \delta \Sigma(\bs{k}, i\omega)\right] \,.
\end{equation}
The insight from Ref.~\cite{rech2006_FM1storder} is that the singular $|\bs{q}|^{3/2}$ contributions to $\Pi^{(\rm SE, 1)}$ and $\Pi^{(\rm SE,2)}$ can be obtained by keeping the momentum-dependent part $\delta \Sigma(\bs{k}, i\omega)$ in the numerator and dropping the momentum-dependent part in the denominator. This means that the dressed fermion leg can be replaced with 
\begin{equation}\label{eq:Chubukov_reduction}
    \left[\frac{1}{-\epsilon(\bs{k}) - \Sigma_0(i\omega)}\right]^2 \delta \Sigma(\bs{k}, i\omega) \approx \frac{1}{-\epsilon(\bs{k}) - \Sigma_0(i\omega) - \delta \Sigma(\bs{k}, i \omega)} - \frac{1}{-\epsilon(\bs{k}) - \Sigma_0(i\omega)} \,.
\end{equation}
This argument shows that the sum of $\Pi^{(\rm SE, 1)}$ and $\Pi^{(\rm SE,2)}$ within the fundamental large $N$ expansion is equivalent to a single diagram $\Pi^{(0)}$ with internal fermion propagators replaced by the right-hand-side of Eq.~\eqref{eq:Chubukov_reduction}. One can furthermore show that the second term in the right-hand-side of Eq.~\eqref{eq:Chubukov_reduction} gives no singular contribution. Replacing the fermion propagators in $\Pi^{(0)}$ with only the first term in Eq.~\eqref{eq:Chubukov_reduction} corresponds precisely to the Eliashberg approximation, with $\epsilon$ na\"{i}vely extrapolated to $\epsilon = 1$. This explains why the $\alpha(\epsilon)$ coefficient reduces to $\alpha$ in Eq.~\eqref{eq:Rech_integrals} when $\epsilon = 1$.\footnote{We thank Andrey Chubukov for explaining how this subtle point enters the approach in Ref.~\cite{rech2006_FM1storder}.}

The remaining $\mathcal{O}(\epsilon)$ effects in the double expansion correspond to the Aslamasov-Larkin and Maki-Thompson diagrams, 
$\Pi^{(\rm MT)}$, $\Pi^{(\rm AL, 1)}$, and $\Pi^{(\rm AL, 2)}$, since fermion self-energy corrections have already been resummed above. 
After evaluating the relevant integrals, we arrive at a simple generalization of the results obtained by Ref.~\cite{rech2006_FM1storder}
\begin{equation}
    I_{\rm MT}(\bs{q}, i\Omega = 0) = \alpha(\epsilon) |\bs{q}|^{1+ \epsilon/2} \,, \quad I_{\rm AL1}(\bs{q}, i\Omega = 0) = - I_{\rm AL2}(\bs{q}, i\Omega = 0) = - \beta(\epsilon) |\bs{q}|^{1+\epsilon/2} \,,
\end{equation}
where $\alpha(\epsilon), \beta(\epsilon)$ again extrapolate back to $\alpha, \beta$ in \eqref{eq:Rech_integrals} when $\epsilon = 1$. Using these integrals and the spin structure factors, we obtain the following total boson self-energy
\begin{equation}
    \Pi_G(\bs{q}, i\Omega=0) = |\bs{q}| + \epsilon \left( |\bs{q}| \log |\bs{q}| - \frac{C(G)}{N\epsilon}\,  |\bs{q}| \right) + \mathcal{O}(\epsilon^2) \,,
\end{equation}
where
\begin{equation}
    C(G) = \begin{cases}
        8 \left[\alpha(\epsilon = 0) + 2 \beta(\epsilon = 0) \right] & G = SU(2) \\ 4 \alpha(\epsilon = 0)  & G = U(1) \\ 0 & G = Z_2/I 
    \end{cases} \,.
\end{equation}
Observe that the prefactor of the negative correction exactly matches the augmented Eliashberg 
result for all symmetry classes, up to the replacement of $\alpha(\epsilon), \beta(\epsilon)$ with $\alpha, \beta$. The basic reason for this agreement is that in both procedures, the singular corrections can be mapped to the last five diagrams in Fig.~\ref{fig:RPA_diagrams}, evaluated with internal fermion propagators that retain a frequency-dependent but momentum-independent self-energy.

To recap, the main advantage of the double expansion is that the finite set of diagrams in Fig.~\ref{fig:RPA_diagrams} completely account for the boson self-energy at $\mathcal{O}(\epsilon)$. This is to be contrasted with the fundamental large $N$ expansion, in which an infinite set of additional planar diagrams could make a contribution. However, suppressing these planar diagrams comes at a cost: the starting point of the double expansion has a non-analytic kinetic term proportional to $|\bs{q}|$, which is far away from the natural $|\bs{q}|^2$ kinetic term for order parameter fluctuations. The path from $\epsilon = 0$ to $\epsilon = 1$ need not be smooth, and we give a brief discussion of possible scenarios at the end of Section~\ref{subsec:FL_bilayer_continuity}.

\section{RG analysis of the CFL-\texorpdfstring{$\overline{\text{CFL}}$}{} bilayer}\label{app:RG}

In this Appendix, we work out the one-loop RG flow of gauge couplings in the Coulomb-coupled CFL-$\overline{\text{CFL}}$ bilayer. For the evaluation of one-loop diagrams, it suffices to work in a single patch on the Fermi surface. Let $x$ be the direction perpendicular to the Fermi surface in this patch and let the patch dispersion be $\epsilon(\bs{k}) = v_0 k_x + K_0 k_y^2$, then the patch action takes the form
\begin{equation}
    \begin{aligned}
    S_0 &= \sum_{i = 1,2} \int \psi^{\dagger}_i \left[i\omega - (v_0 k_x + K_0 k_y^2)\right] \psi_i + \frac{1}{2} \int |q_y|^{1+\epsilon_+} |a_{+,x}|^2 + |q_y|^{1+\epsilon_-} |a_{-,x}|^2 \quad  \,, \\
    S_{\rm int} &= g_{+,0}v_0 \int a_{+,x} \left[\psi^{\dagger}_1 \psi_1 + \psi^{\dagger}_2 \psi_2\right] + g_{-,0}v_0 \int a_{-,x} \left[\psi^{\dagger}_1 \psi_1 - \psi^{\dagger}_2 \psi_2 \right] \,.
    \end{aligned}
\end{equation}
Note that $\epsilon_-$ has been introduced as a regulator to facilitate the extraction of $\beta$-functions in the field-theoretic RG procedure. We will set $\epsilon_-$ to zero in the end for the physical problem. Leveraging the layer exchange symmetry, we can introduce renormalized couplings
\begin{equation}
    v_0 = v Z_v \quad g_{+,0} = \mu^{\epsilon_{+}/2} g_{+} Z_{g_{+}} \quad g_{-,0} = \mu^{\epsilon_{-}/2} g_{-} Z_{g_{-}} \quad \psi_i = Z^{1/2} \psi_{i,R} \,,
\end{equation}
in terms of which the action can be written as
\begin{equation}
    \begin{aligned}
    S_0 &= \sum_{i = 1,2} \int \psi^{\dagger}_{i,R} \left[iZ\omega - ZZ_v (v k_x + K k_y^2)\right] \psi_{i,R} + \frac{1}{2} \int |q_y|^{1+\epsilon_+} |a_{+,x}|^2 + |q_y|^{1+\epsilon_-} |a_{-,x}|^2 \,, \\
    S_{\rm int} &= v Z_v Z \\
    &\left(g_+ \mu^{\epsilon_+/2} Z_{g_+} \int a_{+,x} \left[\psi^{\dagger}_{1,R} \psi_{1,R} + \psi^{\dagger}_{2,R} \psi_{2,R}\right] + g_- \mu^{\epsilon_-/2}Z_{g_-} \int a_{-,x} \left[\psi^{\dagger}_{1,R} \psi_{1,R} - \psi^{\dagger}_{2,R} \psi_{2,R}\right]\right) \,.
    \end{aligned}
\end{equation} 
Having established all the notations, we can now extract various counterterms from the fermion self energy. By layer-exchange symmetry, it suffices to consider the self energy for $\psi_1$. There are two one-loop diagrams that describe the scattering of $\psi_1$ off of gauge fields $a_+$ and $a_-$. After adding up these two contributions (which have been evaluated in Ref.~\cite{Nayak1994}), we get
\begin{equation}
    \Sigma(i\omega) = \left[\frac{g_+^2 v}{2\pi^2 \epsilon_+}  + \frac{g_-^2 v}{2 \pi^2 \epsilon_-}\right] \cdot i\omega \,.
\end{equation}
Since the one-loop self energy has no momentum dependence, $Z_v = Z^{-1}$ to this order. Demanding that $Z$ cancels the divergent terms in $\Sigma(\omega)$, we find that
\begin{equation}
    Z_v = 1 + \frac{g_+^2 v}{2\pi^2 \epsilon_+} + \frac{g_-^2 v}{2 \pi^2\epsilon_-} = 1 + \frac{2\alpha_+}{\epsilon_+} + \frac{2\alpha_-}{\epsilon_-} = Z^{-1} \,.
\end{equation}
where we defined the dimensionless running coupling constants $\alpha_{\pm} = \frac{g_{\pm}^2 v}{4\pi^2}$. By the $U(1) \times U(1)$ Ward identity, $Z_{g_+} = Z_{g_-} = 1$. As a result, the running couplings can be simplified to
\begin{equation}
    \alpha_{+,0} = \alpha_{+} \mu^{\epsilon_{+}} Z_{g_{+}}^2 Z_v = \alpha_{+} \mu^{\epsilon_{+}} Z_v \,, \quad \alpha_{-,0} = \alpha_- \mu^{\epsilon_{-}}Z_{g_{-}}^2 Z_v = \alpha_- Z_v \,. 
\end{equation}
Finally, note that the bare couplings $\alpha_{\pm, 0}$ are invariant under RG. Differentiating $\alpha_{\pm, 0}$ with respect to the RG scale $\ln \mu$ and setting the result to zero, we obtain the $\beta$-functions up to $\mathcal{O}(\alpha_{\pm}^2)$
\begin{equation}
    \begin{aligned}
    \beta(\alpha_{+}) &= \epsilon_{+} \alpha_{+} + \alpha_{+} \frac{d \ln Z_v}{d \ln \mu} = \epsilon_+ \alpha_+ - 2 \alpha_+ \left[\epsilon_+^{-1} \beta(\alpha_+) + \epsilon_-^{-1} \beta(\alpha_-)\right] = \alpha_+ \left[\epsilon_+ - 2 (\alpha_+ + \alpha_-)\right] \\
    \beta(\alpha_-) &= \epsilon_- \alpha_- + \alpha_- \frac{d \ln Z_v}{d \ln \mu} = \epsilon_- \alpha_- - 2 \alpha_- \left[\epsilon_+^{-1} \beta(\alpha_+) + \epsilon_-^{-1} \beta(\alpha_-)\right] = \epsilon_- \alpha_- -2 \alpha_- \left[\alpha_+ + \alpha_-\right]  \,.
    \end{aligned}
\end{equation}
where we defined $\beta(\alpha_{\pm}) = - \frac{\partial \alpha_{\pm}}{\partial \ln \mu}$. Setting $\epsilon_- \rightarrow 0$ recovers the RG equations quoted in Section~\ref{subsec:ACFLbilayer_RGstability}. Note that unlike for the decoupled CFL bilayer, the coupled RG equation contains cross terms proportional to $\alpha_+ \alpha_-$. Nevertheless, we show in Section~\ref{sec:BilayerACFL} that these cross terms do not modify the location and stability of the fixed point at $\alpha_+ = \frac{\epsilon_+}{2}, \alpha_- = 0$. 

\section{Symmetries and anomalies}
\label{app:gifts}

In this appendix, we flesh out the nonperturbative symmetry and anomaly arguments briefly mentioned in Sections~\ref{subsec:FLbilayer_nonpert} and Section~\ref{subsec:ACFLbilayer_pert}. We start with the simpler case of the excitonic QCP in a Fermi liquid bilayer and then generalize to the QCP between the bilayer CFL-$\overline{\text{CFL}}$ phase and the EI* phase.

\subsection{Fermi liquid bilayer}

As pointed out in Refs.~\cite{shi2022_gifts,shi2023_loop}, the emergent symmetries and anomalies of generalized Hertz-Millis models can be efficiently analyzed within a mid-IR effective action
\begin{equation}\label{eq:midIRaction_FLbilayer}
    \begin{aligned}
        S[c_1,c_2,\phi] &= \sum_{n=1,2} \int_{\bs{k}, \tau} \sum_{\theta} c^{\dagger}_{\theta,n}(\bs{k},\tau) \left[\partial_{\tau} - \epsilon_{\theta}(\bs{k})\right]c_{\theta,n}(\bs{k},\tau) \\
        &+ \int_{\bs{x}, \tau} \sum_{\theta} \frac{1}{2} g(\theta) \left(\phi_1 \left[c^{\dagger}_{\theta,1} c_{\theta,2} + c^{\dagger}_{\theta,2} c_{\theta,1}\right] + \phi_2 \left[-i c^{\dagger}_{\theta,1} c_{\theta,2} + i c^{\dagger}_{\theta,2} c_{\theta,1}\right]\right) \\
        &+ \frac{1}{2} \int_{\bs{q}, \Omega} \sum_{m=1}^2 \left[\Omega^2 + |\bs{q}|^{1+\epsilon}\right] \left|\phi_m(\bs{q},\Omega)\right|^2  \,.
    \end{aligned}
\end{equation}
Here $c_{\theta,i}$ is the annihilation operator for fermions in layer $i$ with momentum constrained to an angular patch labeled by $\theta$. Each patch has a finite width $\Lambda$ in the angular direction, but is infinite in the radial direction. $\epsilon_{\theta}(\bs{k})$ is an approximate dispersion relation for fermions in patch $\theta$. In the mid-IR action one usually keeps the leading two terms $\epsilon_{\theta}(\bs{k}) \sim v_F(\theta) k_{\perp} + \kappa_{\theta} k_{||}^2$ where $k_{\perp}, k_{||}$ are momentum coordinates perpendicular/parallel to the Fermi surface at patch $\theta$. In passing to the mid-IR action, UV processes in which bosons mediate large-momentum scattering between fermions on the Fermi surface are neglected. The intuition is very simple: the most singular bosonic fluctuations are centered around $\bs{q} = 0$, and large-$\bs{q}$ scattering processes can be integrated out at the cost of finite renormalizations of model parameters with no singular effect in the IR limit. 

We are now in a position to analyze the symmetries and anomalies of this mid-IR action. The quadratic fermionic action enjoys a $U(2)^{N_{\rm patch}}$ symmetry that rotates between the two layers within each momentum patch $\theta$. The interlayer interactions can be written as minimal couplings between $\phi$ and generators of the $U(2)$ algebra. To elucidate this algebraic structure, it is helpful to introduce $U(2)$ generators in layer space 
\begin{equation}
    L_0 = \frac{I}{2} \,,\quad L_1 = \frac{\tau_x}{2}\,, \quad L_2 = \frac{\tau_y}{2}\,, \quad L_3 = \frac{\tau_z}{2}\,,
\end{equation}
where $\tau_i$ with $i = x, y, z$ are the Pauli matrices. Using these generators, we can decompose the $U(2)$ current operators in a patch $\theta$ into these components
\begin{equation}
    \begin{aligned}
    n_{\mu,\theta} &= \begin{pmatrix} c^{\dagger}_{\theta,1} & c^{\dagger}_{\theta,2} \end{pmatrix} L_\mu \begin{pmatrix} c_{\theta,1} \\ c_{\theta,2} \end{pmatrix} \,, \\
    J^{\perp}_{\mu,\theta} &= v(\theta) \begin{pmatrix} c^{\dagger}_{\theta,1} & c^{\dagger}_{\theta,2} \end{pmatrix} L_\mu \begin{pmatrix} c_{\theta,1} \\ c_{\theta,2} \end{pmatrix} \,,  \\
    J^{\parallel}_{\mu,\theta} &= \kappa(\theta) \begin{pmatrix} c^{\dagger}_{\theta,1} & c^{\dagger}_{\theta,2} \end{pmatrix} L_\mu (-i \partial_{\parallel}) \begin{pmatrix} c_{\theta,1} \\ c_{\theta,2} \end{pmatrix} \,. 
    \end{aligned}
\end{equation}
One of the essential insights from Ref.~\cite{shi2022_gifts} is that the $U(2)$ currents in each patch carry a t'Hooft anomaly when coupled to an external $U(2)$ gauge field $A = A_{\mu} L^{\mu}$. The $U(2)$ anomaly equation can be written as deformations of the current conservation equation within each patch
\begin{equation}\label{eq:U(2)anomaly_FLbilayer}
    \begin{aligned}
    \partial_t n_{0,\theta} + \nabla_{\alpha} J^{\alpha}_{0,\theta} &= - \frac{1}{2} \frac{\Lambda(\theta)}{(2\pi)^2} \partial_{t} A_{\perp, 0} \,, \\
    \partial_t n_{i,\theta} + \nabla_{\alpha} J^{\alpha}_{i,\theta} + i f_i^{jk} A_{\alpha,j} J^{\alpha}_{k,\theta} &= - \frac{1}{2} \frac{\Lambda(\theta)}{(2\pi)^2} \partial_t A_{\perp, i} \,,
    \end{aligned}
\end{equation}
where factors of $\frac{1}{2}$ on the RHS comes from the normalization choice $\Tr L_i L_j = \frac{1}{2} \delta_{ij}$. 

We now make an important observation: within the mid-IR effective action of the Fermi liquid bilayer, the critical boson plays the role of an effective $U(2)$ gauge field that couples to the $U(2)$ fermion currents in each patch. If we reexpress the interaction term in the mid-IR action as
\begin{equation}
    S \supset \int_{\bs{r}, \tau} \sum_{\theta} g(\theta) \left[\phi_1 n_{1,\theta} + \phi_2 n_{2,\theta}\right] \,,
\end{equation}
then the effective $U(2)$ gauge fields can be identified as
\begin{equation}
    A_{\perp,0} = 0 \,, \quad A_{\perp, 1} = \frac{g(\theta)}{v(\theta)} \phi_1 \,, \quad A_{\perp, 2} = \frac{g(\theta)}{v(\theta)} \phi_2 \,, \quad A_{\perp, 3} = 0 \,, \quad A_{\parallel, i} = 0 \,.
\end{equation}
Plugging these identifications into the anomaly equations in Eq.~\eqref{eq:U(2)anomaly_FLbilayer} and setting $\bs{q} \rightarrow 0$, we obtain
\begin{equation}\label{eq:U(2)_anomaly_q=0_FLbilayer}
    \begin{aligned}
    \partial_t n_{0, \theta}(\bs{q}=0, t) &= 0 \,,  \\
    \partial_t n_{1,\theta}(\bs{q}=0, t)  - i g(\theta) \phi_2 n_{3,\theta}(\bs{q}=0, t)  &= - \frac{1}{2} \frac{\Lambda(\theta)}{(2\pi)^2} \frac{g(\theta)}{v(\theta)} \partial_t \phi_1(\bs{q}=0, t)  \,,\\
    \partial_t n_{2,\theta}(\bs{q}=0, t)  - i g(\theta) \phi_1 n_{3,\theta}(\bs{q}=0, t)  &= -\frac{1}{2} \frac{\Lambda(\theta)}{(2\pi)^2} \frac{g(\theta)}{v(\theta)} \partial_t \phi_2(\bs{q}=0, t) \,, \\
    \partial_t n_{3,\theta}(\bs{q}=0, t)  + i g(\theta) \left[\phi_1 n_{2,\theta} + \phi_2 n_{1,\theta}\right](\bs{q}=0, t) &= 0 \,. 
    \end{aligned}
\end{equation}
The four equations above completely characterize the anomaly constraints at $\bs{q} = 0$. An additional set of bosonic constraints come from the equations of motion for $\phi_m$. After adding a source term for $\phi_m$, we find that 
\begin{equation}\label{eq:phi_eom_FLbilayer}
    \begin{aligned}
    \left[\Omega^2 + |\bs{q}|^{1+\epsilon}\right] \phi_m &= - \sum_{\theta} g(\theta) n_{m,\theta} + h_m  \,, \quad m = 1, 2\,.
    \end{aligned}
\end{equation}
We emphasize that Eq.~\eqref{eq:U(2)_anomaly_q=0_FLbilayer} and Eq.~\eqref{eq:phi_eom_FLbilayer} are exact within the mid-IR effective theory defined by Eq.~\eqref{eq:midIRaction_FLbilayer}. 
However, these anomaly equations are far less constraining than Eq.~\eqref{eq:EM_anomaly}. Take for example the last line of Eq.~\eqref{eq:U(2)_anomaly_q=0_FLbilayer}. This equation relates the time-derivative of $n_{3,\theta}(\bs{q}=0)$ to the sum of $(\phi_1\, n_{2,\theta})(\bs{q} = 0)$ and $(\phi_2\, n_{1,\theta})(\bs{q}=0)$, which are convolutions of two bosonic fields
\begin{equation}
    \lim_{\bs{q} \rightarrow 0} \int d^2 \bs{x}\, e^{i \bs{q} \cdot \bs{x}} \,\phi_1(\bs{x}, t) \,n_{2,\theta}(\bs{x},t) = \int \frac{d^2 \bs{q}}{(2\pi)^2} \, \phi_1(\bs{q}, t) \,n_{2,\theta}(-\bs{q}, t) \,.
\end{equation}
This convolution structure implies that the anomaly equations for $n_{a, \theta}, \phi_m$ are no longer a closed set of equations. 
As a result, both the fermionic and bosonic contributions to counterflow are allowed to have universal scaling contributions. Whether or not this incoherent quantum critical transport is realized can only be determined through a controlled diagrammatic calculation, which is the content of Section~\ref{subsec:FLbilayer_pert}.

\subsection{CFL-\texorpdfstring{$\overline{\text{CFL}}$}{} bilayer}

Next, we generalize to the CFL-$\overline{\text{CFL}}$ bilayer. The mid-IR action is almost identical to the FL bilayer action in Eq.~\eqref{eq:midIRaction_FLbilayer} except for the inclusion of additional fluctuating gauge fields $a_{\pm}$
\begin{equation}\label{eq:midIRaction_ACFLbilayer}
    \begin{aligned}
        S[c,b,a_+, a_-] &= \int_{\tau, \bs{k}} \sum_{\theta} c^{\dagger}_{\theta,1}(\bs{k},\tau) \left[\partial_{\tau} - i (a_{+,0} + a_{-,0}) - \epsilon_{\theta}(\bs{k}) + \bs{v}(\theta) \cdot (\bs{a}_+ + \bs{a}_-)\right]c_{\theta,1}(\bs{k},\tau) \\
        &+\int_{\tau, \bs{k}} \sum_{\theta} c^{\dagger}_{\theta,2}(\bs{k},\tau) \left[\partial_{\tau} - i (a_{+,0} - a_{-,0})- \epsilon_{\theta}(\bs{k}) + \bs{v}(\theta) \cdot (\bs{a}_+ - \bs{a}_-)\right]c_{\theta,2}(\bs{k},\tau)\\
        &+ \int_{\tau, \bs{r}} \sum_{\theta} \frac{1}{2} g(\theta) \left(\phi_1 \left[c^{\dagger}_{\theta,1} c_{\theta,2} + c^{\dagger}_{\theta,2} c_{\theta,1}\right] + \phi_2 \left[-i c^{\dagger}_{\theta,1} c_{\theta,2} + i c^{\dagger}_{\theta,2} c_{\theta,1}\right]\right) \\
        &+ \frac{1}{2} \int_{\tau, \bs{r}} \sum_m \phi_m(\bs{r},\tau) \left[\partial_{\tau} - i a_{-,0} - \mu_b - (\nabla - i \bs{a}_-)^2\right] \phi_m(\bs{r},\tau) \\
        &+ \frac{1}{2} \int_{\tau, \bs{q}} \left|\bs{a}_{+}(\bs{q}, \tau)\right|^2 |\bs{q}|^2 + \frac{1}{2} \int_{\tau, \bs{q}} \left|\bs{a}_{-}(\bs{q}, \tau)\right|^2 |\bs{q}| + \frac{1}{4\pi} a_+ d a_-   \,.
    \end{aligned}
\end{equation}
In writing down the mid-IR action, we have expanded the coupling between fermions and the gauge fields $\epsilon_{\theta}(\bs{k} + \bs{a})$ to linear order in $\bs{a}$, since higher order terms are strongly irrelevant upon scaling towards the Fermi surface. We have also dropped the frequency dependence of the bare gauge field propagator, as it is always overwhelmed by Landau damping effects. 

We now examine the symmetries and anomalies of the mid-IR action. The quadratic action for the composite fermions again enjoys a $U(2)$ symmetry that rotates between the two layers. The various interactions can be written as minimal couplings between $\phi, a_+, a_-$ and generators of the $U(2)$ algebra
\begin{equation}
    S \supset \int_{\bs{r}, \tau} \sum_{\theta} g(\theta) \left[\phi_1 n_{1,\theta} + \phi_2 n_{2,\theta}\right] + \int_{\bs{r}, \tau} \sum_{\theta} 2\bs{a}_+ \cdot \bs{J}_{0,\theta} + \int_{\bs{r}, \tau} \sum_{\theta} 2\bs{a}_- \cdot \bs{J}_{3,\theta} \,,
\end{equation}
where we adopt the notations for patch charge densities and currents introduced earlier in this Appendix. When $U(2)$ currents are coupled to external gauge fields, there is again a t'Hooft anomaly which induces a set of deformed Ward identities as in Eq.~\eqref{eq:U(2)anomaly_FLbilayer}. In the mid-IR action for the CFL-$\overline{\text{CFL}}$ bilayer, the bosonic fields $a_+, a_-, \phi$ can be regarded as $U(2)$ gauge fields if we make the following identifications
\begin{equation}
    A_{\perp,0} = 2\bs{a}_+ \cdot \hat v(\theta) \,, \quad A_{\perp, 1} = \frac{g(\theta)}{v(\theta)} \phi_1 \,, \quad A_{\perp, 2} = \frac{g(\theta)}{v(\theta)} \phi_2 \,, \quad A_{\perp, 3} = 2\bs{a}_- \cdot \hat v(\theta) \,, \quad A_{\parallel, i} = 0 \,.
\end{equation}
The anomaly equations at $\bs{q} \rightarrow 0$ then take a slightly more complicated form compared with Eq.~\eqref{eq:U(2)_anomaly_q=0_FLbilayer}
\begin{equation}\label{eq:U(2)_anomaly_q=0_ACFLbilayer}
    \begin{aligned}
    \partial_t n_{0, \theta} &= - \frac{\Lambda(\theta)}{(2\pi)^2} \partial_t \bs{a}_+ \cdot \hat v(\theta) \,,  \\
    \partial_t n_{1,\theta} - i g(\theta) \phi_2 n_{3,\theta} - i \bs{a}_- \cdot \bs{v}(\theta) n_{2,\theta} &= - \frac{1}{2} \frac{\Lambda(\theta)}{(2\pi)^2} \frac{g(\theta)}{v(\theta)} \partial_t \phi_1 \,,\\
    \partial_t n_{2,\theta} - i g(\theta) \phi_1 n_{3,\theta} + i \bs{a}_- \cdot \bs{v}(\theta) n_{1,\theta} &= -\frac{1}{2} \frac{\Lambda(\theta)}{(2\pi)^2} \frac{g(\theta)}{v(\theta)} \partial_t \phi_2\,, \\
    \partial_t n_{3,\theta} + i g(\theta) \left[\phi_1 n_{2,\theta} + \phi_2 n_{1,\theta}\right] &= -\frac{\Lambda(\theta)}{(2\pi)^2} \partial_t \bs{a}_- \cdot \hat v(\theta) \,. 
    \end{aligned}
\end{equation}
Another set of bosonic constraints are provided by the equations of motion for $\phi, a_+, a_-$. After adding a source term $h_{\phi}, h_+, h_-$ for $\phi, a_+, a_-$, we find 
\begin{equation}\label{eq:aphi_eom_ACFLbilayer}
    \begin{aligned}
    |-i\nabla|^2 a_{+,m} + \sum_n \epsilon_{mn} \partial_t a_{-,n} - 2 \sum_{\theta} J_{0,\theta,m} &= h_+ \,, \\
    |-i\nabla| a_{-,m} - \sum_n \epsilon_{mn} \partial_t a_{+,n} - 2\sum_{\theta} J_{3,\theta,m} - J_{\phi,m} - a_{-,m} \sum_n \phi^*_n \phi_n &= h_- \,, \\
    \left[\partial_{\tau} - i a_0 - \mu - (\nabla - i \bs{a}_-)^2\right] \phi_i &= - \sum_{\theta} g(\theta) n_{i,\theta} + h_{\phi} \,,
    \end{aligned}
\end{equation}
where we defined the current operator for $\phi$
\begin{equation}
    J_{\phi,m} = \sum_n \phi_n^*(\bs{r},\tau) (-i\nabla_m) \phi_n(\bs{r}, \tau)  \,.
\end{equation}
We now contrast Eq.~\eqref{eq:U(2)_anomaly_q=0_ACFLbilayer} and Eq.~\eqref{eq:aphi_eom_ACFLbilayer} with the analogous equations Eq.~\eqref{eq:U(2)_anomaly_q=0_FLbilayer} and Eq.~\eqref{eq:phi_eom_FLbilayer} for the Fermi liquid bilayer. First of all, the inclusion of two emergent gauge fields breaks the exact conservation of $n_{0,\theta}$. This is to be expected because gauge fields couple directly to $n_{0,\theta}$ and deform their Ward identities, in line with the arguments of Ref.~\cite{shi2022_gifts}. Second of all, the non-Abelian Ward identities manifest themselves as a direct coupling between $a_-$ and $n_{1,\theta}, n_{2,\theta}$ in the anomaly equations, despite the fact that $a_-$ only couples to $n_{0,\theta}, n_{3,\theta}$ in the action. These facts complicate the transport calculations. 

Nevertheless, we now demonstrate that the charge conductivity at the IR fixed point can be determined by these non-perturbative constraints. To calculate the charge conductivity, we need the correlators of the total charge current $\bs{J} = 2\sum_{\theta} \bs{v}(\theta) n_{0,\theta}$. By the anomaly equations in Eq.~\eqref{eq:U(2)_anomaly_q=0_ACFLbilayer}, we can rewrite the current-current correlator as
\begin{equation}
    \begin{aligned}
    G_{J^p J^p}(\bs{q}= 0,\Omega) &= 4 \sum_{\theta, \theta'} v^p(\theta) v^p(\theta') G_{n_{0,\theta}, n_{0,\theta'}}(\bs{q} = 0, \Omega) \\
    &= 4 \sum_{\theta, \theta'} v^p(\theta) v^p(\theta') \frac{\Lambda(\theta) \Lambda(\theta')}{(2\pi)^4} G_{a_{+,n} a_{+,m}}(\bs{q}=0,\Omega) \hat v^m(\theta) \hat v^n(\theta')  \,.
    \end{aligned}
\end{equation}
Invoking the anomaly equation for $n_{0,\theta}$ in Eq.~\eqref{eq:U(2)_anomaly_q=0_ACFLbilayer} allows us to simplify the equations of motion for $a_+$ in Eq.~\eqref{eq:aphi_eom_ACFLbilayer}
\begin{equation}
    \begin{aligned}
    2\sum_{\theta} J_{0, \theta, m}(\bs{q}=0, \Omega) &= 2\sum_{\theta} v(\theta) n_{0,\theta}(\bs{q}=0,\Omega) \\
    &= - \sum_{\theta} \frac{2 \Lambda(\theta)}{(2\pi)^2 v(\theta)} \sum_j v^n(\theta) v^m(\theta) a_{+,n}(\bs{q}=0,\Omega) = h a_{+,m}(\bs{q}=0,\Omega) \,.
    \end{aligned}
\end{equation}
From the last equation, we directly infer the self-energy for $a_+$ in the $\bs{q} \rightarrow 0$ limit which is exact in the mid-IR theory
\begin{equation}
    \Pi_{nm}(\bs{q}=0,\Omega) = \sum_{\theta} \frac{2\Lambda(\theta)}{(2\pi)^2 v(\theta)} v^n(\theta) v^m(\theta) \,, \quad G_{a_{+,n} a_{+,m}}(\bs{q}=0,\Omega) = -\left(\Pi^{-1}\right)_{nm}(\bs{q}=0, \Omega) \,. 
\end{equation}
Plugging this exact Green's function into the current-current correlator yields
\begin{equation}\label{eq:totalcharge_conductivity}
    \begin{aligned}
        G_{J^pJ^p}(\bs{q}=0,\Omega) &= \sum_{n,m} \Pi_{pn}(\bs{q}=0,\Omega) G_{a_{+,n} a_{+,m}}(\bs{q}=0,\Omega) \Pi_{mp}(\bs{q}=0,\Omega) \\
        &= -\left(\Pi \cdot \Pi^{-1} \cdot \Pi \right)_{p,p}(\bs{q}=0,\Omega) = -\Pi_{p,p}(\bs{q}=0,\Omega) = -2\Pi_0 \,,
    \end{aligned}
\end{equation}
where we have defined $\Pi_0$, a positive constant independent of $i, \Omega$, which is the self-energy for a single gauge field coupled to a spinless Fermi surface. This factor of 2 makes sense because $\bs{a}_+$ couples to both layers. As a result, $G_{J^p J^p}(\bs{q}=0, \Omega)$ is also a constant and gives rise to a Drude weight $2\pi \Pi_0$ in the conductivity with no incoherent corrections
\begin{equation}
    \sigma_{\rm charge}(\bs{q}=0, \omega) = \frac{2i \Pi_0}{\omega} \,.
\end{equation}
The presence of Chern-Simons terms can give frequency-independent corrections to the conductivity, but cannot generate a nontrivial incoherent term. 

\section{Ioffe-Larkin rule for the CFL-\texorpdfstring{$\overline{\text{CFL}}$}{} bilayer}\label{app:ioffe_larkin}

In low energy effective theories with partons and emergent gauge fields, it is often the case that different partons carry different charges under emergent and external vector potentials. In these scenarios, the physical conductivity is given by the response of the physical current to the external gauge field, but parton conductivities (i.e. the response of each parton current to the emergent and external fields to which it couples) are typically easier to calculate. Therefore, the first step in any transport calculation is to derive a relationship between the physical conductivity and the parton conductivities, often referred to as ``Ioffe-Larkin composition rules''. In what follows, we will briefly review how the Ioffe-Larkin rule works for a single layer CFL, before generalizing to the coupled CFL-$\overline{\text{CFL}}$ bilayer. 

For the single layer CFL, the standard HLR action can be written in the following partonic form
\begin{align}
    S_{\rm CFL} &= \int_{\bs{x},t} c^{\dagger}[i\partial_t + a_t + A_t - \epsilon(\bs{k} + \bs{a} + \bs{A}) ] c + \int_{\bs{x},\bs{x}',t} \rho(\bs{x},t) V(\bs{x}-\bs{x}') \rho(\bs{x}',t) \nonumber\\
    &\quad +\int_{\bs{x},t}\left(-\frac{1}{2\pi} a db - \frac{2}{4\pi} b db-a_t\overline\rho\right) \,.
\end{align}
To obtain this action, one can fractionalize the microscopic electron into $\psi = \Phi^* c$ such that $\Phi, c$ both couple to an emergent $U(1)$ gauge field $a$. The Chern-Simons terms for $a, b$ define a topological field theory that corresponds to putting $\Phi$ in a $\nu = 1/2$ bosonic Laughlin state. 

Let us assign a unit of physical charge to $c$ and $\psi$. This means that $\Phi$ is neutral under the external gauge field. On the other hand, under the emergent gauge field, $\psi$ is neutral and $c, \Phi$ carry unit charge. Given these charge assignments, we can define the following parton conductivities:
\begin{equation}
    \ev{\bs{J}^{(c)}}_{\bs{A}} = \Pi^{(c)} (\bs{A} + \ev{\bs{a}}_{\bs{A}}) + \mathcal{O}(A^2) \,, \quad \ev{\bs{J}^{(b)}}_{\bs{A}} = \Pi^{(b)} \ev{\bs{a}}_{\bs{A}} + \mathcal{O}(A^2) \,.
\end{equation}
In writing down these equations, we have fixed a gauge and assumed that in the linear response regime (i.e. small $|\bs{A}|$), $\ev{\bs{a}}_{\bs{A}}$ is linear in $\bs{A}$. On the other hand, since only $c$ carries charge under the external gauge field, the total physical conductivity is given by 
\begin{equation}
    \ev{\bs{J}^{(c)}}_{\bs{A}} = \Pi \bs{A} + \mathcal{O}(A^2)  \,.
\end{equation}
Now given the gauge constraint $\bs{J}^{(b)} + \bs{J}^{(c)} = 0$, we can solve for $\ev{a}_{\bs{A}}$ in terms of $\bs{A}$ as
\begin{equation}
    -\Pi^{(b)} \ev{\bs{a}}_{\bs{A}} = \Pi^{(c)} (\bs{A} + \ev{\bs{a}}_{\bs{A}}) \quad \rightarrow \quad \ev{a}_{\bs{A}} = -\frac{\Pi^{(c)}}{\Pi^{(b)} + \Pi^{(c)}} \bs{A} \,,
\end{equation}
which immediately implies that 
\begin{equation}
    \ev{\bs{J}}_{\bs{A}} = \Pi^{(c)} \left(\bs{A} - \frac{\Pi^{(c)}}{\Pi^{(b)} + \Pi^{(c)}} \bs{A}\right) = \frac{\Pi^{(b)} \Pi^{(c)}}{\Pi^{(b)} + \Pi^{(c)}} \bs{A} \,. 
\end{equation}
By inverting this relationship and recalling that $\bs{A} = - \partial_t \bs{E}$, we arrive at the familiar form of the Ioffe-Larkin rule
\begin{equation}
    \sigma^{-1} = \sigma_b^{-1} + \sigma_{c}^{-1} \,.
\end{equation}
In the CFL, the bosonic parton is in a $\nu=1/2$ bosonic fractional quantum Hall state. Therefore $\sigma_b$ is purely off-diagonal and quantized. On the other hand, $\sigma_{c}$ is the conductivity of the composite fermion which reduces to the free fermion conductivity in the RPA approximation. 

Next we generalize to the coupled CFL-$\rm \overline{CFL}$ bilayer. We now have two species of internal gauge fields $a_1, a_2$ and two species of external gauge fields $A_1, A_2$. The action that includes all of these contributions is
\begin{equation}
    S = S_{\rm CFL} + S_{\rm \overline{CFL}} + S_{\rm int}
\end{equation}
where 
\begin{align}
    S_{\rm CFL}[c_1, a_1, b_1, A_1] &= \int_{\bs{x},t} c^{\dagger}_1 \left[i \partial_t + a_{1,t} + A_{1,t} - \epsilon(\bs{k} + \bs{a}_1 + \bs{A}_1)\right] c_1 \nonumber\\ &\quad+ \int_{\bs{q},\Omega}\frac{|\bs{q}|}{2g^2} \left|a_1^T(\bs{q}, \Omega)\right|^2 - \int_{\bs{x},t} \left(\frac{1}{2\pi} a_1 d b_1 + \frac{2}{4\pi} b_1 d b_1\right)   \,, \\
    S_{\rm \overline{CFL}}[c_2, a_2, b_2, A_2] &= \int_{\bs{x},t} c^{\dagger}_2 \left[i \partial_t + a_{2,t} + A_{2,t} - \epsilon(\bs{k} + \bs{a}_2 + \bs{A}_2)\right] c_2 \nonumber\\&\quad+ \int_{\bs{q},\Omega}\frac{|\bs{q}|}{2g^2}\left|a_2^T(\bs{q}, \Omega)\right|^2 + \left(\frac{1}{2\pi} a_2 d b_2 + \frac{2}{4\pi} b_2 d b_2\right)  \,,\\
    S_{\rm int}[a_1,a_2] &= - \frac{1}{2g^2} \int_{\bs{q},\Omega} \left[a_1^T(\bs{q}, \Omega) a_2^T(-\bs{q}, - \Omega) + a_2^T(\bs{q}, \Omega) a_1^T(-\bs{q}, - \Omega)\right] |\bs{q}| e^{-|\bs{q}| d} \,.
\end{align}
Now an additional complication arises: the layers are coupled together via a flux-flux interaction. Therefore we need to introduce general parton response matrices with indices in layer space
\begin{equation}
    \ev{\bs{J}^{(c)}_n} = \Pi^{(c)}_{nm} (\bs{A}_m + \ev{\bs{a}_m}) \,, \quad \ev{\bs{J}^{(b)}_n} = \Pi^{(b)}_{nm} \ev{\bs{a}_m} \,.
\end{equation}
The gauge constraints $\bs{J}_{c_n} + \bs{J}_{b_n} = 0$ can be written as a matrix equation
\begin{equation}
    \left(\Pi^{(c)} + \Pi^{(b)} \right)_{nm} \ev{\bs{a}_m} = - \Pi^{(c)}_{nm} \bs{A}_m \quad \rightarrow \quad \ev{\bs{a}_n} = -\left[\left(\Pi^{(c)} + \Pi^{(b)} \right)^{-1} \Pi^{(c)}\right]_{nm} \bs{A}_m  \,.
\end{equation}
Assuming that the matrix is invertible, the physical response function can be written as
\begin{equation}
    \Pi = \left[1 - \Pi^{(c)} \left(\Pi^{(c)} + \Pi^{(b)} \right)^{-1}\right] \Pi^{(c)} = \Pi^{(b)} \left(\Pi^{(c)} + \Pi^{(b)} \right)^{-1} \Pi^{(c)} \,,
\end{equation}
which implies a generalized Ioffe-Larkin rule 
\begin{equation}
    \sigma^{-1} = \sigma_c^{-1} + \sigma_b^{-1} \,,
\end{equation}
where $\sigma_c, \sigma_b$ are both matrices in layer space.  

Finally, we come to the quantum critical point associated with the phase transition from the CFL-$\overline{\text{CFL}}$ bilayer metal to the EI* phase. 
The critical theory has an emergent gapless boson $\phi \sim c^{\dagger}_2 c_1$, which carries $+1$ charge under $\bs{a}_1, \bs{A}_1$ and $-1$ charge under $\bs{a}_2, \bs{A}_2$. This charge assignment implies new terms in the effective action 
\begin{equation}
    S \supset \int \phi\, c^{\dagger}_1 c_2 + h.c. + \int \phi^* \left[i\partial_t + a_{1t} - a_{2t} + A_{1t} - A_{2t} - \frac{(q + \bs{a}_1 - \bs{a}_2 + \bs{A}_1 - \bs{A}_2)^2}{2}\right] \phi \,.
\end{equation}
These additional terms modify the gauge constraints to
\begin{equation}
    \phi^* \phi + c^{\dagger}_1 c_1 - \frac{\nabla \times \bs{b}_1}{2\pi} = 0 \,, \quad -\phi^* \phi + c^{\dagger}_2 c_2 + \frac{\nabla \times \bs{b}_2}{2\pi} = 0 \,,
\end{equation}
\begin{equation}
    \bs{J}^{(\phi)}_n + \bs{J}^{(c)}_n + \bs{J}^{(b)}_n = 0 \,, \quad \bs{J}^{(\phi)}_1 = - \bs{J}^{(\phi)}_2 = \int \phi^* (-i \nabla) \phi \,.
\end{equation}
Note that although $\bs{J}^{(\phi)}_1, \bs{J}^{(\phi)}_2$ are redundant, it is useful to use the vector notation to streamline the matrix inversion later. Also note that the parton currents for $b$ are defined with opposite signs as $\bs{J}^{(b)}_1 = -\frac{1}{2\pi} \epsilon \partial_t \bs{b}_{1}$ and $\bs{J}^{(b)}_2 = \frac{1}{2\pi} \epsilon \partial_t \bs{b}_{2}$. Now we can introduce parton response functions 
\begin{equation}
    \ev{\bs{J}^{(c)}_n} = \sum_m \Pi^{(c)}_{nm} (\bs{A}_m + \ev{\bs{a}_m}) \,, \hspace{0.2cm} \ev{\bs{J}^{(b)}_n} = \sum_m \Pi^{(b)}_{nm} (\ev{\bs{a}_m})\,, \hspace{0.2cm} \ev{\bs{J}^{(\phi)}} = \Pi^{(\phi)}_{nm} (\bs{A}_{m} + \ev{\bs{a}_m}) \,.
\end{equation}
The gauge constraints reduce to a simple matrix equation
\begin{equation}
    \left(\Pi^{(c)} + \Pi^{(b)} + \Pi^{(\phi)}\right)_{nm} \ev{\bs{a}_m} = - \left(\Pi^{(c)} + \Pi^{(\phi)}\right)_{nm} \bs{A}_m \,.
\end{equation}
Inverting this relation, we obtain the physical response function in matrix notation
\begin{equation}
    \begin{aligned}
    \Pi &= \left[\Pi^{(c)} + \Pi^{(\phi)}\right] \cdot \left[1 - \left(\Pi^{(c)} + \Pi^{(\phi)}\right) \left(\Pi^{(c)} + \Pi^{(\phi)} + \Pi^{(b)}\right)^{-1} \left(\Pi^{(c)} + \Pi^{(\phi)}\right)\right]  \\
    &= \left[\Pi^{(c)} + \Pi^{(\phi)}\right] \left(\Pi^{(c)} + \Pi^{(\phi)} + \Pi^{(b)}\right)^{-1} \Pi^{(b)} \,,
    \end{aligned} 
\end{equation}
from which the Ioffe-Larkin rule follows
\begin{equation}
    \sigma^{-1} = \sigma_b^{-1} + (\sigma_c + \sigma_{\phi})^{-1} \,.
\end{equation}
As a sanity check, note that in the absence of the emergent Chern-Simons gauge field, $\sigma_b^{-1} = 0$ and $\sigma = \sigma_c + \sigma_{\phi}$, which is the formula for the excitonic QCP in the Fermi liquid bilayer. 

\section{Technical details in the quantum critical transport calculation}\label{app:transport}

\subsection{Diagrammatic calculation of the conductivity at the excitonic QCP of electrons}\label{app:diagram_FLbilayer}

As explained in Section~\ref{subsec:FLbilayer_pert}, the complete layer-resolved conductivity matrix can be inferred from $G_{J_1J_1}(\bs{q}=0,\Omega)$, the current-current correlation function for fermions in layer 1. In this section, we evaluate this correlation function and derive the result Eq.~\eqref{eq:maintext_FLbilayer_JJ} stated in Section~\ref{subsec:FLbilayer_pert}. 

\begin{figure}[h!]
    \centering
    \includegraphics[width = \textwidth]{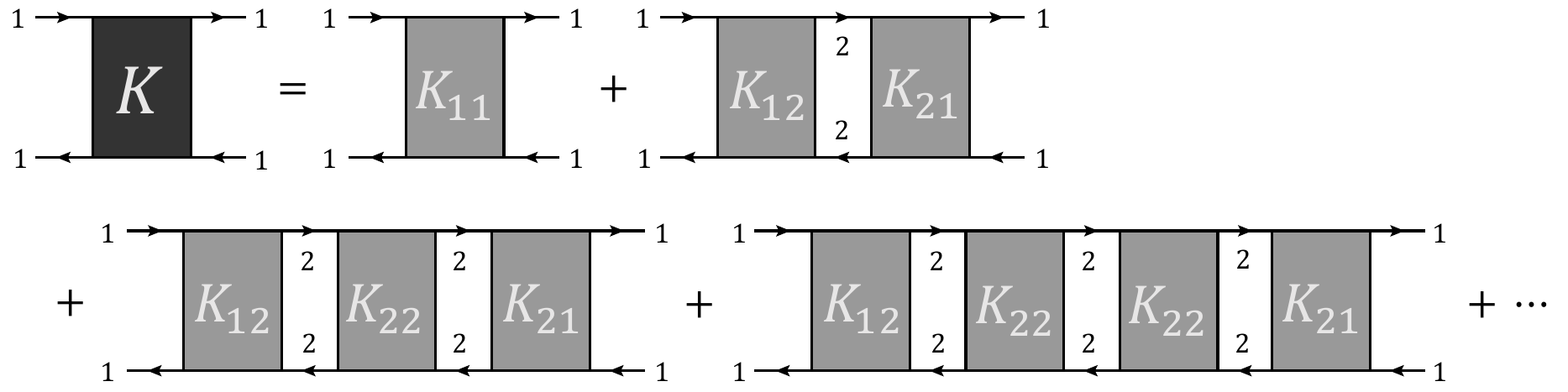}
    \includegraphics[width = \textwidth]{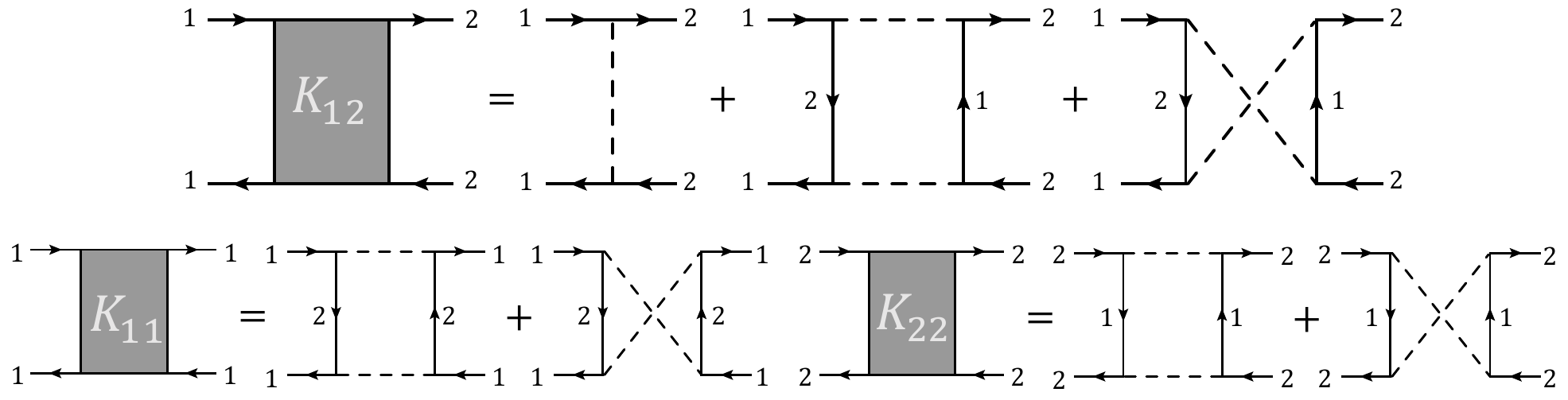}
    \caption{Diagrams that contribute to the kernel $K$ in Fig.~\ref{fig:ph_reducible_FLbilayer} at leading order in the double expansion of Ref.~\cite{Mross2010}. Note that $K_{21}$ is just the transpose of $K_{12}$ and hence not explicitly defined.}
    \label{fig:K_Kphi_FLbilayer}
\end{figure}

Generally, $G_{J_1J_1}(\bs{q}=0,\Omega)$ can be obtained by summing a geometric series of the particle-hole irreducible kernel $K$ which is represented in Fig.~\ref{fig:ph_reducible_FLbilayer}. Within the double expansion scheme, the kernel can be written as
\begin{equation}
    K = \sum_{n=0}^{\infty} N^{-n} K^{(n)} \,.
\end{equation}
where the set of diagrams that contribute to $K^{(n)}$ can be classified by an explicit recipe in Ref.~\cite{Shi2023_mross}. To leading order, the kernel takes the form $K^{(0)} = K_{11} + K_{12} (1 - K_{22})^{-1} K_{21}$ where $K_{ij}$ are defined graphically in Fig.~\ref{fig:K_Kphi_FLbilayer}. 
When the external legs are contracted with fermion propagators, the three terms in $K_{12}, K_{21}$ can be identified as the familiar Maki-Thompson and Aslamazov-Larkin diagrams. In contrast, due to the layer index structure, $K_{11}, K_{22}$ only contain Aslamazov-Larkin diagrams.

Guided by the pictorial representation, we can write down the geometric sum in equation form, using a compact braket notation. Let each ket vector $\ket{F}$ represents a pair of outgoing particle-hole lines with two momentum coordinates and two frequency coordinates. In the simplest case of an external current vertex insertion at momentum $\bs{q}$ and frequency $\Omega$, the component representation is
\begin{equation}
    \bra{\bs{k}, \bs{q}, \omega, \Omega}\ket{J_{\bs{\bar q}=0,\bar \Omega}} = v_F(\bs{k}) \,\delta^2(\bs{q} - \bs{\bar q}) \,\delta(\Omega - \bar \Omega) \,. 
\end{equation}
Each particle-hole kernel is a matrix acting on ket vectors $\ket{F}$ and hence carry four momentum and four frequency components. For example, the one-loop fermion bubble diagram $\Pi_{\rm 1-loop}$ for layer 1 can be represented as the expectation value of the simplest particle-hole kernel $W_G$
\begin{equation}
    \Pi_{\rm 1-loop}(\bs{q} = 0, \Omega) = \bra{J_{1, \bs{q}=0, \Omega}} W_G \ket{J_{1, \bs{q}=0, \Omega}} \,, 
\end{equation}
where
\begin{equation}
    \begin{aligned}
    \bra{\bs{k},\bs{q}, \omega, \Omega}W_G\ket{\bs{k}',\bs{q}',\omega', \Omega} &= \delta^2(\bs{k} - \bs{k}') \,\delta^2(\bs{q} - \bs{q'})\, \delta(\omega - \omega') \,\delta(\Omega - \Omega') \\
    &\cdot G(\bs{k} + \bs{q}/2, i\omega + i\Omega/2) \, G(\bs{k} - \bs{q}/2, i\omega-i\Omega/2) \,,
    \end{aligned}
\end{equation}
is nothing but the product of two fermion propagators running in parallel.

Using this braket notation, the geometric sum of diagrams can be compactly represented as
\begin{equation}
    G_{J_1 J_1}(\bs{q}=0,\Omega) = \bra{J_{1, \bs{q}=0, \Omega}} W_G \left[I - K^{(0)}\right]^{-1} \ket{J_{1, \bs{q}=0, \Omega}} \,, \quad K^{(0)} = K_{11} + K_{12} (1 - K_{22})^{-1} K_{21} \,,
\end{equation}
where, following the graphical representation in Fig.~\ref{fig:K_Kphi_FLbilayer}, we decompose the subkernels $K_{ij}$ as
\begin{equation}\label{eq:Kform_FLbilayer}
    K_{12} = K_{21} = K_{\rm MT, \phi} + K_{\rm AL1, \phi} + K_{\rm AL2, \phi} \,, \quad K_{11} = K_{22} = K_{\rm AL1, \phi} + K_{\rm AL2, \phi} \,.
\end{equation}
The integral representations of these kernels immediately follow from the diagrams. When acting on a particle-hole vertex $F$ with total energy-momentum $(\omega_1, \bs{k}_1)$ and relative energy-momentum $(\Omega, \bs{q} = 0)$, we have
\begin{equation}
    \begin{aligned}
    \left(K_{\rm MT, \phi} F\right)(\bs{k}_1, \omega_1, \Omega) &= g_{\phi}^2 \,G(\bs{k}_1, i\omega_1 + i\Omega/2) \,G(\bs{k}_1, i\omega_1 - i\Omega/2) \\
    & \hspace{2cm} \int_{\bs{k}_2, \omega_2} D_{\phi}(\bs{k}_1 - \bs{k}_2, i\omega_1 - i\omega_2)\, F(\bs{k}_2, \omega_2, \Omega) \,,
    \end{aligned}
\end{equation}
\begin{equation}
    \begin{aligned}
    &\left(K_{\rm AL1, \phi} F + K_{\rm AL2, \phi} F\right)(\bs{k}_1, \omega_1, \Omega) = g_{\phi}^4\, G(\bs{k}_1, i\omega_1 + i\Omega/2)\, G(\bs{k}_1, i\omega_1 - i\Omega/2) \\
    &\hspace{1cm} \int_{\bs{q}_1, \Omega_1} G(\bs{k}_1-\bs{q}_1, i\omega_1 + i\Omega/2 - i \Omega_1)\, D_{\phi}(\bs{q}_1, i\Omega_1)\, D_{\phi}(\bs{q}_1, i\Omega_1 - i\Omega) \\
    &\hspace{1cm} \int_{\bs{k}_2, \omega_2} \left[G(\bs{k}_2 - \bs{q}_1, i\omega_2 + i\Omega/2 - i \Omega_1) + G(\bs{k}_2 + \bs{q}_1, i\omega_2 - i\Omega/2 + i \Omega_1)\right]\, F(\bs{k}_2, \omega_2, \Omega) \,.
    \end{aligned}
\end{equation}
To compute the most singular contributions to the layer-1 conductivity, we need to find soft modes (i.e. near-zero eigenvalues) of $1 - K^{(0)} = 1 - K_{11} - K_{12} (1-K_{22})^{-1} K_{21}$. Generally, there are infinitely many such soft modes $\ket{\theta}$ originating from the infinitely many emergent conserved charges $n_{\theta}$ localized to every angle $\theta$ on the Fermi surface~\cite{shi2022_gifts,shi2023_loop,guo2022_largeN,guo2023migdal,guo2023fluctuation}. Passing to an angular momentum representation $\ket{l} = \int_0^{2\pi} e^{i l \theta} \ket{\theta}$, we can rewrite the most singular part of the current-current correlation function as
\begin{equation}
    G_{J_1J_1}(\bs{q}=0,\Omega) \approx \sum_{l\in \mathbb{Z}} \bra{J_{1,\bs{q}=0,\Omega}} W_G \ket{l} (1 - k^{(0)}_l)^{-1} \bra{l}\ket{J_{1,\bs{q}=0,\Omega}} \,,
\end{equation}
where the complex numbers $k^{(0)}_l$ are defined by the eigenvalue equation $K^{(0)} \ket{l} = k^{(0)}_l \ket{l}$. 

For a generic Fermi surface shape, $J_1$ is proportional to the Fermi velocity $v_F(\theta)$ and hence overlaps with all modes with odd angular momentum. The contributions of these different modes to $G_{J_1J_1}(\bs{q}=0, \Omega)$ have identical frequency scaling but distinct numerical prefactors. Since we are only concerned with frequency scaling, we impose rotational invariance so that $J_1$ only overlaps with a single mode $\ket{F_1} = \frac{1}{\sqrt{2}} [\ket{l = 1} + \ket{l = -1}]$ which can be expanded as a function of frequency and momentum
\begin{equation}
    \bra{\bs{k}, \omega, \Omega}\ket{F_1} = \cos \theta_{\bs{k}} \, \left[G(\bs{k}, i\omega + i\Omega/2) - G(\bs{k}, i\omega - i\Omega/2)\right] \,.
\end{equation}
It is easy to compute the action of various subkernels on $\ket{F_1}$ and extract the frequency scalings:
\begin{equation}
    \bra{\bs{k}, \omega, \Omega}K_{\rm MT, \phi} \ket{F_1} \approx \frac{i C_{\phi} \left[\sgn(\omega + \frac{\Omega}{2})|\omega+\frac{\Omega}{2}|^{\frac{2}{2+\epsilon}} - \sgn(\omega - \frac{\Omega}{2})|\omega-\frac{\Omega}{2}|^{\frac{2}{2+\epsilon}}\right]}{i \Omega + i C_{\phi}\left[\sgn(\omega + \frac{\Omega}{2})|\omega+\frac{\Omega}{2}|^{\frac{2}{2+\epsilon}} - \sgn(\omega - \frac{\Omega}{2})|\omega-\frac{\Omega}{2}|^{\frac{2}{2+\epsilon}}\right]} \bra{\bs{k}, \omega, \Omega}\ket{F_1}  \,,
\end{equation}
\begin{equation}
    \bra{\bs{k}, \omega, \Omega} (K_{\rm AL1, \phi} + K_{\rm AL2,\phi}) \ket{F_1} \sim |\Omega|^{\frac{2}{2+\epsilon}}  \bra{\bs{k}, \omega, \Omega}\ket{F_1}\,.  
\end{equation}
From these equations, we see that in the IR limit $\Omega \rightarrow 0$, $\ket{F_1}$ becomes a simultaneous eigenvector of $K_{ij}$ for $i,j = 1, 2$ with eigenvalue 0 for $K_{11}, K_{22}$ and eigenvalue 1 for $K_{12}, K_{21}$. For small but nonzero $\Omega$, the corrections to these eigenvalues can be computed from first order perturbation theory (see Ref.~\cite{guo2022_largeN,shi2023_loop} for details). The end result turns out to be
\begin{equation}
    k_{11} = k_{22} \approx \alpha \, |\Omega|^{\frac{2}{2+\epsilon}} \,, \quad k_{12} = k_{21} \approx 1 - \beta |\Omega|^{\frac{\epsilon}{2+\epsilon}} \,,
\end{equation}
for some real constant $\alpha,\beta$. This implies that
\begin{equation}
    \begin{aligned}
    k^{(0)}_1 &= \alpha |\Omega|^{\frac{2}{2+\epsilon}} + \left[1 - \beta |\Omega|^{\frac{\epsilon}{2+\epsilon}}\right] \left(1 - \alpha |\Omega|^{\frac{2}{2+\epsilon}}\right)^{-1} \left[1 - \beta |\Omega|^{\frac{\epsilon}{2+\epsilon}}\right] \\
    &= 1 - 2 \beta |\Omega|^{\frac{\epsilon}{2+\epsilon}} + \beta^2 |\Omega|^{\frac{2\epsilon}{2+\epsilon}} + \mathcal{O}\left(|\Omega|^{\frac{2}{2+\epsilon}}\right) \,.
    \end{aligned}
\end{equation}
Therefore, the most singular contributions to the current-current correlator takes the form
\begin{equation}\label{eq:FLbilayer_JJ_correlator}
    \begin{aligned}
    G_{J_1J_1}(\bs{q}=0, i\Omega) &\approx \frac{\bra{J_{1,\bs{q}=0,\Omega}} W_G \ket{F_1} \bra{F_1}\ket{J_{1,\bs{q}=0,\Omega}}}{1 - k^{(0)}_1} = \frac{\bra{J_{1,\bs{q}=0,\Omega}} W_G \ket{F_1} \bra{F_1}\ket{J_{1,\bs{q}=0,\Omega}}}{2\beta |\Omega|^{\frac{\epsilon}{2+\epsilon}} \left[1 - \beta/2 |\Omega|^{\frac{\epsilon}{2+\epsilon}}\right] + \mathcal{O}(|\Omega|^{\frac{2}{2+\epsilon}})} \,.
    \end{aligned}
\end{equation}
The numerator in Eq.~\eqref{eq:FLbilayer_JJ_correlator} has been evaluated in the single layer transport calculations of Ref.~\cite{guo2022_largeN, shi2023_loop,guo2023migdal,guo2023fluctuation}
\begin{equation}
    \bra{J_{1,\bs{q}=0,\Omega}} W_G \ket{F_1} \bra{F_1}\ket{J_{1,\bs{q}=0,\Omega}} = -\beta \Pi_0 |\Omega|^{\frac{\epsilon}{2+\epsilon}} \,.
\end{equation}
Plugging this result back into the current-current correlator implies that 
\begin{equation}
    G_{J_1J_1}(\bs{q}=0, i\Omega) \approx \frac{- \beta \Pi_0 |\Omega|^{\frac{\epsilon}{2+\epsilon}}}{2 \beta |\Omega|^{\frac{\epsilon}{2+\epsilon}}} \left[1 + \frac{\beta}{2} |\Omega|^{\frac{\epsilon}{2+\epsilon}}\right] = - \frac{\Pi_0}{2} - \frac{\beta \Pi_0}{4} |\Omega|^{\frac{\epsilon}{2+\epsilon}} \,,
\end{equation}
which is precisely the functional form in Eq.~\eqref{eq:maintext_FLbilayer_JJ}.

\subsection{Diagrammatic calculation of the conductivity at the excitonic QCP of composite fermions}\label{app:diagram_ACFLbilayer}

\begin{figure}[h!]
    \centering
    \includegraphics[width = 0.8\textwidth]{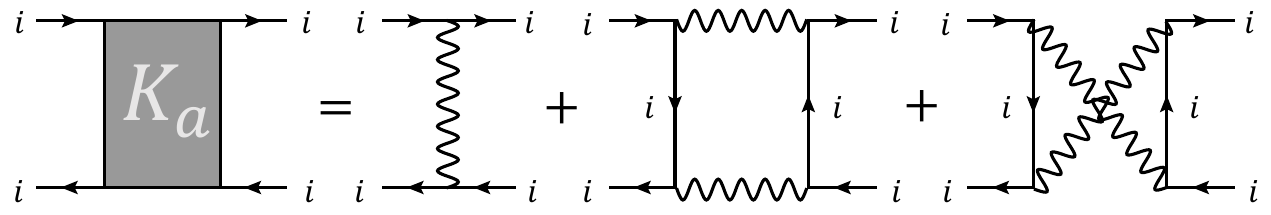}
    \caption{The subkernel $K_a$ connects four fermion lines in the same layer ($i = 1, 2$ are both allowed). The wavy lines are gauge field propagators to be contrasted with the dashed lines in Fig.~\ref{fig:K_Kphi_FLbilayer} which are critical boson propagators.}
    \label{fig:Ka_subkernels}
\end{figure}

In this appendix, we flesh out the calculations that lead to Eq.~\eqref{eq:maintext_ACFLbilayer_JJ} in Section~\ref{subsec:ACFLbilayer_pert}. Using the braket notation introduced in Appendix~\ref{app:diagram_FLbilayer}, the conductivity for the layer-1 composite fermions can be written as a geometric sum of the particle-hole irreducible kernel $K$
\begin{equation}
    G_{J_1 J_1}(\bs{q}=0,\Omega) = \bra{J_{1,\bs{q}=0,\Omega}} W_G \frac{I}{I - K} \ket{J_{1,\bs{q}=0,\Omega}} \,,
\end{equation}
where $W_G$ is again the product of two fermion propagators and $J_1$ is the bare layer-1 current vertex. Following the classification scheme in Ref.~\cite{Shi2023_mross}, we can determine the structure of the kernel $K^{(0)}$ which is not a simple sum of gauge field and critical boson contributions:
\begin{equation}
    K^{(0)} = K_a + K_{11} + K_{12} \frac{I}{I - K_a - K_{22}} K_{21} \,, 
\end{equation}
where $K_{ij}$ are subkernels that already appeared in Appendix~\ref{app:diagram_FLbilayer} and 
\begin{equation}
    K_a = K_{\rm MT, a} + K_{\rm AL, a} \,.
\end{equation}
A diagrammatic representation of these two equations can be found in Fig.~\ref{fig:K_Kphi_FLbilayer} and Fig.~\ref{fig:Ka_subkernels}.

The formula for the kernel $K^{(0)}$ is a simple generalization of the formula Eq.~\eqref{eq:Kform_FLbilayer} for the Fermi liquid bilayer, with the replacements
\begin{equation}
    K_{11} \rightarrow K_{11} + K_a \,, \quad K_{22} \rightarrow K_{22} + K_a \,.
\end{equation}
This replacement makes sense as the gauge field merely provides a set of additional channels that connect particle-hole vertices in the same layer. 

The integral representations of the subkernels $K_a, K_{ij}$ immediately follow from the diagrammatic representation. When acting on a particle-hole vertex $F$ with total energy-momentum $(\omega_1, \bs{k}_1)$ and relative energy-momentum $(\Omega, \bs{q} = 0)$, we have 
\begin{equation}
    \begin{aligned}
    \left(K_{\rm MT, \,a/\phi} F\right)(\bs{k}_1, \omega_1, \Omega) &= g_{a/\phi}^2\, G(\bs{k}_1, i\omega_1 + i\Omega/2)\, G(\bs{k}_1, i\omega_1 - i\Omega/2) \\
    & \hspace{4cm} \int_{\bs{k}_2, \omega_2} D_{a/\phi}(\bs{k}_1 - \bs{k}_2, i\omega_1 - i\omega_2) \,F(\bs{k}_2, \omega_2, \Omega) \,.
    \end{aligned}
\end{equation}
\begin{equation}
    \begin{aligned}
    &\left(K_{\rm AL,\, a/\phi} F\right)(\bs{k}_1, \omega_1, \Omega) = g_{a/\phi}^4\, G(\bs{k}_1, i\omega_1 + i\Omega/2)\, G(\bs{k}_1, i\omega_1 - i\Omega/2) \\
    &\hspace{1cm} \int_{\bs{q}_1, \Omega_1} G(\bs{k}_1-\bs{q}_1, i\omega_1 + i\Omega/2 - i \Omega_1)\, D_{a/\phi}(\bs{q}_1, i\Omega_1)\, D_{a/\phi}(\bs{q}_1, i\Omega_1 - i\Omega) \\
    &\hspace{1cm} \int_{\bs{k}_2, \omega_2} \left[G(\bs{k}_2 - \bs{q}_1, i\omega_2 + i\Omega/2 - i \Omega_1) + G(\bs{k}_2 + \bs{q}_1, i\omega_2 - i\Omega/2 + i \Omega_1)\right]\, F(\bs{k}_2, \omega_2, \Omega) \,.
    \end{aligned}
\end{equation}
The most singular contributions to the layer-1 conductivity again come from near-unit eigenvalues of $K^{(0)}$. From the calculation in Appendix~\ref{app:diagram_FLbilayer}, we know that the contributions from $K_{11}, K_{22}$ are subleading in the IR limit. Therefore, we will consider the simplified kernel 
\begin{equation}
    K^{(0)} \approx K_a + K_{12} \frac{I}{I - K_a} K_{21} \,. 
\end{equation}
Now we observe that $K_a, K_{12}$ are essentially identical up to numerical coefficients in the Landau damping term of $D_{a/\phi}$. Therefore, it suffices to look for eigenvectors of $K_a$ and $K_{12}$ with eigenvalues $k_a, k_{12}$ that satisfy 
\begin{equation}
    1 \approx k_a + \frac{k_{12}^2}{1-k_a} \quad \rightarrow \quad 1 \approx k_a + k_{12} \,.
\end{equation}
This trick reduces the problem to finding soft modes (near-zero eigenvalues) of $1 -K_a - K_{12}$ in the IR limit. Like in Appendix~\ref{app:diagram_FLbilayer}, we specialize to the case with rotational invariance, so that the only soft mode overlapping with the current operator $J_1$ takes the form
\begin{equation}
    \bra{\bs{k}, \omega, \Omega}\ket{F_1} = \cos \theta_{\bs{k}} \left[G(\bs{k}, \omega + \Omega/2) - G(\bs{k}, \omega - \Omega/2)\right] \,.
\end{equation}
One can easily verify that 
\begin{equation}
    \begin{aligned}
    &\bra{\bs{k},\omega,\Omega}K_{a/12} \ket{F_1} \\
    &\approx \frac{i C_{a/\phi} \sgn(\omega + \frac{\Omega}{2})|\omega+\frac{\Omega}{2}|^{\frac{2}{2+\epsilon}} - i C_{a/\phi} \sgn(\omega - \frac{\Omega}{2})|\omega-\frac{\Omega}{2}|^{\frac{2}{2+\epsilon}}}{i \Omega + i (C_a + C_{\phi}) \sgn(\omega + \frac{\Omega}{2})|\omega+\frac{\Omega}{2}|^{\frac{2}{2+\epsilon}} - i (C_a + C_{\phi}) \sgn(\omega - \frac{\Omega}{2})|\omega-\frac{\Omega}{2}|^{\frac{2}{2+\epsilon}}} \bra{\bs{k},\omega,\Omega}\ket{F_1} \,,
    \end{aligned}
\end{equation}
which implies that in the IR limit $\Omega \rightarrow 0$, $\ket{F_1}$ is a simultaneous eigenvector of $K_a$, $K_{12}$, and $K_a + K_{12}$ with eigenvalues $\frac{C_a}{C_a + C_{\phi}}$, $\frac{C_{\phi}}{C_a + C_{\phi}}$, and $1$ respectively. For small but nonzero $\Omega$, the leading $\Omega$-dependent corrections to these eigenvalues can be obtained from first order perturbation theory (see Ref.~\cite{guo2022_largeN,shi2023_loop,guo2023migdal,guo2023fluctuation})
\begin{equation}
    k_a + k_{12} = 1 - \beta |\Omega|^{\frac{2}{2+\epsilon}}\,, \quad \frac{k_a}{k_a + k_{12}}= \frac{C_a}{C_a + C_{\phi}}\,, \quad \frac{k_{12}}{k_a + k_{12}} = \frac{C_{\phi}}{C_a + C_{\phi}} \,.
\end{equation}
After some simple algebraic manipulations, we end up with
\begin{equation}\label{eq:layer1_JJ_correlator_ACFLbilayer}
    \begin{aligned}
    G_{J_1J_1}(\bs{q}=0,\Omega) &= \bra{J_{1,\bs{q}=0,\Omega}} W_G \ket{F_1} \bra{F_1}\ket{J_{1,\bs{q}=0,\Omega}} \left[1 - k_a - \frac{k_{12}^2}{1 - k_{a}} \right]^{-1} \\
    &= \bra{J_{1,\bs{q}=0,\Omega}} W_G \ket{F_1} \bra{F_1}\ket{J_{1,\bs{q}=0,\Omega}} \frac{1 - k_a}{(1-k_a)^2 - k_{12}^2} \\
    &=\bra{J_{1,\bs{q}=0,\Omega}} W_G \ket{F_1} \bra{F_1}\ket{J_{1,\bs{q}=0,\Omega}} \frac{1 - k_a - k_{12} + k_{12}}{(1-k_a - k_{12})^2 + 2(1-k_a - k_{12}) k_{12}} \\
    &= \frac{\bra{J_{1,\bs{q}=0,\Omega}} W_G \ket{F_1} \bra{F_1}\ket{J_{1,\bs{q}=0,\Omega}}}{2 \beta |\Omega|^{\frac{\epsilon}{2+\epsilon}}} \frac{1 + (1-k_a - k_{12})/k_{12}}{1 + (1-k_a - k_{12})/(2k_{12})} \,.
    \end{aligned}
\end{equation}
Finally invoking the single layer transport result 
\begin{equation}
    \bra{J_{1,\bs{q}=0,\Omega}} W_G \ket{F_1} \bra{F_1}\ket{J_{1,\bs{q}=0,\Omega}} = - \beta \Pi_0 |\Omega|^{\frac{\epsilon}{2+\epsilon}} \,,
\end{equation}
we obtain a compact formula for $G_{J_1J_1}$
\begin{equation}
    G_{J_1J_1}(\bs{q}=0,\Omega) \approx - \frac{\Pi_0}{2} \left(1 + \frac{1-k_a - k_{12}}{2k_{12}}\right) \approx -\frac{\Pi_0}{2} \left(1 + \frac{(C_a + C_{\phi})\beta}{2C_{\phi}} |\Omega|^{\frac{\epsilon}{2+\epsilon}}\right) \,.
\end{equation}
The main result Eq.~\eqref{eq:maintext_ACFLbilayer_JJ} in Section~\ref{subsec:ACFLbilayer_pert} is recovered once we define the constant $\beta' = \frac{\beta (C_a + C_{\phi})}{C_{\phi}}$.

\subsection{Subleading contributions to the conductivity from the bosonic sector}\label{app:bosonic_conductivity}

In both Section~\ref{subsec:FLbilayer_pert} and~\ref{subsec:ACFLbilayer_pert}, we neglected the bosonic contribution to the conductivity $\sigma_{\phi}$. Here, we justify this omission by estimating the scaling of leading diagrams that contribute to the boson current-current correlation function. 

We start by identifying the boson current operator in the Fermi liquid bilayer. Within the double expansion, coupling to an external gauge field $\bs{A}_-$ modifies the non-local kinetic term as
\begin{equation}
    |\bs{q} + \bs{A}_-|^{1+\epsilon} \approx \left[q^2 + \bs{q} \cdot \bs{A}_- + A_-^2\right]^{\frac{1+\epsilon}{2}} \approx q^{1+\epsilon} \left[1 + \frac{\bs{q} \cdot \bs{A}_-}{q^2} + \frac{A_-^2}{q^2}\right]^{\frac{1+\epsilon}{2}} \approx q^{1+\epsilon} + \frac{1+\epsilon}{2} q^{\epsilon - 1} \bs{q} \cdot \bs{A}_- \,.
\end{equation}
Therefore, the current operator is identified as
\begin{equation}
    \bs{J}_{\phi}(\bs{q}) = \frac{1+\epsilon}{2} \int_{\bs{q}'} \phi^*(\bs{q}' + \bs{q}/2)\, |\bs{q}'|^{\epsilon-1}\, \bs{q}'\, \phi(\bs{q}' - \bs{q}/2) \,.
\end{equation}
Knowing the IR propagators $D_{\phi}$ for the critical boson $\phi$, we can immediately estimate the scaling of $G_{J_{\phi} J_{\phi}}$ from the one-loop bubble diagram, using $z = 2 + \epsilon$ dynamical scaling
\begin{equation}\label{eq:app_boson_conduct1}
    \begin{aligned}
    G_{J_{\phi} J_{\phi}}(\bs{q}=0, \Omega) &\sim \int_{\bs{q}', \Omega'} |\bs{q}'|^{2\epsilon} D_{\phi}(\bs{q}', i\Omega' + i \Omega/2) D_{\phi}(\bs{q}', i\Omega' - i \Omega/2) \\
    &= \int_{\bs{q}', \Omega'} |\bs{q}'|^{2\epsilon} \frac{1}{|\bs{q}'|^{1+\epsilon} + \frac{|\Omega' - \Omega/2|}{|\bs{q'}|}} \cdot \frac{1}{|\bs{q}'|^{1+\epsilon} + \frac{|\Omega' + \Omega/2|}{|\bs{q'}|}}  \\
    &\sim \Omega^{\frac{2}{2+\epsilon} + 1} \cdot \Omega^{\frac{2\epsilon}{2+\epsilon}} \cdot \Omega^{- \frac{2+2\epsilon}{2+\epsilon}} \sim \Omega^{\frac{2+\epsilon}{2+\epsilon}} = \Omega \,.
    \end{aligned}
\end{equation}
The scaling above is indeed subleading relative to the fermionic contribution computed in Section~\ref{subsec:FLbilayer_pert}. 

For the CFL-$\overline{\text{CFL}}$ bilayer in Section~\ref{subsec:ACFLbilayer_pert}, the bosonic current operator includes an additional contribution from the gauge field $\bs{a}_-$. By coupling to the external gauge field $\bs{A}$, we find that
\begin{equation}
    \begin{aligned}
    &|\bs{q} + \bs{a}_- + \bs{A}_-|^{1+\epsilon} \\
    &\approx |\bs{q} + \bs{a}_-|^{1+\epsilon} + \frac{1+\epsilon}{2} |\bs{q} + \bs{a}_-|^{\epsilon-1} (\bs{q}+\bs{a}_-) \cdot \bs{A}_- \\
    &= |\bs{q} + \bs{a}_-|^{1+\epsilon} + \frac{1+\epsilon}{2} (q^2 + \bs{q} \cdot \bs{a}_- + a_-^2)^{\frac{\epsilon-1}{2}} (\bs{q} + \bs{a}_-) \cdot \bs{A}_- \\
    &= |\bs{q} + \bs{a}_-|^{1+\epsilon} + \frac{1+\epsilon}{2} q^{\epsilon-1} \bs{q} \cdot \bs{A}_-  \\
    &\hspace{2cm} + \frac{1+\epsilon}{2} q^{\epsilon-1} \bs{a}_- \cdot \bs{A}_- + \frac{(1+\epsilon)(\epsilon-1)}{4} q^{\epsilon-3} (\bs{q} \cdot \bs{a}_-) (\bs{q} \cdot \bs{A}_-) + \mathcal{O}(a_-^2, A_-^2)\,. 
    \end{aligned}
\end{equation}
The last line identifies the gauge-invariant current operator as
\begin{equation}
    \begin{aligned}
    &\bs{J}^{(\rm gauge)}_{\phi} \\
    &= \bs{J}_{\phi} + \int_{\bs{q}'} \phi^*(\bs{q'}+\bs{q}/2) \left[\frac{1+\epsilon}{2} |\bs{q}'|^{\epsilon-1} \bs{a}_- + \frac{(1+\epsilon)(\epsilon-1)}{4} |\bs{q}'|^{\epsilon-3} (\bs{q'} \cdot \bs{a}_-) \bs{q'}\right] \phi(\bs{q'}-\bs{q}/2) \,.
    \end{aligned}
\end{equation}
The scaling of $G_{J_{\phi} J_{\phi}}$ can be inferred from the Fermi liquid bilayer result. We therefore focus on the $\bs{a}_-$-dependent contribution $G_{J^{\rm gauge} - J_{\phi}, J^{\rm gauge} - J_{\phi}}$. The leading contribution to this correlator is a two-loop diagram with two $\phi$ propagators and one $\bs{a}_-$ propagator. The scaling can be estimated as
\begin{equation}\label{eq:app_boson_conduct2}
    \begin{aligned}
    &G_{J^{\rm gauge} - J_{\phi}, J^{\rm gauge} - J_{\phi}}(\bs{q}=0, \Omega) \\
    &\sim \int_{\bs{q}_1, \Omega_1} \int_{\bs{q}_2, \Omega_2} |\bs{q}_1 + \bs{q}_2|^{2\epsilon-2} D_{\phi}(\bs{q}_1, i\Omega_1)\, D_{\phi}(\bs{q}_2, i\Omega_2)\, D_{a_-}(\bs{q}_1 + \bs{q}_2, i\Omega_1 + i\Omega_2 -i \Omega) \\
    &= \int_{\bs{q}_1, \Omega_1} \int_{\bs{q}_2, \Omega_2} |\bs{q}_1 + \bs{q}_2|^{2\epsilon-2}\, \frac{1}{|\bs{q}_1|^{1+\epsilon} + \frac{|\Omega_1|}{|\bs{q}_1|}} \cdot \frac{1}{|\bs{q}_2|^{1+\epsilon} + \frac{|\Omega_2|}{|\bs{q}_2|}} \cdot \frac{1}{|\bs{q}_1+\bs{q}_2| + \frac{|\Omega_1+\Omega_2 - \Omega|}{|\bs{q}_1+\bs{q}_2|}} \\
    &\sim \Omega^{\frac{4}{2+\epsilon} + 2} \cdot \Omega^{\frac{2\epsilon-2}{2+\epsilon}} \cdot \Omega^{-\frac{1+\epsilon}{2+\epsilon}} \cdot \Omega^{-\frac{1+\epsilon}{2+\epsilon}} \cdot \Omega^{-\frac{1}{2+\epsilon}} \sim \Omega^{\frac{3 + \epsilon}{2+\epsilon}} \,.
    \end{aligned}
\end{equation}
Eq.~\ref{eq:app_boson_conduct1} and Eq.~\ref{eq:app_boson_conduct2} constitute the scaling results quoted in Sections~\ref{subsec:FLbilayer_pert} and~\ref{subsec:ACFLbilayer_pert}. 

\nocite{apsrev41Control}
\bibliographystyle{apsrev4-1}
\bibliography{bibilayer}

\end{document}